\documentclass[hyper,11pt,letterpaper]{JHEP3}

\usepackage{ifpdf}
\usepackage{graphicx}
\usepackage{amsmath, amssymb}
\usepackage{amsfonts}
\usepackage{indentfirst}\usepackage[enableskew]{youngtab}
\usepackage{slashed} 
\usepackage{amsbsy}

\newcommand{\ba}{\begin{align}}
\newcommand{\ea}{\end{align}}
\newcommand{\bea}{\begin{eqnarray}}
\newcommand{\eea}{\end{eqnarray}}
\newcommand{\be}{\begin{eqnarray}}
\newcommand{\ee}{\end{eqnarray}}
\newcommand{\nn}{\nonumber}
\newcommand{\bn}{\begin{enumerate}}
\newcommand{\en}{\end{enumerate}}

\def\eg{{\it e.g.~}}
\newcommand{\ie}{{\it i.e.~}}
\newcommand{\one}{\mathbf{1}}
\newcommand{\two}{\mathbf{2}}

\newcommand{\eo}{\epsilon_1}
\newcommand{\et}{\epsilon_2}
\newcommand{\bp}{\mathbf{+}}
\newcommand{\bm}{\mathbf{-}}

\newcommand{\qSp}{q_{Sp(1)}}
\newcommand{\qU}{q_{U(2)}}

\newcommand{\qSO}{q_{SO(4)}}
\newcommand{\PH}{P_{\CH_\phi}}

\newcommand{\qCFT}{q}
\newcommand{\Ccover}{\mathbf{\tilde C}}
\newcommand{\Cbase}{\mathbf{ C}}

\def\IC{\mathbb{C}}

\def\IN{\mathbb{N}}
\def\IP{\mathbb{P}}
\def\IR{\mathbb{R}}
\def\IZ{\mathbb{Z}}

\def\CE{{\cal E}}
\def\CF{{\cal F}}

\def\CH{{\cal H}}

\def\CL{{\cal L}}
\def\CM{{\cal M}}
\def\CN{{\cal N}}
\def\CO{{\cal O}}

\def\CS{{\cal S}}

\def\CV{{\cal V}}
\def\CW{{\cal W}}

\def\e{\epsilon}

\def\i{\iota}

\def\m{\mu}

\def\w{\omega}
\def\Tr{{\rm Tr}}

\def\det{{\rm det}}

\def\Ch{{\rm Ch}}
\def\Ind{{\rm Ind}}
\def\Td{{\rm Td}}

\title{ From SO/Sp instantons to W-algebra blocks}

\author{Lotte Hollands\footnote{hollands@theory.caltech.edu}\ , Christoph
  A.~Keller\footnote{ckeller@theory.caltech.edu}\ , Jaewon Song\footnote{jaewon@theory.caltech.edu}
\\
\\
California Institute of Technology, Pasadena, CA 91125, USA}

\abstract{We study instanton partition functions for
  $\CN=2$  superconformal $Sp(1)$ and $SO(4)$ gauge theories. 
We find that they
  agree with the corresponding $U(2)$ instanton partitions functions
  only after 
  a non-trivial mapping of the microscopic gauge couplings, since
  the instanton counting involves different renormalization
  schemes. 
Geometrically, this mapping relates the Gaiotto
  curves of the different realizations as double coverings.
We then formulate an AGT-type correspondence between $Sp(1)/SO(4)$
instanton partition functions and chiral blocks with an underlying
$\CW(2,2)$-algebra symmetry. This form of the correspondence
eliminates the need to divide out extra $U(1)$ factors.
Finally, to check this correspondence for linear quivers, we
compute expressions for the $Sp(1) \times SO(4)$ half-bifundamental. 
}

\preprint{
CALT-68-2795
}

\begin{document}


\section{Introduction}

Over the last year a substantially deeper understanding  has been
obtained of S-duality in four-dimensional supersymmetric gauge 
theories and of the relation to two-dimensional geometry and conformal
field theory (see for instance \cite{Gaiotto:2009we,AGT,Nekrasov:2009rc,Nekrasov:2010ka,Dijkgraaf:2009pc}).   

The simplest example of S-duality 
appears in $\CN=4$ supersymmetric gauge 
theory. Such a gauge theory is characterized purely
by the choice of a 
gauge group $G$ and a value for the complexified gauge
coupling   
$\tau$. The theory has long been known to be invariant under
$SL(2,\IZ)$ transformations of the coupling $\tau$, which can be
geometrically realized as M\"obius transformations of a two-torus
$T^2$ with complex structure parameter $\tau$.  

S-duality in $\CN=4$ gauge theory translates into modular properties of
its instanton partition function. In
particular, the $\CN=4$ instanton
partition function for gauge group $G$ is argued to be a character for
the affine Lie algebra $\hat{\mathfrak{g}}$
\cite{Vafa:1994tf,Nakajima:affine,Nakajima:ALE}. This character also 
appears as the partition function of a two-dimensional conformal field
theory with
current algebra $\hat{\frak{g}}$ on the two-torus $T^2$. This implies
a close relation between $\CN=4$ gauge theory and two-dimensional
conformal field theory
on $T^2$. Such a connection is most naturally understood by embedding
the $\CN=4$ gauge theory on a set of M5-branes wrapping the
four-manifold times the two-torus $T^2$ \cite{Dijkgraaf:2007sw,Witten:2009at}.

Supersymmetric $\CN=2$ gauge theories are much richer than their
$\CN=4$ counterparts. Not only can different kinds of matter multiplets
be added to the theory, but even for conformal theories 
the gauge coupling receives a finite renormalization.
In the low-energy limit the gauge theory is characterized by a
two-dimensional Seiberg-Witten curve, whose geometry captures
the prepotential of the gauge theory as well as the masses of BPS
particles \cite{Seiberg:1994rs,Seiberg:1994aj}. 
Using a brane
realization of the $\CN=2$ gauge theory in string or
M-theory, the 
Seiberg-Witten curve naturally comes about as a 
branched covering over yet another two-dimensional curve
\cite{Witten:1997sc}. We will refer to this base curve as the Gaiotto
curve  (or G-curve). Mathematically, the $\CN=2$
geometry is encoded in a ramified Hitchin system. 

It was realized recently that it is important to not
forget about the additional information contained in the covering
structure of the Seiberg-Witten curve. In particular, it is
conjectured that the complex moduli 
space of the Gaiotto curve is the parameter space of the exactly marginal
couplings of the superconformal $\CN=2$ gauge
theory \cite{Gaiotto:2009we}.
Analogous to the $\CN=4$ example, this suggests good transformation
properties of the $\CN=2$ partition function under mapping
class group transformation of the Gaiotto curve. Furthermore, it hints at an
extension of the relation between four-dimensional gauge theories and
two-dimensional conformal field theories to $\CN=2$ supersymmetric gauge theories.

Indeed, such a relation between $\CN=2$ gauge theory and
two-dimensional conformal field theory has been found \cite{AGT}, and
is referred to 
as the AGT correspondence. In particular, an
equivalence was discovered between $U(2)$ instanton partition
functions and Virasoro conformal blocks. This was extended to asymptotically free theories
\cite{Gaiotto:2009ma}, to $U(N)$ theories \cite{Wyllard:2009hg} and to
inclusion of surface 
operators (see amongst others
\cite{Gaiotto:2009fs,Alday:2009fs,Kozcaz:2010af,Dimofte:2010tz,Alday:2010vg,Kozcaz:2010yp,BravermanfiniteAGT,Wyllard:2010vi,Bruzzo:2010fk}).

Nonetheless, a few important open questions remain, such as a
good understanding through the M5-brane picture, the extension
to generalized quivers and general gauge groups, and a better
explanation of
 the spurious $U(1)$ factor which appears in the AGT
correspondence. In this paper we start by tackling the 
last question, and gradually gain more insight in the correspondence
between four-dimensional $\CN=2$ gauge theories and two-dimensional
conformal field theory. 

In \cite{AGT} it was found that after a suitable
identification of parameters the conformal block
agrees with an instanton partition function for gauge group $U(2)$,
whose Coulomb branch
parameters are specialized to $SU(2)$ values,
up to a spurious factor. This spurious
factor closely resembles the partition function
of a $U(1)$ gauge theory.
The interpretation given was therefore
that the partition function of $U(2)$ factorizes
into an $SU(2)$ part which corresponds to the
conformal block, and a $U(1)$ part which decouples \cite{Alday:2010vg}.

There is no direct way to compute
instanton partition functions for $SU(N)$ theories.
What one does instead is to consider the construction
for $U(N)$ and then impose tracelessness of the
Coulomb branch parameters in the end. 
From this
point of view one might expect the appearance of an additional
$U(1)$ factor, which corresponds to the overall
$U(1)$ that somehow decouples.

For $SU(2)$ however there is a
direct computation of the partition function,
which uses the fact that $SU(2) = Sp(1)$.
A similar situation arises for $SO(4) \cong SU(2)\times SU(2)$.
In both cases we can thus obtain the partition function
either by a direct computation,
or by computing the corresponding $U(2)$ partition
functions and splitting off $U(1)$ factors.
The computation of the $Sp(1)$ and $SO(4)$ partition
functions are technically more difficult than the $U(2)$ 
case. We describe a procedure for the computation
in appendix~\ref{app:instPoles}, which we use
to compute the $Sp(1)$ result up to instanton number $k=6$.
Somewhat surprisingly, however, we find 
that for the conformal version of the theories with hypermultiplets,
the computations naively look completely different. In particular, it
does not seem 
possible to split off a $U(1)$ factor to make
them agree.

We argue in this article that this difference is
due to the fact that in the two computations one implicitly chooses a
different renormalization scheme. 
More precisely, from our computation of the
instanton contributions to the prepotential
of these theories we can read off the relation
between the microscopic gauge couplings $q$ in the UV 
and the gauge couplings $\tau$ in the IR.
Once we express the instanton partition functions in terms of infrared
variables, 
it turns out that both expressions agree. The
different choices of renormalization schemes 
therefore correspond
to different parametrizations of the moduli space
of the theory in terms of microscopic gauge couplings.
In the case at hand we find in fact a very simple
relation between the two parametrizations.
We also the discuss in detail the situation for
$SO(4)$ as opposed to $U(2)\times U(2)$, which
leads to similar results, and we argue that we can expect such a relation
for any two different appearances of the same physical gauge theory.

The fact that we find such simple relations between
the UV gauge couplings implies that there should be
a geometric interpretation. In section~\ref{sec:N=2geometry} we thus
turn to a string theory embedding of those
gauge theories. More precisely, we discuss
the underlying Gaiotto curve of the theories,
whose complex structure moduli are given by
the UV couplings $q$. From a string theory point
of view $U(2)$ and $Sp(1)$  naturally lead to
two different G-curves. In fact we show that
the $U(2)$ curve is the double cover of 
the $Sp(1)$ curve, and their complex structure moduli
are related in exactly the way we obtained
from the instanton computation. Again,
the situation for $SO(4)$ and
$U(2)\times U(2)$ is completely analogous.

\begin{figure}[h!]
\begin{center}
\includegraphics[width=1.5in]{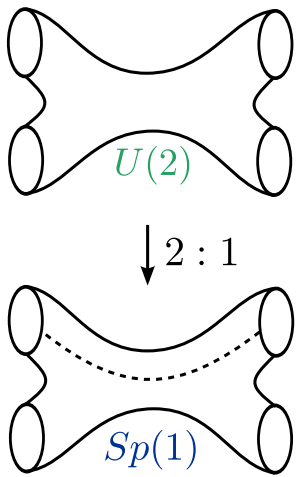}
\caption{The Gaiotto curve of the $U(2)$ gauge theory coupled to
4 hypers as a double cover over the Gaiotto curve of 
the $Sp(1)$ theory. The dotted line is a branch cut.}
\end{center}
\end{figure}

Finding a geometric interpretation of the 
relation between those theories leads us to the second focus
of this article. We want to find an AGT-like relation
for more general gauge groups, like $Sp(1)$ and $SO(4)$, \ie we want
to find a configuration of a conformal field theory whose chiral block
agrees directly with 
the $Sp(1)$ or $SO(4)$ partition function.

For general reasons we expect the symmetry of the CFT
to be related to the gauge group $G$. More precisely, we argue that the
Lie algebra structure of the Hitchin system 
 is reflected in the $\CW$-algebra that 
describes the symmetry of the CFT, in such a way that the generators
of the $\CW$-algebra are determined
by the Casimirs of the Lie algebra $\mathfrak{g}$. In
section~\ref{sec:cft},  
we find configurations whose $\CW$-blocks
agree with the $Sp(1)$ and $SO(4)$ partition functions,
which we check for the first few orders by explicit computation.
As expected, no additional $U(1)$ factor appears
for this kind of AGT-like relation.

We also find a natural interpretation for the double
cover map found geometrically from the conformal field theory side: The 
CFT configurations we consider involve
$\IZ_2$-twist fields and lines. 
The standard method to compute such correlators is to
map them to a double cover, which gives a configuration
without any twist fields that is straightforward to 
compute. For $Sp(1)$ and $SO(4)$ it turns out that 
the configuration on the cover is exactly the Virasoro
conformal block on the four punctured sphere and
the two punctured torus, respectively, which were found
in \cite{AGT} to correspond to the conformal $U(2)$
and $U(2)\times U(2)$ gauge theories.

Lastly, in section~5 we then formulate what the correspondence
will look like for linear quivers. That
is, quivers with alternating $Sp/SO$
gauge groups coupled by bifundamental half-hypermultiplets. 
We write down the instanton partition function for
these bifundamental fields and check that the result 
indeed agrees with the chiral block of the 
corresponding CFT configuration.

Additional information can be found in the appendices. 
In appendix~\ref{app:Instanton} we review the Nekrasov method of
instanton counting in 
detail and 
give the derivations of the formulae that we use in the main body of
the paper. In appendix~\ref{app:instPoles}, we give a detailed
explanation of how we 
evaluate the (refined) instanton partition function for $Sp/SO$
theories up to order 6 in the instanton parameter. Finally 
in appendix~\ref{app:SW}, we compare $SU(2)$ Seiberg-Witten curves from different perspectives.  \\

\emph{Notation:} Let us explain our conventions for naming gauge
groups to avoid confusion. We will say $Sp(1)$ if we refer to
using the $Sp(1)$ renormalization scheme, and $U(2)$
if we refer to the $U(2)$
renormalization scheme with Coulomb branch parameters
specialized to $a=a_1=-a_2$.

\section{Instanton counting for $Sp/SO$ versus $U$
 gauge groups}\label{sec:gauge} 

In this section we explore $\CN=2$ instanton partition functions for
 symplectic and orthogonal gauge groups. We compare this to instanton
 counting for unitary gauge groups in cases when there exist
 both a unitary and a symplectic/orthogonal way of counting instantons.
 Examples of such specific instances are $Sp(1)$ versus 
 $U(2)$, and $SO(4)$ versus  $U(2) \times U(2)$. 
The two ways of counting
 instantons are not
 obviously the same, since they are based on a 
 different realization of the instanton moduli space. Nonetheless,
 the two ways should give physically equivalent results in order for instanton counting for
 general gauge groups to make 
 sense. 

In this section we compare instanton partition functions for such
specific instances. It turns out that 
the instanton partition functions for conformal gauge theories with
these gauge groups are not equal on the nose. More precisely, they cannot
even be related by a factor that is independent of the gauge theory
moduli. We resolve this apparent disagreement by carefully examining
the dependence of the instanton partition function on the microscopic
versus physical gauge couplings. We find that instanton counting for
different appearances of a gauge group are in general related by the
choice of an inequivalent renormalization scheme.   

One of the examples that we study in detail is $Sp(1)$ versus $U(2)$
instanton counting.  We perform the
instanton computations directly for $Sp(1)$, and then compare
the results to the much better understood $U(2)$ instanton partition
functions.  Our motivation for studying this particular
example is to gain a better understanding of the spurious factors that
appear in the relation between $U(2)$ instanton partition functions
and Virasoro blocks in the AGT correspondence. We will return to this
issue in section~\ref{sec:cft}, which also contains a summary of
the AGT correspondence.

\subsection{Instanton counting}\label{subsec:insta2}

At low energies the four-dimensional $\CN=2$ gauge theory is
governed by the prepotential $\CF_0$, which determines the metric on
the Coulomb branch of the gauge theory. Classically, the metric on the
Coulomb branch is flat and the prepotential 
\be
\CF_0^{\rm clas} = 
 2\pi i  \, \tau_{UV} \, \vec{a} \cdot \vec{a},
\ee
is proportional to the microscopic coupling constant $\tau_{\rm
  UV}$. At the quantum level the prepotential receives both
one-loop and non-perturbative instanton corrections, which give
corrections to the metric on the
Coulomb moduli space. The instanton corrections to the prepotential
can be computed as equivariant integrals over the instanton moduli
space \cite{Nekrasov:2002qd}. Let 
us briefly sketch how 
this comes about. We refer to appendix~\ref{app:Instanton} for a
detailed discussion. 

Instantons on $\IR^4$ are solutions of the self-dual instanton equation
\be
F^+_A=0. 
\ee
The instanton moduli space $\CM^G$ parametrizes
these solutions up to gauge transformations that leave the fiber at
infinity fixed. The components $\CM^G_k$ of the instanton moduli space
are labeled by the  
topological instanton number $k = 1/8\pi^2  \int F_A \wedge F_A$. 
The instanton corrections to the prepotential for the pure $\CN=2$
gauge theory are captured by the instanton partition function
\be \label{instpartfunctionlocpure}
Z^{\rm inst} = \sum_k q^k \oint_{\CM^{G}_k} 1,
\ee
where $\oint 1$ formally computes the volume of the moduli space. 
The parameter $q$ can be considered as a
formal parameter which counts the number of instantons. Physically, it
is identified with a power $q = \Lambda^{b_0}$ of the dynamically
generated scale $\Lambda$, when the gauge theory is asymptotically
free. The power $b_0$ is determined by the one-loop
$\beta$-function. It is 
identified with an exponent $q = \exp(2 \pi i \tau_{\rm 
  UV})$ of the microscopic 
coupling $\tau_{\rm UV}$ when the beta-function of the gauge 
theory vanishes.  

If we introduce hypermultiplets to the pure $\CN=2$ gauge theory, the
instanton correction to the prepotential are instead determined by
solutions of the monopole equations 
\begin{align}
F_{A,\mu \nu}^+ + \frac{i}{2} \ \overline{q}_{\alpha}
\Gamma^{\hspace{3.5mm}\alpha}_{\mu \nu \hspace{2mm} \beta}
q^{\beta}  &= 0, \\  
\sum_{\mu} \Gamma^{\mu}_{\dot{\alpha} \alpha} D_{A,\mu} q^{\alpha} &=0.  \notag
\end{align}
In these equations $\Gamma^{\mu}$ are the Clifford matrices and $\sum_{\mu}
\Gamma^{\mu} D_{A,\mu}$ is the Dirac operator in the instanton
background for the gauge field $A$. 
Although there are no positive
chirality solutions to the Dirac equation, the vector space of negative
chirality solutions is $k$-dimensional. Because this vector space 
depends on the gauge background $A$, it is useful to view it as a 
$k$-dimensional vector bundle over the instanton moduli space
$\CM^{G}_k$. We will
call this vector bundle $\CV$. More precisely, since the solutions to
the Dirac equations are naturally twisted by the half-canonical line bundle 
$\CL$ over $\IR^4$ we will denote it by $\CV \otimes \CL$. 

Instanton corrections to the $\CN=2$ gauge theory,
with $N_f$ hypermultiplets in the fundamental representation of the
gauge group, are computed by the instanton partition function
\be \label{instpartfunctionloc}
Z^{\rm inst} = \sum_k q^k \oint_{\CM^{G}_k} e(\CV \otimes \CL \otimes M),
\ee
which is the integral of the Euler class of the vector bundle $\CV \otimes
\CL$ of solutions to the Dirac equation over the moduli space
$\CM^G_k$. The flavor vector space $M = \IC^{N_f}$ encodes the number of
hypermultiplets in the gauge theory. 

A difficulty in the evaluation of the instanton partition functions
(\ref{instpartfunctionlocpure}) and (\ref{instpartfunctionloc}) is
that the instanton moduli space $\CM^G_{k}$ both suffers from an
UV and an IR non-compactness. Instantons can become arbitrary small,
as well as move away to infinity in $\IR^4$. The IR
non-compactness can be solved by introducing the $\Omega$-background, which
refers to the action of the torus   
\be 
\mathbf{T}^2_{\e_1,\e_2}  = U(1)_{\e_1} \times U(1)_{\e_2}
\ee
on $\IR^4 = \IC \oplus \IC$ by a rotation $(z_1,z_2) \mapsto (e^{i
  \e_1} z_1, e^{i \e_2} z_2)$ around the origin with parameters $\e_1,
\e_2 \in \IC$. 
If we localize the instanton partition function
equivariantly with respect to the $\mathbf{T}^2_{\e_1,\e_2}$-action, only
instantons at the fixed origin will contribute, so that we can ignore
the instantons that run off to infinity.  
The UV non-compactness can be cured for gauge group $U(N)$ by turning
on an FI parameter. For $Sp$ and $SO$ gauge groups it is shown
in \cite{Nekrasov:2004vw} how to evaluate the
instanton integrals, while implicitly curing the UV
non-compactness of the instanton moduli space.
Note that this effectively means that we have introduced
a renormalization scheme.

Apart from the torus $\mathbf{T}^2_{\e_1,\e_2}$ there are
a few other groups that act on the instanton moduli space $\CM^G_k$. 
Their actions can be understood best from the famous ADHM
construction of the instanton moduli space \cite{ADHM}. 
This construction gives the moduli space as the quotient
of the solutions of the ADHM equations by
the so-called dual group $G^D_k$ with
Cartan torus $\mathbf{T}^k_{\phi_i}$
whose weights we will call $\phi_i$.
There is a also natural action of the Cartan torus $\mathbf{T}^N_{\vec{a}}$
of the framing group $G$ on the ADHM solution space, whose weights
are given by the 
Coulomb branch parameters $\vec{a}$. 
Last, if the theory contains hypermultiplets,
there is furthermore an action of the Cartan $\mathbf{T}^{N_f}_{\vec{m}}$
of the flavor symmetry group acting on $M$, whose weights correspond
to the masses $\vec{m}$ of the hypers.

In total, we want to compute the partition function equivariantly with
respect to the torus
\be
\mathbf{T} = \mathbf{T}^2_{\e_1,\e_2} \times \mathbf{T}^N_{\vec{a}} \times
\mathbf{T}^k_{\phi_i} \times \mathbf{T}^{N_f}_{\vec{m}},
\ee
which comes down to computing 
the equivariant character of the action of those four tori. This
results in a rational function $\mathbf{z}^k(\phi_i,\vec{a}, \vec{m}, \e_1, \e_2)$
of the weights. 
From the construction of the Dirac bundle it is clear
that $\textbf{z}^k$ factorizes if there are multiple hypers.
Finally, we need to take into
account the ADHM quotient.
This we do by integrating over the dual group $G^D_k$.
In total the instanton partition function is given by the
integral 
\be \label{contour}
Z^{\rm inst}_k = \int \prod_i d\phi_i \, \mathbf{z}^k_{\textrm{gauge}} (\phi_i,
\vec{a}, \e_1, \e_2) \, \mathbf{z}^k_{\textrm{matter}} (\phi_i,\vec{a}, \vec{m}, \e_1, \e_2),
\ee
where all the $\CN=2$ multiplets in the gauge theory give a separate
contribution. 
The instanton partition function of in principle any $\CN=2$ gauge theory
with a Lagrangian prescription can be computed in this way. Explicit
expressions can be found in 
appendix~\ref{app:Instanton}. 


The integrand of (\ref{contour}) will have poles on the real axis.
To cure this we will introduce small positive imaginary parts
for the equivariance parameters. At least for
asymptotically free theories we can then convert (\ref{contour}) into
a contour integral, so that the problem reduces to 
enumerating poles and
evaluating their residues.
For $U(N)$ theory the poles  are labeled by $N$ Young diagrams 
with in total $k$
boxes \cite{Nekrasov:2002qd, Nakajima:2003pg, NakajimaHilb}: one way
of phrasing this is that the $U(N)$ instanton splits into $N$
non-commutative $U(1)$ instantons. 

For the $Sp(N)$ or $SO(N)$ theory it is not that simple to enumerate
the poles of the contour integrals. Furthermore, not only the 
fixed points of the gauge multiplet are more complicated, but 
(in contrast to the $U(N)$ theory) also matter multiplets contribute
additional poles.  As an 
example, in appendix~\ref{app:instPoles} we devise a technique to
enumerate all the poles for an $Sp(N)$ gauge multiplet. Each pole
can still be expressed as a generalized diagram with signs, but the
prescription is much more involved than in the $U(N)$ case.

The instanton partition function $Z^{\rm inst}$ in the 
$\Omega$-background obviously depends on the equivariant
parameters $\e_1$ and $\e_2$. In fact, it is rather easy to see that the series
expansion of $\log (Z^{\rm inst})$ starts out with a term proportional to 
$\frac{1}{\e_1 \e_2}, $
which is the regularized volume of the $\Omega$-background. Even
better, $Z^{\rm inst}$ has a series 
expansion\footnote{To find this expansion in merely even powers of
  $\hbar$ 
  it is crucial to study  
  the twisted kernel of the Dirac operator, in contrast to the kernel
  of the Dolbeault operator. This twist is ubiquitous in the theory of
  integrable systems. Mathematically, it has been emphasized in this
  setting in 
  \cite{Gottsche:2010}. Physically, it corresponds to a mass shift
  $m \to m + \frac{\e_1 + \e_2}{2}$. This mass shift was studied in
  several related contexts, see \ie \cite{Okuda:2010ke,Krefl:2010jb}. An
  exception to the above expansion is the 
$U(N)$ theory which has a 
  non-vanishing contribution $\frac{1}{\hbar} \CF_{1/2}^{\rm
    inst}(\beta)$.  }    
\begin{align}
Z^{\rm inst} = \exp \CF^{\rm inst} = 
\exp \left(  
\sum_{g=0}^{\infty} \hbar^{2g-2} \CF^{\rm
    inst}_{g} (\beta)
\right), 
\end{align}
in terms of the parameter $\hbar^2= - \e_1 \e_2$ and $\beta =
-\frac{\e_1}{\e_2}$. We call the exponent of the instanton partition
function the instanton free energy~$\CF^{\rm inst}$.  
As our notation  
suggests, we recover the non-perturbative instanton contribution to
the prepotential $\CF_0$ from the leading contribution of the exponent
when $\hbar \to 0$. This has been showed in \cite{Nekrasov:2003rj,
  Nakajima:2003pg, 
  Braverman:2004cr}. Let us emphasize that the prepotential $\CF_0$
does not depend on the parameter $\beta$. The higher genus 
free energies $\CF_{g \ge 1}(\beta)$ compute 
gravitational couplings to the $\CN=2$ gauge theory, and play an
important role in for example (refined) topological string theory.

To recover the full prepotential, we need to add
classical and 1-loop contributions to the instanton partition
function. We call the complete partition function 
\begin{align}\label{eqn:Nekpartfunc}
Z^{\rm Nek}=  Z^{\rm clas} Z^{\rm
  1-loop} Z^{\rm inst}
\end{align}
the \emph{Nekrasov partition function}.

\subsection{Infrared versus ultraviolet}\label{subsec:N=2gauge}

Let us now turn to the goal of this section, which is comparing
Nekrasov partition functions for gauge theories whose gauge group
can be represented in two ways, possibly differing by a $U(1)$
factor. Think for instance of $Sp(1)$ versus $U(2)$ 
or $SO(4)$ versus $U(2) \times U(2)$. With a view on the AGT
correspondence we are particularly keen on comparing $Sp(1)$ and $U(2)$ partition
functions for conformally invariant theories. Naively, we would expect 
that the difference simply reproduces the ``$U(1)$ factor''. As we report in
subsection~\ref{sec:instantonexamples}, however, the Nekrasov
partition functions of the $Sp(1)$ and $U(2)$ gauge theory coupled to
four hypermultiplets are not at all related in such a simple
way.

To find a resolution of this disagreement, we should keep in mind the
difference between infrared and ultraviolet quantities. Whereas the
Nekrasov partition function $Z^{\rm Nek} (q)$ computes low-energy
quantities, such as the prepotential $\CF_0$, it is defined in terms of
a series expansion in the exponentiated microscopic gauge coupling
$q = \exp(2 \pi i \tau_{UV})$.  The gauge coupling $\tau_{UV}$, however, is sensitive to
the choice of the renormalization scheme and therefore cannot be assigned a
physical (low-energy) meaning. As we have pointed out in section~\ref{subsec:insta2},
the renormalization schemes in the two instanton computations indeed
differ. 
This means in particular that we should not
identify the microscopic gauge couplings for the $U(2)$ and the $Sp(1)$
gauge theory. Instead, 
we should only expect to find agreement between the $Sp(1)$ and $U(2)$
Nekrasov partition functions when we express them in terms of
physical low
energy variables.

\begin{figure}[h!]
\begin{center}
\includegraphics[width=3.7in]{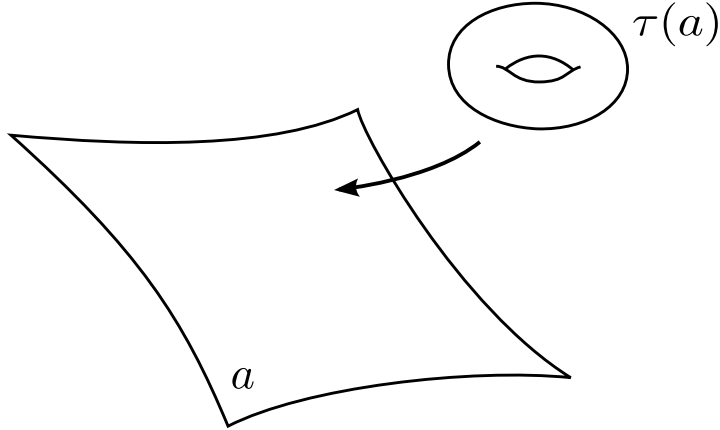}
\caption{The period matrix $\tau_{\rm IR,ij}$ of the Seiberg-Witten curve
is equal to the second derivative $\partial_{a_i} \partial_{a_j} \CF_0$ of the
prepotential with respect to the Coulomb parameter $a$. The imaginary
part of $\tau_{\rm IR}$ determines the metric on the Coulomb
branch.}\label{fig:Coulomb}   
\end{center}
\end{figure}

What are such low energy variables? Recall that in the low energy
limit of the 
$\CN=2$ gauge theory the
Coulomb branch opens up, which is
classically parametrized by the Casimirs of the gauge 
group.  The prepotential $\CF_0$ of the $\CN=2$ gauge theory
determines the corrections to the metric on the Coulomb branch, whose
imaginary part in turn prescribes the period 
 matrix $\tau_{\rm IR}$ of the so-called
Seiberg-Witten curve \cite{Seiberg:1994rs,Seiberg:1994aj}. As the
Seiberg-Witten curve changes along with the Coulomb moduli $a$, its
Jacobian defines a torus-fibration over the Coulomb branch (see
Figure~\ref{fig:Coulomb}).   
 Whereas for asymptotically free gauge theories the Seiberg-Witten
curve depends on the dynamically generated scale $\Lambda$,  
for conformally invariant gauge theories the Seiberg-Witten curve is
dependent on the value of the microscopic gauge couplings
$\tau_{UV}$. Conformally invariant theories are characterized by a 
moduli space for the UV gauge couplings, from each element of which a
Coulomb moduli space emanates in the low energy limit.  

Since the Seiberg-Witten curve determines the masses of BPS particles
in the low-energy limit of the $\CN=2$ gauge theory, its period matrix
$\tau_{\rm IR}$ is a physical quantity that should be independent of the
chosen renormalization scheme. In contrast, the microscopic gauge
couplings $\tau_{\rm UV}$ are characteristics of the chosen
renormalization scheme. 

To be more concrete, we can use the prepotential $\CF_0$
computed from the Nekrasov partition function to find
the relation between $\tau_{\rm IR}$ and $\tau_{\rm UV}$
by 
\begin{align}\label{eqn:IR-UV}
2 \pi i \tau_{\rm IR}=\frac{1}{2} \partial_a^2 \CF_0 (\tau_{\rm UV},a)=
\frac{1}{2} \partial_a^2 (\CF_{0,\rm pert} + \CF_{0,\rm inst})
(\tau_{\rm UV},a) \,.
\end{align}
Here $\CF_{0,\rm pert}$ contains the classical 
as well as 1-loop contribution to the prepotential, which for instance
can be found in \cite{Shadchin:2005mx}\footnote{ Note that there is a typo in the
  expression for the gauge contribution in \cite{Shadchin:2005mx}.}.
In particular, if two prepotentials that are computed
using two different schemes differ
by an $a$-dependent term, then the corresponding relations between
$\tau_{\rm IR}$ and $\tau_{\rm UV}$ differ as well. If we invert the
relation~(\ref{eqn:IR-UV}), and express both Nekrasov 
partition functions $Z^{\rm Nek}$ in terms of the period matrix 
$\tau_{\rm IR}$, we expect that they should agree up to a possible
spurious factor that doesn't depend on the Coulomb 
parameters.  This says that the two ways of instanton counting
correspond to two distinct renormalization schemes. 

\begin{figure}[h!]
\begin{center}
\includegraphics[width=1.1in]{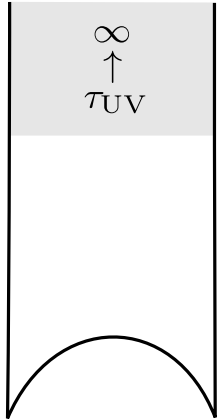}
\caption{The marginal coupling $\tau_{\rm UV}$ in the Nekrasov
  partition function defines a local coordinate
  on the moduli space of the $\CN=2$ conformal gauge theory near a
  weak-coupling point where $\tau_{\rm UV} \to \infty$.}\label{fig:tauUV}    
\end{center}
\end{figure}

In fact, it is not quite obvious that the full Nekrasov partition
functions, in contrast to just the prepotential, should agree when expressed in the period matrix $\tau_{\rm
  IR}$. It would have been possible
that the relation between the prepotential $\tau_{\rm IR}$ and the
microscopic couplings $\tau_{\rm UV}$ gets quantum corrections in terms
of the deformation parameters $\e_1$ and $\e_2$, in such a way that
only when expressed in terms of a 
quantum period matrix $\tau_{\rm IR}(\e_1,\e_2)$ the Nekrasov partition
functions do agree. In subsection~\ref{sec:instantonexamples} we will
find however that this is not the case. The Nekrasov partition
function agree when expressed in terms of the classical period
matrix $\tau_{\rm IR}$. One possible argument for this is that the
higher genus free energies $\CF_{g \ge 1}$ 
are uniquely determined given the prepotential $\CF_0$. In other
words, that when the
prepotentials (and thus Seiberg-Witten curves plus differentials) for
two gauge theories agree we also expect the higher genus free 
energies to match up. This is reasonable to
expect from several points of view, \ie the interpretation of the
$\CF_g$'s as free energies in an integrable hierarchy \cite{Aganagic:2003qj,eynardorantin}.   

The Nekrasov partition function is computed as a series expansion in
$q = \exp(2 \pi i \tau_{UV})$. The microscopic coupling $\tau_{UV}$
thus corresponds to a choice of local coordinate on the moduli space
of microscopic gauge couplings near a weak-coupling point (see
Figure~\ref{fig:tauUV}).  
An inequivalent renormalization scheme corresponds to 
a different choice of coordinate in that neighborhood. 
In particular, given two different renormalization schemes, by
combining their respective IR-UV relations, we
can obtain the relation between the two different
microscopic couplings, and thus find the explicit coordinate
transformation 
on the moduli space. 
Explicitly, we identify the infra-red couplings of two
related theories by  
\be\label{eqn:UV-UV}
 \tau_{\rm IR} = \frac{1}{2} \partial_a^2 \CF_0^{A} (\tau_{\rm UV}^{A},a) = \frac{1}{2} \partial_a^2 \CF_0^{B} (\tau_{\rm UV}^{B},a). 
\ee
By inverting the IR-UV relation for the gauge theory characterized by the
microscopic coupling $\tau^A_{\rm UV}$, we find the relation
between the microscopic couplings $\tau^{A}_{\rm UV}$ and
$\tau^{B}_{\rm UV}$ of both gauge 
theories.  Since this is a relation between quantities in  
the ultra-violet, we expect it to be independent of infra-red parameters
such as the masses and Coulomb branch parameters. Indeed, in all
examples that we study in subsection~\ref{sec:instantonexamples}, we
will find that the moduli-independent UV-UV relation that
follows from equation~(\ref{eqn:UV-UV}) relates the Nekrasov partition
functions up to a spurious factor that is independent of the Coulomb
parameters.\footnote{Yet another example of a renormalization scheme
  for 
  the four-dimensional $Sp(1)$ gauge theory with four flavors
  is found by 
  counting string instantons in 
  a system of D3 and D7 branes in Type I'  
  \cite{billo1,billo2,billo3}.}  

We will often consider gauge theories as being
  embedded in string theories. Different models of the same
  gauge theory give different embeddings in string theory,
  which means that the results will differ when expressed
  in terms of UV variables, even though the IR results
  agree.   
   
Take as an example the string theory realization of a
supersymmetric $\CN=2$ $SU(2)$ gauge theory. The unitary point of view leads 
to a construction of D4, NS5 and D6-branes in type IIA theory
\cite{Witten:1997sc}, whereas 
the symplectic point of view introduces an orientifold in this
picture and mirror images for all D4-branes \cite{Evans:1997hk,Landsteiner:1997vd,Brandhuber:1997cc}. Clearly, these are
different realizations of the $SU(2)$ gauge theory. Nevertheless, both
descriptions should give the same result in the infra-red.

Indeed, the two aforementioned string theory embeddings, based on either a $U(2)$
or a $Sp(1)$ gauge group, determine a physically equivalent Seiberg-Witten
curve. For instance, the brane embedding of the pure $Sp(1)$ gauge
theory determines the curve \cite{Landsteiner:1997vd}
\begin{align}\label{eqn:Sp1curve}
s^2 - s \left( v^2 (v^2 + u) +2 \Lambda^4 \right) +\Lambda^8 = 0,
\end{align}
in terms of the covering space variables $s \in \mathbb{C}^*$, $v
\in \mathbb{C}$ and the gauge invariant coordinate $u = \Tr (\Phi^2)$ on
the Coulomb branch. This is merely a double cover
\cite{Argyres:1995fw}  of the more familiar
parametrization of the $SU(2)$ Seiberg-Witten curve
\begin{align}\label{eqn:U2curve}
\Lambda^2 t^2 - t \left( w^2 +u \right) + \Lambda^2 =0,
\end{align}
with $t \in \mathbb{C}^*$ and $w \in \mathbb{C}$, which follows from
the unitary brane construction \cite{Witten:1997sc}.

In fact, the choice for an instanton renormalization scheme is closely
related to the choice for a brane embedding, as the precise
parametrizations of the Seiberg-Witten curves (\ref{eqn:Sp1curve}) and
(\ref{eqn:U2curve}) can be recovered in a thermodynamic (classical)
limit by a saddle-point approximation of the $Sp(1)$ and the $U(2)$
Nekrasov partition  
functions respectively \cite{Nekrasov:2003rj,Nekrasov:2004vw}.

\subsection{Examples}\label{sec:instantonexamples}

Let us illustrate the above theory by a selection of examples. We
start with comparing $Sp(1)/SO(4)$ and $U(2)$ partition functions in
gauge theories with a single gauge group, and extend this to partition
functions for more general linear and cyclic quivers. In particular, we find
the identification of $Sp(1)/SO(4)$ and $U(2)$ instanton partition functions
expressed in low-energy moduli and the relation between the $Sp(1)/SO(4)$
and $U(2)$ microscopic gauge couplings.

\subsubsection{$Sp(1)$ versus $U(2)$ : the asymptotically free case}

First of all, let us consider the $Sp(1)$ theory with a single gauge
group coupled to $N_f$ massive hypermultiplets, where $N_f$ runs from 1 to
4. In the asymptotically free theories, with $N_f \le 3$, we find that
the $Sp(1)$ Nekrasov partition function equals the $U(2)$ Nekrasov
partition function -- with Coulomb parameters $(a,-a)$ -- up to a
factor that doesn't depend on the Coulomb parameter and only
contributes to the low genus refined free energies $\CF_{0,\frac{1}{2},1}$. In
the following we will call a factor with these two properties a
\emph{spurious} factor. 

More precisely, we compute that \footnote{In this
  section we denote the Nekrasov partition function $Z^{\rm Nek}$ by $Z$. In the
  equations~(\ref{ZSp1Nf0})-(\ref{ZSp1Nf3}) there is an agreement for the instanton partition
  functions as well.}  
\begin{eqnarray}
 &&Z^{N_f = 0}_{U(2)}(q )= Z^{N_f = 0}_{Sp(1)}(q) \label{ZSp1Nf0}\\
&&Z^{N_f = 1}_{U(2)}(q) = Z^{N_f = 1}_{Sp(1)}(q) \label{ZSp1Nf1} \\
 &&Z^{N_f = 2}_{U(2)}(q) = Z^{N_f = 2}_{Sp(1)}(q) 
 [Z_{U(1)}^{\tilde{N}_f=0}(q)]^{1/2} \label{ZSp1Nf2} \\
&&Z^{N_f = 3}_{U(2)}(q) = Z^{N_f = 3}_{Sp(1)}(q)
[Z_{U(1)}^{\tilde{N}_f=1}(q)]^{1/2}  \exp\left( -\frac{q^2}{ 32 \e_1
    \e_2}\right),  \label{ZSp1Nf3} 
\end{eqnarray}
up to degree six in the $q = \Lambda^{4-N_f}$ expansion, where $Z_{U(1)}^{\tilde{N}_f}$ is
the instanton partition function of the 
$U(1)$ gauge theory coupled to $\tilde{N}_f$ hypermultiplets with masses
$m_1$ up to $m_{\tilde{N}_f}$. Explicitly,
\begin{align}
 Z_{U(1)}^{\tilde{N}_f=0}(q) = \exp \left(-\frac{q}{\e_1 \e_2} \right), \\
 Z_{U(1)}^{\tilde{N}_f=1}(q) = \exp \left(-\frac{m q}{\e_1 \e_2} \right), 
\end{align}
with $q = \Lambda^{2-\tilde{N}_f}$, $m = \mu_1 + \mu_2 + \mu_3 +
\e_1+\e_2 $ where $\m_i$ being the masses of the fundamental hypermultiplets in equation~(\ref{ZSp1Nf3}).  Note that the
equalities (\ref{ZSp1Nf0})--(\ref{ZSp1Nf3}) can equally well be written down for any combination of
hypers in the fundamental and anti-fundamental representation of the
$U(2)$ gauge group. The contribution of a fundamental hypermultiplet
just 
differs from that of an anti-fundamental hypermultiplet by mapping $\mu
\mapsto -\mu$. 

Let us make two more remarks about the formulas
(\ref{ZSp1Nf0})--(\ref{ZSp1Nf3}).
First, the form of the spurious factor in the equalities
  (\ref{ZSp1Nf0})--(\ref{ZSp1Nf3}) is close to what is called the
  $U(1)$ factor in the AGT correspondence: they agree for $N_f=2$ and
  differ slightly for the $N_f=3$ theory. In particular, both factors
  don't depend on the Coulomb parameters and only
  contribute to the lowest genus contributions $\CF_{0,\frac{1}{2},1}$
 of the refined free energy.
Second, since the $U(2)$ and $Sp(1)$ Nekrasov
partition functions coincide up to moduli-independent terms,
the relation between IR and UV couplings is the same. It follows
that they will
agree up to spurious factors even when written in terms of IR couplings.

\subsubsection{$Sp(1)$ versus $U(2)$ : the conformal case}

\begin{figure}[h!]
\begin{center}
\includegraphics[width=4.9in]{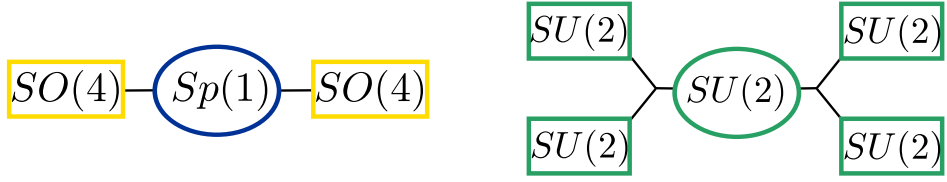}
\caption{On the left: Quiver of the $Sp(1)$ gauge theory coupled to
  two fundamental and two anti-fundamental hypermultiplet. Since the
  (anti-)fundamental representation of $Sp(1)$ is pseudo-real, the
  flavor symmetry group of two hypermultiplets enhances to
  $SO(4)$. On the right: Quiver of the $SU(2)$ gauge
  theory coupled to two fundamental and two anti-fundamental
  hypermultiplets. The flavor symmetries of the hypermultiplets is
  enhanced to $SU(2)$.}\label{fig:sp1u2quiver}  
\end{center}
\end{figure}

Comparing the Nekrasov partition functions for conformal $Sp(1)$ and
the $U(2)$ gauge 
theories, both coupled to four hypermultiplets, yields a substantially
different result.\footnote{The four flavor
  partition function is nevertheless perfectly consistent 
  with the partition functions for fewer flavors. When we send the
  masses of the hypermultiplets to infinity, we do
  find the corresponding $N_f<4$ partition
  functions. }$^{,}$\footnote{The 
  expression for the ratio proposed in \cite{Wyllard:2009hg} does not
  hold beyond instanton number $k=1$.}   (The quivers of the respective gauge theories are
illustrated in 
Figure~\ref{fig:sp1u2quiver} for later reference.) The $Sp(1)$
Nekrasov partition function 
does not agree with the $U(2)$  Nekrasov partition function
up to a spurious factor, when expressed in the
UV gauge couplings with the identification $q_{Sp(1)} = q_{U(2)}$.
In particular, the prepotentials $\CF_0$ differ, leading
to a different relation between $\tau_{\rm IR}$ and
$q_{Sp(1)}$ than between $\tau_{\rm IR}$ and $q_{U(2)}$.

When all the hypers are massless, or equivalently when
sending the
Coulomb parameter 
$a \to \infty$, we find that the map between
UV gauge couplings and the period matrix $\tau_{\rm IR}$ does not
depend on $a$ and is given
by \footnote{All $Sp(1)$ results in this subsection have been checked up
  to order 6 in the $Sp(1)$ instanton parameter.}$^,$\footnote{Equation (\ref{eqn:U2UV-IR})  was first found in
  \cite{Grimm:2007tm}.} 
\begin{align}
q_{Sp(1)}^2 & =  16 \, \frac{\theta_2(q_{\rm IR}^2)^4}{\theta_3(q_{\rm
    IR}^2)^4}  \label{eqn:Sp1UV-IR}\\
q_{U(2)} & = \frac{\theta_2(q_{\rm IR})^4}{\theta_3(q_{\rm IR})^4}, \label{eqn:U2UV-IR}
\end{align}
where we define $q_{\rm IR} = \exp(2 \pi i \tau_{\rm
  IR})$.\footnote{Here we
  use a convention different from appendix B of \cite{AGT}.} The
$Sp(1)$ and $U(2)$ mappings
just differ by doubling 
the value of the  microscopic gauge coupling as well as the infra-red
period matrix.

If we express the massless partition functions in terms of
the low-energy variables by using (\ref{eqn:Sp1UV-IR}) and
(\ref{eqn:U2UV-IR}), they agree up to a spurious factor (which is
independent of the Coulomb parameter $a$ and only contributes to the
lower genus refined free energies $\CF_{0,\frac{1}{2},1}$).  In fact, it turns
out that even if we re-express the \emph{massive} 
partition functions using the \emph{massless}
 UV-IR mappings (\ref{eqn:Sp1UV-IR}) and (\ref{eqn:U2UV-IR}), we still
find agreement up to a 
spurious factor. 

On the other hand, even if we use the \emph{massive} $Sp(1)$ and $U(2)$ IR-UV
mappings which do depend on the Coulomb parameter $a$ and the masses
of the hypers,  
we find that the $Sp(1)$ and $U(2)$ renormalization scheme are related by
the transformation 
\begin{align}\label{eqn:UV-UV-U2Sp1}
\boxed{q_{U(2)} = q_{Sp(1)} \left(1+ \frac{q_{Sp(1)}}{4}\right)^{-2},}
\end{align}
which as expected
is not dependent on the Coulomb
branch moduli.

For completeness let us give the expression for the spurious
factor once we express both full partition functions
in terms of $q_{Sp(1)}$. For the unrefined case 
$\hbar = \e_1 = -\e_2$ we find  
\begin{align}\label{Sp1U2spurious}
\frac{Z_{U(2)}(q_{U(2)}(q_{Sp(1)}))}{Z_{Sp(1)} (q_{Sp(1)})} = \left( 1
  + \frac{q_{Sp(1)}}{4} \right)^{M+N}
\left( 1 - \frac{q_{Sp(1)}}{4}\right)^{N-M},
\end{align}
where $M = \frac{1} {\hbar^2} \sum_{i<j} \mu_i \mu_j$ and $N = -\frac{1}{2 \hbar^2} \sum_i
\mu_i^2 + \frac{1}{8}$. Here, we emphasize that this relation is
between the full Nekrasov partition functions including the perturbative
pieces and not just between the instanton parts.  
Notice that this spurious factor is quite close to, yet more
complicated than the square-root of
the unrefined $U(1)$ partition function
\begin{align*}
 Z_{U(1)}^{N_f=2}(q) = \left( 1-q \right)^{-\frac{m_1 m_2}{\hbar^2} },
\end{align*}
of the $U(1)$ gauge theory coupled to two hypermultiplets with masses
$m_1$ and $m_2$, the square of which entered the AGT correspondence as
the ``$U(1)$ factor''.  Similarly, we interpret the spurious
factor~(\ref{Sp1U2spurious}) as a decoupled $U(1)$ factor.

\subsubsection{$SO(4)$ versus $U(2) \times U(2)$ instantons}

The instanton partition function for the pure $SO(4)$ gauge theory
agrees with that of the pure $U(2)\times U(2)$ theory \footnote{We
  checked the $SO(4)$ results in this subsection up to order 2 for the
  refined $SO(4)$ partition functions and up to order 6 for the unrefined ones.}
\be
Z_{SO(4)}^{N_f=0}(q) = Z_{U(2) \times U(2)}^{N_b = 0} (q)\ , 
\ee
if we make the
identifications 
\begin{align} \label{so4tou2u2}
q_{U(2),1} = q_{U(2),2} = 16 \ q_{SO(4)} \quad \mbox{and} \quad
(b_1,b_2) = (a_1+a_2,a_1-a_2)\ .
\end{align}
Here, $b_{1,2}$ are the Coulomb parameters of the $SO(4)$ gauge
theory and $a_{1.2}$ those of the $U(2) \times U(2)$ gauge
theory. The second relation follows
simply from the embedding of $\mathfrak{su}(2) \times \mathfrak{su}(2)$
in $\mathfrak{so}(4)$. 

When we couple the $SO(4)$ theory to a single massive hypermultiplet, its instanton
partition function matches with that of the $U(2) \times U(2)$
theory coupled to a massive bifundamental up to a spurious factor
\begin{align}
Z_{SO(4)}^{N_f=1}(q) = Z_{U(2) \times U(2)}^{N_{b}=1}(q)
\exp\left(-\frac{4q}{\hbar^2} \right), 
\end{align}
for the unrefined case under the same identification \eqref{so4tou2u2}.

\begin{figure}[h!]
\begin{center}
\includegraphics[width=4.9in]{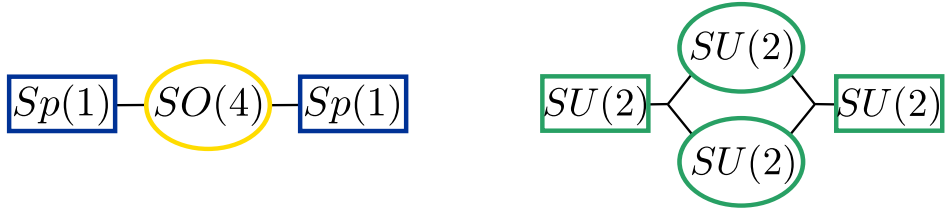}
\caption{On the left: Quiver representation of the $SO(4)$ gauge theory coupled to
  one fundamental and one anti-fundamental hypermultiplet. Since the
  (anti-)fundamental representation of $SO(4)$ is real, the flavor
  symmetry group of each hypermultiplet enhances to $Sp(1)$. On the
  right: Quiver representation of the $SU(2)\times SU(2)$ gauge 
  theory coupled to two bi-fundamental
  hypermultiplets. The flavor symmetry of the bifundamental field is
  enhanced to $SU(2)$.}\label{fig:so4u2quiver}
\end{center}
\end{figure}

Now, let us consider the conformal case. Naively comparing the
Nekrasov partition function of the 
conformal $SO(4)$ gauge theory coupled to two massive hypermultiplets
with that of the $U(2)\times U(2)$ theory coupled by two
massive bifundamentals shows a serious disagreement. (Their quivers
are illustrated in Figure~\ref{fig:so4u2quiver}.) However, if we
follow the same strategy as explained in the conformal $Sp(1)$
example, we see that it is once more simply a matter of different
renormalization schemes. The $SO(4)$ and the $U(2)$ gauge
theory are related by the change of marginal couplings
\begin{align}\label{eqn:SOUVUV}
 \boxed{q_{SO(4)} 
= \frac{\theta_2(q_{U(2)})^4}{\theta_3(q_{U(2)})^4},} 
\end{align}
when we identify $q_{U(2)} = q_{U(2),1} = q_{U(2),2}$. Using this
UV-UV relation and the relation between the Coulomb branch parameters \eqref{so4tou2u2}, the $SO(4)$ and
$U(2)\times U(2)$ Nekrasov partition functions agree up to a
spurious factor 
\begin{align}
\frac{Z_{U(2)\times U(2)}(q_{U(2)}(q_{SO(4)}))}{Z_{SO(4)} (q_{SO(4)})} 
= 1-\frac{4 M}{\hbar ^2} q  + \frac{8 M^2 + 2  N \hbar^2}{\hbar^4} q^2+ \ldots 
\end{align}
that is similar to the $Sp(1)-U(2)$ spurious factor
in equation~(\ref{Sp1U2spurious}),  
with now $M = m_1^2 - m_1 m_2 + m_2^2$ and  $N = 3m_1^2 + m_1 m_2 + 3 
m_2^2$. 

\subsubsection{$Sp(1) \times Sp(1)$ versus $U(2) \times U(2)$ instantons}

Next, we analyze $Sp(1) \times Sp(1)$ quiver gauge theories coupled to
at most 4 massive hypermultiplets. The bifundamental multiplet, that couples
the two $Sp(1)$ gauge groups, introduces new poles in the theory,
similar to the adjoint multiplet in the $\CN=2^*$ gauge theory. 

As expected, we find immediate
agreement between the $Sp(1)$ and $U(2)$ 
instanton partition functions up to a spurious factor, when we couple
fewer than two hypers to each multiplet. 

If more than two hypers are
coupled to one of the gauge groups, we need to express the partition
function in terms of the physical period matrix $\tau_{\rm IR}$. Notice that the
$Sp(1)$ (as well as $U(2)$) instanton partition function is a function of two
UV gauge couplings, whereas the period matrix is a symmetric
$3\times3$ matrix. It is nevertheless easy to find a bijective relation between
the two diagonal entries of the period matrix and the two UV gauge
couplings. The off-diagonal entry in the period matrix represents a
mixing of the two gauge groups, and can be expressed in terms of the
diagonal entries.   
  
Let us consider the conformal linear quiver with two $Sp(1)$ gauge groups
as an example. We couple the two $Sp(1)$'s by a bifundamental
and add two extra hypermultiplets to the first and to the second 
gauge group. The \emph{moduli-independent} UV-IR relation for $Sp(1)$ has a series
expansion \footnote{Here we have rescaled $q_{\rm IR}
  \to q_{\rm IR}/16$.} 
\begin{align}
q_{Sp(1),i} &=  q_{\mathrm{IR},ii} - \frac{1}{64} q_{\mathrm{IR},ii} ^3 +
\frac{1}{32} \ q_{\mathrm{IR},ii}   q_{\mathrm{IR},jj}^2 + \CO(q_{\mathrm{IR}}^4), \label{Sp1UVIRbifund}
\end{align}
whereas the one for $U(2)$ has the form
\begin{align}
q_{U(2),i} &= q_{\mathrm{IR},ii} - \frac{1}{2} q_{\mathrm{IR},ii}^2 + \frac{1}{2} q_{\mathrm{IR},ii}  q_{\mathrm{IR},jj}  +\frac{11}{64}  q_{\mathrm{IR},ii} ^3 \nn
\\ & \quad -\frac{1}{2}  q_{\mathrm{IR},ii} ^2  q_{\mathrm{IR},jj}+\frac{3}{32}  q_{\mathrm{IR},ii}  q_{\mathrm{IR},jj}^2+ \CO(q_{\mathrm{IR}}^4), \label{U2UVIRbifund} 
\end{align}
for $i \in \{1,2\}$ and $i \neq j$.

As before we use the \emph{moduli-independent}  UV-IR mappings
(\ref{Sp1UVIRbifund}) and (\ref{U2UVIRbifund}) to evaluate
the \emph{massive} partition function as a function of the physical IR
moduli $\tau_{\rm IR,11}$ and $\tau_{\rm IR, 22}$.  Again this shows
agreement of the $Sp(1)$ and $U(2)$ partition functions up to
a spurious factor in the lower genus free energies.    

Composing the \emph{moduli-dependent} mappings between UV-couplings and
the period matrix, we find that the two renormalization schemes are
related by 
\begin{align} \label{UVUV5pt}
q_{Sp(1),i} &= q_{U(2),i} + \frac{1}{2} q_{U(2),i}^2 - \frac{1}{2} q_{U(2),i} q_{U(2),j} + \frac{5}{16}
q_{U(2),i}^3 \nn \\ & \quad  - \frac{1}{16}
q_{U(2),i} q_{U(2),j}^2+ \CO(q_{U(2)}^4), 
\end{align}
for $i \in \{1,2\}$ and $i \neq j$. 
Note that this mapping is independent of the Coulomb branch moduli and
the mass parameters,
as it should be. Notice as well that a mixing amongst the two gauge
groups takes places, so that we cannot simply use the UV-UV mapping
for a single gauge group 
twice. Substituting this relation into the $Sp(1)$ partition function
indeed turns brings it into the form of the $U(2)$ partition function
up to a spurious AGT-like factor.  

The above procedure can be applied to any linear or cyclic
quiver.\footnote{For example, we also tested it for the $\CN=2^*$
  $Sp(1)$ gauge theory.}  Most importantly, the $Sp(1)$ and $U(2)$
partition function 
agree (up to a spurious factor) when expressed in IR coordinates,
and, the mapping between $Sp(1)$ and $U(2)$
renormalization schemes is independent of the moduli in the gauge
theory and
mixes the gauge groups. Moreover, the spurious factors that we find are
more complicated than the ``$U(1)$-factors'' that appear in the AGT
correspondence.

 \section{$\CN=2$ geometry}\label{sec:N=2geometry}

In section~\ref{sec:gauge} we found in which way
Nekrasov partition functions for different models of the
same underlying physical gauge theory are related, \ie comparing the
$U(2)$ versus the $Sp(1)$ method of instanton counting.  
The $\CN=2$ instanton counting defines a renormalization
scheme for 
each model, such that, when expressed in terms of physical
infra-red variables, the Nekrasov partition functions of two such models
agree up to a spurious 
factor. When expressed in terms of the microscopic couplings, however,
the Nekrasov partition functions are related by a non-trivial
mapping. Our goal in this section is to explain this mapping
geometrically.  

The previous section contained a brief review of the low energy data
contained in the Seiberg-Witten
curve. As expected we were able to verify that physically relevant
quantities agree for different models of the same low energy gauge
theory. In \cite{Gaiotto:2009we} it was shown how to extract additional information
from the Seiberg-Witten curve beyond the low energy data. This information is
encoded in a realization of the Seiberg-Witten curve as a branched
cover over yet another punctured Riemann surface, which we will refer to
as the Gaiotto curve (or G-curve). 

As we review in more detail in a moment, the branched cover
realization can be read off from the brane 
construction of the $\CN=2$ gauge theory. The number of D4-branes that
is necessary to engineer the gauge theory determines the degree of the
covering. Branch points of the covering (which are punctures on the
Gaiotto curve) correspond to poles in the Seiberg-Witten differential, and
their residues are associated to the bare mass parameters in the gauge
theory.  

In \cite{Gaiotto:2009we}  the complex structure moduli of the Gaiotto curve
are identified with the exactly marginal couplings of the conformal
$\CN=2$ gauge theory. The
Gaiotto curve therefore not only captures the infra-red data of the gauge theory, but
also information about the chosen renormalization scheme. Different
renormalization schemes are  
related by non-trivial mappings of exactly marginal
couplings. Geometrically, it was argued that a choice of renormalization
scheme corresponds to a choice of local coordinates on the complex
structure moduli space  of the Gaiotto curve. 

A given physical gauge theory of course admits a countless number of
renormalization schemes. On the other hand, our focus in this paper is
only to find a geometric interpretation of a small subset of such
choices. We expect to find such an interpretation when different
models can both be embedded by a brane construction in string
theory. In such examples we can construct the corresponding Gaiotto
curves and a mapping between them. This mapping geometrizes the
mapping between the exactly marginal couplings. 

Most of the examples
we considered in the previous section are of this type. However, one
of them is not. This is the comparison of the $SU(2) \times SU(2)$
gauge theory with the $Sp(1) \times Sp(1)$ gauge theory.\footnote{The
  same holds for the comparison of the $\CN=2^*$ theory with gauge
  group $Sp(1)$ versus $U(2)$.}
  Since we are not aware of a brane embedding of the latter theory,
  there is no 
immediate reason to expect
an (obvious) geometric interpretation of the mapping between
exactly marginal couplings in that example. In the other examples, for
which brane constructions are well-known, we do expect to find a
geometric explanation for the UV-UV mappings.

As a last comment, let us emphasize that the validity of the mappings
between exactly marginal couplings extends beyond the prepotential
$\CF_0$. As we found in the last section, the full Nekrasov partition functions $Z^{\rm
  Nek}$ are 
related by the mapping between UV couplings, which does not receive
corrections in the  
deformation parameters $\e_1$ and $\e_2$. Geometrically, this implies
that there are no quantum corrections to the mappings between
the Gaiotto curves for different models of the same physical  gauge
theory.   

In this section we start by reviewing the construction of Gaiotto
curves for $\CN=2$ gauge theories with a classical gauge group. We
explain how the $\CN=2$ geometry is encoded in a ramified Hitchin
system whose base is the Gaiotto curve. Furthermore, we make some
comments on its six-dimensional origin. We then compare geometric
realizations for unitary and symplectic/orthogonal gauge theories. In
particular, we argue how different models of the same theory are
related geometrically. We explicitly verify this in some examples,
where we compare the geometry underlying $Sp(1)$ 
and $SO(4)$ quiver gauge theories to those of $U(2)$ quivers. As
expected, we find  
a geometric explanation for the relation between the marginal 
couplings that we computed in the previous section. 

\subsection{G-curves and Hitchin systems}

Let us start this section with reviewing the constructions of 
Gaiotto curves for $SU(N)$ and $Sp(N)/SO(N)$ gauge groups and their
embedding in a ramified Hitchin system. We then explain how to compare them in specific cases.  

\subsubsection{Unitary gauge group}

\begin{figure}[h!]
\begin{center}
\begin{tabular}{lr}
$\vcenter{\hbox{\includegraphics[width=2.8in]{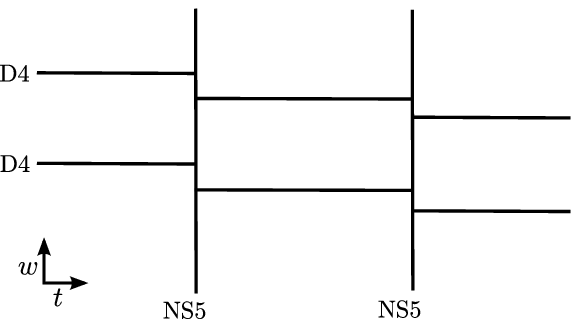}}}$ & \qquad $\vcenter{\hbox{\includegraphics[width=2.0in]{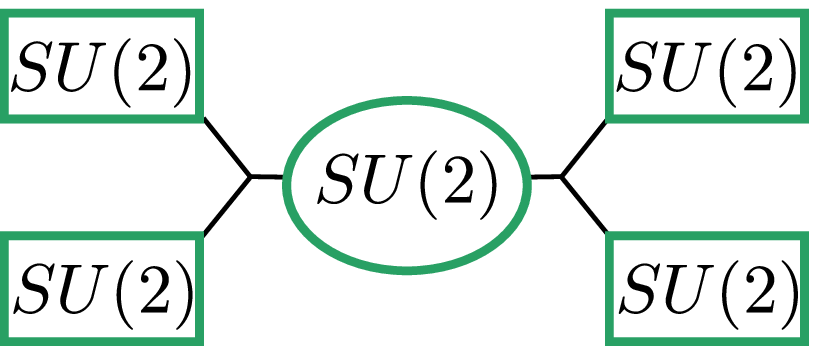}} }$
\end{tabular}
\caption{Illustrated on the left is an example of a D4/NS5 brane construction
  realizing the $SU(2)$ quiver gauge theory illustrated on the
  right. The Coulomb and mass 
  parameters of the $SU(2)$ gauge theory parametrize the separation of the
  D4-branes, while the separation of the NS5-branes determines the 
  microscopic coupling~$\tau_{\rm UV}$. The Seiberg-Witten curve for this
  $SU(2)$ gauge theory is a torus with complex structure
  parameter $\tau_{\rm IR}$. }\label{U2brane}
\end{center}
\end{figure}

A special unitary quiver gauge theory can be realized in type IIA string
theory using a D4/NS5 brane embedding \cite{Witten:1997sc}. See
Figure~\ref{U2brane} for the brane embedding of the $SU(2)$ gauge
theory coupled to four hypermultiplets. 
From such a brane
embedding one can read off the Seiberg-Witten curve $\Sigma$ of the quiver
gauge theory. It is, roughly speaking,
a fattening of the D4/NS5 graph, as it is for instance illustrated in
Figure~\ref{U2brane}. 

The D4/NS5 brane embedding can be 
lifted to an M5-brane embedding in M-theory. 
The resulting ten-dimensional M-theory
background is
\begin{align}\label{U(N)Mbackground}
\IR^4 \times T^* \tilde{C} \times \IR^2 \times S^1,
\end{align}
where we introduced a possibly punctured Riemann surface $\tilde{C}$
and its cotangent bundle $T^*\tilde{C}$. 
We insert a stack of $N$ M5-branes
that wraps the six-dimensional manifold $\IR^4 \times \tilde{C}$. 
The positions of these M5-branes in the cotangent
bundle determine the Seiberg-Witten curve $\Sigma$ as a subspace of
 $T^*\tilde{C}$. In this perspective the Seiberg-Witten curve is
given 
\cite{Gaiotto:2009we}   
\begin{align}\label{eqn:SWbranched}
0 = \det (w -\boldsymbol\phi_{U}) = w^N + w^{N-2} \phi_2 + w^{N-3} \phi_3 + \ldots + \phi_N
\end{align}
as a branched degree $N$ covering over the so-called Gaiotto curve
$\tilde{C}$. The holomorphic
differential $w$ parametrizes the fiber direction of the cotangent
bundle $T^*\tilde{C}$, whereas $\boldsymbol\phi_U$ is an $SU(N)$-valued
differential on the curve $\tilde{C}$ of degree 1. 

The degree $d$
differentials $\phi_{d} = \Tr(\boldsymbol\phi_U^d)$ encode the
classical vev's of the $SU(N)$ Coulomb
branch  operators of  dimension~$d$. 
They are allowed to have poles at the punctures
of the Gaiotto curve. The coefficients at these poles encode 
the bare mass parameters of the gauge theory. To take care of these
boundary conditions in the M-theory set-up, we need to insert
additional M5-branes at the punctures of the Gaiotto 
curve. These M5-branes should intersect the Gaiotto curve
transversally, and thus locally wrap the fiber of the cotangent
bundle at the puncture \cite{Gaiotto:2009hg}. 

Equation~(\ref{eqn:SWbranched}) 
determines the Seiberg-Witten curve $\Sigma$ as an $N$-fold branched
covering over the 
Gaiotto curve $\tilde{C}$. The Seiberg-Witten differential is simply 
\begin{align}
\lambda = w. 
\end{align}
The $\CN=2$ geometry
for unitary gauge groups is thus encoded in a ramified $A_N$ Hitchin system on
the punctured  Gaiotto curve $\tilde{C}$, whose spectral curve is the
Seiberg-Witten curve~$\Sigma$ and whose canonical 1-form is equal
to the Seiberg-Witten differential \cite{Hitchinsystem}.\footnote{A
  detailed discussion of 
  boundary conditions for this Hitchin system can be found in
  \cite{Gaiotto:2009hg} and references therein.}

\begin{figure}[h!]
\begin{center}
\begin{tabular}{lr}
$\vcenter{\hbox{\includegraphics[width=2.1in]{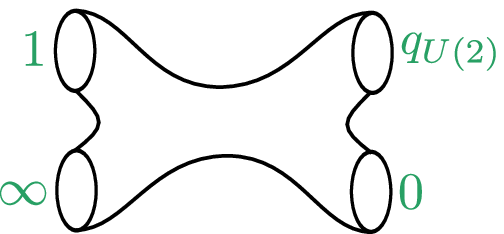}}}$ & \qquad \qquad $\vcenter{\hbox{\includegraphics[width=2.1in]{u2quiver}} }$
\end{tabular}
\caption{The left Figure illustrates the Gaiotto curve $\tilde{C}$ of
  the conformal $SU(2)$ quiver gauge theory that is illustrated on the
  right. The Gaiotto curve is 
  a four-punctured sphere with complex structure parameter
  $q_{U(2)}$. The
  differential $\phi_2$ has second order poles at the four
  punctures. The $SU(2)$ flavor symmetries are encoded in the
  coefficients of the differential $\phi_2$ at these poles. }\label{U2Gcurve}
\end{center}
\end{figure} 

The topology of the Gaiotto curve is fully determined by the corresponding
quiver diagram. A gauge group translates into a tube of the Gaiotto curve, whereas a
flavor group turns into a puncture. This is illustrated in
Figure~\ref{U2Gcurve} for the $SU(2)$ gauge theory coupled to four
flavors. The poles of the differentials $\phi_d$ determine the
branch points of the fibration~(\ref{eqn:SWbranched}). Their
coefficients encode the flavor symmetry of the 
quiver gauge theory. For gauge group $SU(N)$ the degree of the poles is
integer. As we will see shortly this is not true for $Sp(N)$ and $SO(N)$
gauge theories.

\subsubsection{Symplectic/orthogonal gauge group}

\begin{figure}[h!]
\begin{center}
\begin{tabular}{lr}
$\vcenter{\hbox{\includegraphics[width=3in]{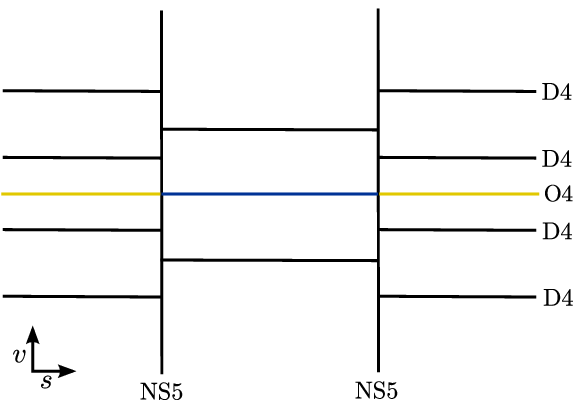}}}$ & \quad $\vcenter{\hbox{\includegraphics[width=2.2in]{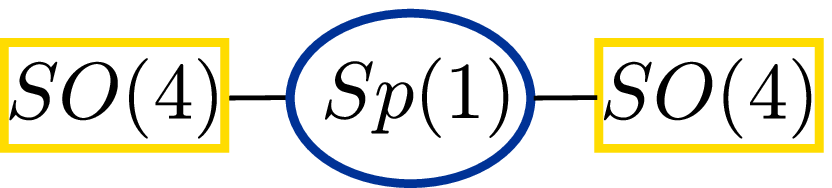}} }$
\end{tabular}
\caption{Illustrated on the left is an example of a D4/NS5 brane
    construction with O4$^{\pm}$ orientifold branes realizing the $Sp(1)$
    quiver gauge theory illustrated on the right. The O4$^-$ branes (in
  yellow) ensure that both flavor symmetry groups are $SO(4)$, whereas
  the O4$^+$ brane (in blue) ensures that the gauge symmetry group
  is $Sp(1)$. The brane embedding of the conformal $SO(4)$ gauge
  theory is found by swapping the
  inner and the outer D4 and O4 branes.}\label{SpSObrane}
\end{center}
\end{figure}

For symplectic or orthogonal gauge theories a similar description
exists. Engineering these gauge theories in type IIA requires
orientifold O4-branes in addition to the D4 and NS5-branes
\cite{Evans:1997hk,Landsteiner:1997vd,Brandhuber:1997cc}. The orientifold
branes are parallel to the D4 branes. They act on the string background
as a combination of a worldsheet parity $\Omega$ and a spacetime
reflection in the five dimensions transverse to it. The space-time
reflection introduces a mirror brane for each D4-brane, whereas the
worldsheet parity breaks the space-time gauge group. More precisely, 
there are two kinds of O4-branes, distinguished by the sign of
$\Omega^2 = \pm 1$. The O4$^{-}$ brane breaks the $SU(N)$ gauge
symmetry to $SO(N)$, whereas the O4$^+$ brane breaks it
to $Sp(N/2)$. The brane construction that engineers the
conformal $Sp(1)$ gauge theory is schematically shown in
Figure~\ref{SpSObrane}.\footnote{The $Sp$ and $SO$ brane 
constructions illustrated here can 
  be naturally extended to linear $Sp/SO$ quivers. We will come back 
  to this in section~\ref{sec:linearquiver}.} Notice that there are
two hidden D4-branes on top of the O4$^+$ brane, so that the number of
D4-branes is equal at each point over the base.

From these brane setups we can extract the Seiberg-Witten curve
for the $Sp(N-1)$ and $SO(2N)$ gauge theories coupled to matter. To find the
Gaiotto curve we rewrite the Seiberg-Witten curve in the form
\cite{Tachikawa:2009rb}   
\begin{align}\label{SpSOSWcover}
0= v^{2N} + \varphi_2 v^{2N-2} + \varphi_4 v^{2N-4} + \ldots + \varphi_{2N},
\end{align}
where the differentials $\varphi_k$ encode the Coulomb
parameters and the bare masses.  Equation~(\ref{SpSOSWcover})
defines 
the Seiberg-Witten curve as a branched covering over the $Sp/SO$ Gaiotto
curve. More precisely, the Seiberg-Witten curve is embedded in the
cotangent bundle $T^*C$ of the Gaiotto curve $C$ with
holomorphic differential $v$. The Seiberg-Witten differential is
simply 
\begin{align}
\lambda = v, 
\end{align}
the canonical 1-form in the cotangent bundle $T^*C$. 

Whereas for the $Sp(N-1)$ gauge theory
there is an extra condition saying 
that the zeroes at $v=0$ of the right-hand-side should be double
zeroes, the $SO(2N)$ gauge theory requires these zeroes to be simple
zeroes. These conditions come up somewhat ad-hoc in the type IIA
description, but can be explained from first principles in an M-theory
perspective \cite{Hori:1998iv}. The orientifold brane
construction lifts in M-theory to a stack of M5-branes in a $\IZ_2$-orbifold
background. The orbifold acts on the five dimensions transverse to the
M5-branes, and in particular maps $v \mapsto -v$. 

For the pure $Sp(N-1)$-theory the differential $\varphi_{2N}$ vanishes, so that
a factor $v^2$ in equation~(\ref{SpSOSWcover}) drops out. The
resulting Seiberg-Witten curve can be written in the form
\be
0=\det(v - \boldsymbol\varphi_{Sp}),
\ee
where $\boldsymbol\varphi_{Sp}$ is a
$Sp(N-1)$-valued differential. The non-vanishing differentials
$\varphi_{2k}$ can thus be obtained from the Casimirs
of the Lie algebra 
$\mathfrak{sp}(N-1)$. 
If we include
massive matter to the $Sp(N-1)$ 
gauge theory, however, or consider an $SO(2N)$ gauge theory,
equation~(\ref{SpSOSWcover}) can be reformulated as 
\be
0 = \det(v-\boldsymbol\varphi_{SO})
\ee
where the differential $\boldsymbol\varphi_{SO}$ is
$SO(2N)$-valued. This 
equation is clearly characterized by the Casimirs of the Lie algebra
$\mathfrak{so}(2N)$. More precisely, we recognize the $D_N$-invariants
$\Tr(\Phi^{2k})$ and $\mathrm{Pfaff}(\Phi)$ in the differentials $\varphi_{2k}$ and
$\varphi_{\tilde{N}} = \sqrt{\varphi_{2N}}$, respectively. In general,
the $\CN=2$ geometry for symplectic and orthogonal gauge groups is
thus encoded in a 
ramified $D_N$ Hitchin system based on the
$Sp/SO$ Gaiotto curve $C$, whose 
spectral curve is the Seiberg-Witten curve~(\ref{SpSOSWcover}).  

\begin{figure}[h!]
\begin{center}
\begin{tabular}{lr}
$\vcenter{\hbox{\includegraphics[width=2in]{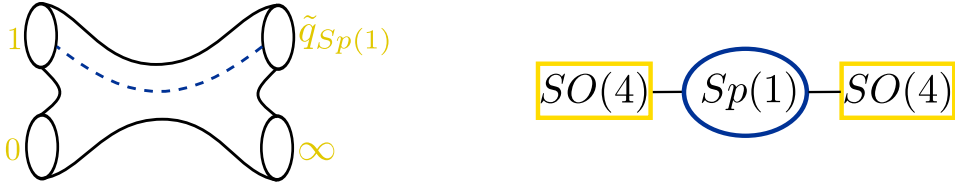}}}$ & \qquad \qquad $\vcenter{\hbox{\includegraphics[width=2.2in]{sp1quiver}} }$
\end{tabular}
\caption{The left Figure illustrates the Gaiotto curve $C$ of the
  conformal $Sp(1)$ quiver gauge theory that is illustrated on the 
  right. The $Sp(1)$ Gaiotto curve differs from the $SU(2)$ Gaiotto
  curve by the 
  $\IZ_2$ twist-line that runs parallel to the tube. We will discuss
  the precise relation between the $Sp(1)$ and the $SU(2)$ Gaiotto
  curve in section~\protect\ref{sec:Gexamples}.}
 \label{Sp1Gcurve}
\end{center}
\end{figure}

\begin{figure}[h!]
\begin{center}
\begin{tabular}{lr}
$\vcenter{\hbox{\includegraphics[width=2in]{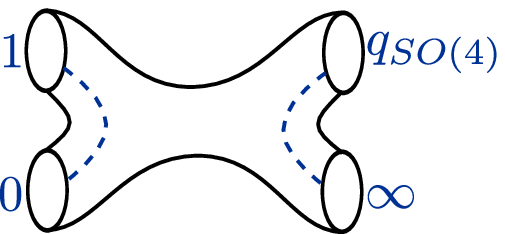}}}$ & \qquad \qquad $\vcenter{\hbox{\includegraphics[width=2.2in]{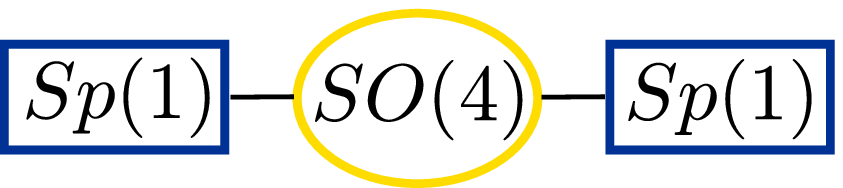}} }$
\end{tabular}
\caption{The left Figure illustrates the Gaiotto curve $C$ of the
  conformal $SO(4)$ quiver gauge theory that is illustrated on the 
  right. The $SO(4)$ Gaiotto curve differs from the $Sp(1)$ Gaiotto
  curve by a 
  different configuration of $\IZ_2$ twist-lines. In particular, the
  twist lines don't run through the tube.} 
 \label{SO4Gcurve}
\end{center}
\end{figure}

The Lie algebra $D_N$ has a $\IZ_2$ automorphism under which the
invariants with exponent $2, \ldots, 2N-2$ are even and the invariant of
degree $N$ is odd. On the level of the differentials $\phi_k$ this
translates into possible half-integer poles for the invariant
$\varphi_{\tilde{N}}$. Going around such a pole the differential
$\varphi_{\tilde{N}}$ has a $\IZ_2$ monodromy. We will see explicitly
in the examples. We call the puncture corresponding to such a pole a
half-puncture. The half-punctures introduce $\IZ_2$ twist-lines  
on the Gaiotto curve \cite{Tachikawa:2009rb}. This is illustrated
for the $Sp(1)$ and $SO(4)$ Gaiotto curve in Figure~\ref{Sp1Gcurve}
and Figure~\ref{SO4Gcurve}.

Lastly, let us make a few remarks on the worldvolume theory on a
stack of M5-branes. In the low energy limit this theory is thought to
be described by 
a six-dimensional conformal  $(2,0)$ theory of type $ADE$. For the M-theory 
background~(\ref{U(N)Mbackground}) it is of type $A$, whereas for the
$\IZ_2$-orbifolded M-theory background it is of type $D$. 
The $(2,0)$ theory has a ``Coulomb branch''
parametrized by the vev's of a subset of chiral operators whose
conformal weights are given by the exponents~$d$ of the Lie algebra
$\frak{g}$. These operators parametrize the
configurations of M5-branes in the
M-theory background. In the Hitchin system they appear as the
degree $d$ differentials.  Boundary conditions at the punctures of the
Gaiotto curve are expected to lift to defect operators in the
M5-brane worldvolume theory. We
refer to \cite{Gaiotto:2009hg} for a more detailed 
description.

\subsubsection{$SO/Sp$ versus $U$ geometries}

Suppose that we have two models for the same physical gauge theory,
who both can be embedded as a ramified Hitchin system in M-theory. How
are these models related geometrically?

First of all, we expect that the mapping between the exactly marginal
couplings is reflected as a mapping between the complex structure
parameters of the corresponding Gaiotto curves. Also, there should
be an isomorphism between the spectral curves of the respective
Hitchin systems, as these correspond to the Seiberg-Witten curves.  
Moreover, the boundary conditions of the Hitchin differentials at the
punctures of both models should be related by the mapping that
identifies the corresponding matter representations. In
particular, 
this relates the eigenvalues of the Seiberg-Witten differential at the
punctures of both models. In total we should
thus find a bijective mapping between the complete ramified Hitchin
system, 
including the Gaiotto curve as well as the Hitchin differentials.  
 
As we will see in detail in the next subsection, it is easy to
come up with such a mapping for $Sp(1)/SO(4)$ versus $SU(2)$ gauge
theories \cite{Tachikawa:2009rb}. We just interpret the $\IZ_2$-twist
lines on the $Sp(1)/SO(4)$ Gaiotto curve as branch-cuts, and its
double cover as the $SU(2)$ Gaiotto curve. The latter curve is thus
equipped with an involution that interchanges the two sheets of the
cover. We can recover the Hitchin differentials on the $Sp(1)/SO(4)$
Gaiotto curve by splitting the Hitchin differential on its
cover into 
even and odd parts under the involution. Indeed, recall that the
$SU(2)$ Seiberg-Witten curve is determined by a single differential
$\phi_2$  of
degree 2, whereas the $Sp(1)/SO(4)$ Seiberg-Witten curve is defined
by two degree 2 differentials $\varphi_2$ and
$\varphi_{\tilde{2}}$, the first one  being even under the
$\IZ_2$-automorphism and the second one odd.  
 
Notice that this double construction doesn't work for any $Sp/SO$
theory, as the differentials on the cover generically do not have a
simple interpretation in terms of a set of differentials of a unitary
theory. Two theories can only be related by a double covering if the
Lie algebra underlying the Hitchin system of one of them splits into
two copies of the Lie algebra underlying the Hitchin system of the
other. Nonetheless, for any two models of the same gauge theory there
should be a corresponding isomorphism of Hitchin systems.

Before going into the example-subsection, let us note that in the
previous section we also encountered a gauge theory without an obvious
brane embedding. This is the $Sp(1) \times Sp(1)$ gauge theory. It is not
possible to realize this theory in the standard manner using
NS5-branes, as the type of the 
orientifold has to differ on either side of the
NS5-brane. Geometrically, this is reflected in the fact that there
doesn't exist an involution on the $SU(2) \times SU(2)$ Gaiotto
curve with the right properties. It would be interesting to find
whether there exists a 
geometric interpretation of the mapping between the exactly marginal
couplings of the $SU(2) \times SU(2)$ and the  $Sp(1) \times Sp(1)$
gauge theory anyway.

\subsection{Examples}\label{sec:Gexamples}

Let us now return to the results of section~\ref{sec:gauge}. First of
all, we can explain the appearance of the modular lambda function 
\begin{align}
\lambda = \frac{\theta_2^4}{\theta_3^4}: \mathbb{H} \to \mathbb{P}^1 \backslash \{0,1,\infty\}
\end{align}
as the relation~(\ref{eqn:U2UV-IR}) between the infra-red coupling
$\tau_{\rm IR}$ and the exactly marginal coupling $q_{U(2)}$ in the conformal
$SU(2)$ gauge theory.
This modular function gives an explicit 
isomorphism between the quotient $\mathbb{H}/\Gamma(2)$ of the upper
half plane $\mathbb{H}$ by the modular group $\Gamma(2)$ and
$ \mathbb{P}^1 \backslash \{0,1,\infty\}$. Whereas the complex structure
modulus $\tau_{\rm IR}$ of the $SU(2)$ Seiberg-Witten curve takes
values in $\mathbb{H}/\Gamma(2)$, the complex structure
modulus $q_{U(2)}$ of the $SU(2)$ Gaiotto curve (which is the
cross-ratio of the four punctures on the G-curve) takes values in
$ \mathbb{P}^1 \backslash  \{0,1,\infty\}$. The modular lambda function thus
determines the double cover map between the Seiberg-Witten curve and
the Gaiotto curve for the conformal $SU(2)$ gauge theory \cite{AGT}.

We continue with studying the $Sp(1)$ and $SO(4)$ geometry in
detail. In appendix~\ref{app:SW} we have summarized various existing
descriptions of the $SU(2)$ geometry, and their relation to the
Gaiotto geometry.

\subsubsection{$Sp(1)$ versus $U(2)$ geometry}

The $Sp(1)$ Seiberg-Witten curve can be derived from the orientifold brane
construction that is illustrated in Figure~\ref{SpSObrane}. In Gaiotto
form it reads
\begin{align}\label{Sp1SWfib}
v^4 & = \varphi_2(s) v^2 + \varphi_4(s) 
\end{align}
with 
\begin{align*}
 \varphi_2(s) &= \frac{(\m_1^2+\m_2^2) s^2 + u(1+\tilde{q}_{Sp(1)})
   s + (\m_3^2+\m_4^2) \tilde{q}_{Sp(1)} 
}{\left(s-1\right)\left(s-\tilde{q}_{Sp(1)}\right)} \left(\frac{ds}{s}\right)^2 \\
  \varphi_4(s) &= -
\frac{\m_1^2 \m_2^2 s^2 + 2  \prod_{i=1}^4 \m_i \sqrt{\tilde{q}_{Sp(1)}}  s +\m_3^2
  \m_4^2
  \tilde{q}_{Sp(1)}}{\left(s-1\right)\left(s-\tilde{q}_{Sp(1)}\right)}
\left(\frac{ds}{s}\right)^4, 
\end{align*}
where $\m_i$ are the bare masses of the
hypermultiplets, and $u$ is 
the classical vev of the adjoint scalar $\Phi$ in the gauge
multiplet. We also introduced a new parameter
$\tilde{q}_{Sp(1)}$
that we will relate to the coupling $q_{Sp(1)}$ in a
moment. The 
differential $\varphi_2$ corresponds to the $D_2$-invariant
$\Tr(\Phi^2)$, 
whereas the square-root $\varphi_{\tilde{2}} = \sqrt{\varphi_4}$
corresponds 
to the $D_2$-invariant $\rm{Pfaff}(\Phi)$.  Note that the differential
$\varphi_4$ vanishes if the masses are set to zero. 

It follows that the $Sp(1)$
Seiberg-Witten curve is a branched fourfold cover over 
the G-curve $\IP^1$ with coordinate $s$. The G-curve has 
four branch points at the positions 
\be
s\in \{ 0, \tilde{q}_{Sp(1)}, 1, \infty\},
\ee
The complex structure of the
$Sp(1)$ Gaiotto curve is parametrized by $\tilde{q}_{Sp(1)}$.
Since the differential $\varphi_{\tilde{2}}$ has a pole of order
half at the punctures at $s=1$ and $s=\tilde{q}_{Sp(1)}$, these are
half-punctures. There is a $\IZ_2$-twist line running between the two
half-punctures, as the differential $\varphi_{\tilde{2}}$ experiences
a $\IZ_2$ monodromy around them. In contrast, the punctures at $s=0$
and $s=\infty$ are full punctures. 

The Seiberg-Witten differential $\lambda$ is a $SO(4,\IC)$-valued
differential. It has nonzero residues at the
poles $s=0$ and $s=\infty$ only. This implies that the $SO(4)$ flavor
symmetry is associated with these punctures. Indeed, the residues
of the differential $\lambda$ at $s=\infty$ are given by $\pm \mu_1$ and
$\pm \mu_2$, whereas at $s=0$ they are $\pm \mu_3$  and $\pm
\mu_4$. So both at $s=0$ and $s=\infty$ the residues parametrize the
Cartan of $\mathfrak{su}(2) \times \mathfrak{su}(2) =
\mathfrak{so}(4)$.   

Summarizing, we have found that the $Sp(1)$ G-curve is a
four-punctured two-sphere with two half-punctures and two full
punctures. The two $SO(4)$-flavor symmetry groups can be associated
to the two full punctures. This is illustrated in the left
picture in Figure~\ref{Sp1Gcurve}. 

\begin{figure}[h!]
\begin{center}
\includegraphics[width=2in]{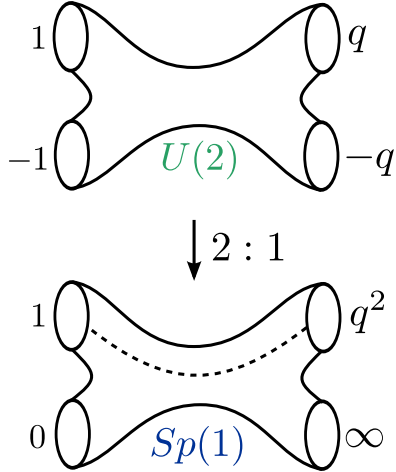}
\caption{The G-curve of the $SU(2)$ gauge theory coupled to
4 hypers is a double cover over the G-curve of 
the $Sp(1)$ gauge theory with 4 hypers. We denote the complex
structure parameter on the $Sp(1)$ Gaiotto curve by $\tilde{q}_{Sp(1)}
= q^2$. }\label{USpGcurve}
\end{center}
\end{figure}

Viewing the $\IZ_2$-twist lines as branch-cuts, and the
half-punctures as branch-points, it is natural to consider
the double cover of the G-curve. 
Let us use the $SL(2,\IC)$ freedom of the $Sp(1)$ theory to
interchange the full-punctures with the half-punctures. Call the
complex structure coordinate of the $Sp(1)$ G-curve $\tilde{q}_{Sp(1)}
= q^2$. The branched
covering map is then simply given by   
\begin{align*}
t^2 = s,
\end{align*}
where $s$ is the coordinate on the $Sp(1)$ G-curve and $t$ the
coordinate on the double cover.
The pre-images of the full punctures on the base 
are at $\pm 1, \pm q$ on the cover,
and there is a $SU(2)$-flavor symmetry attached
to them. The total flavor symmetry at both full-punctures
adds up to $SO(4)$. This is illustrated in Figure~\ref{USpGcurve}.

Note that this double cover of the $Sp(1)$ G-curve has exactly the same
structure as the $SU(2)$ G-curve. The only difference is that the
punctures are at different positions, something which can be taken
care of by 
a M\"obius transformation. Since this leaves the complex structure of
the G-curve invariant, the gauge theory is invariant under such
transformations. In particular the masses of the hypermultiplets,
which are the residues of the 
Seiberg-Witten differential at its poles, remain the same.
We can use the fact that the cross-ratios of the
two configurations have to be equal to read
off the relation between the $U(2)$ and $Sp(1)$ exactly marginal couplings,
\begin{align}\label{eqn:geomUVUV}
\boxed{q_{U(2)} = 4 q \left(1+ q \right)^{-2}.}
\end{align}
Explicitly, the M\"obius transformation
that relates the $SU(2)$ 
G-curve and the double cover of the $Sp(1)$ G-curve is
given by the mapping
\begin{align}\label{eqn:mobiusSp1} 
\gamma(z)=-\frac{z\left(1+q\right)-2q}{z\left(1+q\right)-2}\ ,
\end{align}
that sends the four punctures at positions $\{ 0,1,q_{U(2)},\infty
\}$ to four punctures at the positions $\{ \pm q, \pm 1 \}$.

We can now make contact between the geometry of the $SU(2)$ and
$Sp(1)$ Gaiotto curves and the relation between their exactly marginal
couplings. Indeed, we recover the UV-UV mapping (\ref{eqn:UV-UV-U2Sp1})
from the identification of cross-ratios in
equation~(\ref{eqn:geomUVUV}), when 
we identify 
\begin{align}
\boxed{q^2 = \tilde{q}_{Sp(1)} = \left( \frac{ \ q_{Sp(1)}}{4} \right)^2. }
\end{align}
We should therefore choose the complex structure parameter
$\tilde{q}_{Sp(1)}$ of the 
$Sp(1)$ Gaiotto-curve proportional to the square $q_{Sp(1)}^2$ of the
$Sp(1)$ instanton parameter. The square is related to the
$\IZ_2$-twist line along the $Sp(1)$ Gaiotto curve. The
proportionality constant is merely determined by requiring that 
$
q_{U(2)} = q_{Sp(1)} + \ldots, 
$
which is needed to make the classical contributions to the Nekrasov
partition function agree.

\subsubsection{$SO(4)$ versus $U(2) \times U(2)$
  geometry}\label{sec:so4geometry} 

The $SO(4)$ Seiberg-Witten curve in Gaiotto form reads
\begin{align}\label{SO4SWfib}
v^4 & = \varphi_2(s) v^2 + \varphi_4(s) 
\end{align}
with 
\begin{align*}
 \varphi_2(s) &= \frac{\m_1^2 s^2 - (u_1+u_2)(1+\tilde{q}_{SO(4)})
   s + \m_2^2 \tilde{q}_{SO(4)} 
}{\left(s-1\right)\left(s-\tilde{q}_{SO(4)}\right)} \left(\frac{ds}{s}\right)^2 \\
  \varphi_4(s) &= 
\frac{u_1 u_2 s }{\left(s-1\right)\left(s-\tilde{q}_{SO(4)}\right)}
\left(\frac{ds}{s}\right)^4, 
\end{align*}
The Seiberg-Witten curve is a fourfold cover of the
four-punctured sphere with complex structure parameter
$\tilde{q}_{SO(4)}$. This time there are four 
half-punctures. The differential $\varphi_{\tilde{2}}$ not only has poles
of order $1/2$ at $s=1$ and $s=\tilde{q}_{SO(4)}$, but also poles of order
$3/2$ at $s=0$ and $s=\infty$. The residue of the Seiberg-Witten
differential is only nonzero at $s=0$ and $s=\infty$. Since its nonzero
residues equal $\pm
\mu_1$ at $s=\infty$ and $\pm \mu_2$ at $s=0$, the 
$Sp(1)$ flavor symmetry is associated to these punctures. 

Summarizing, the G-curve corresponding to the conformal $SO(4)$ theory is a
four-punctured sphere with four half-punctures, and $\IZ_2$ twist-lines running
between two pairs of half-punctures.  This is illustrated on the left
in Figure~\ref{SO4Gcurve}.

\begin{figure}[htbp]
\begin{center}
\includegraphics[width=1.5in]{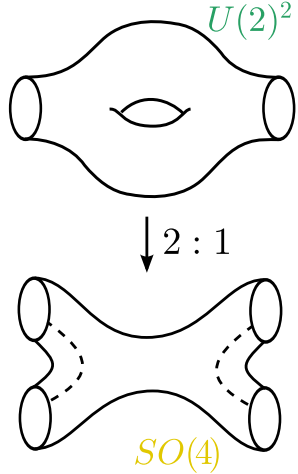}
\caption{The G-curve of the $SU(2) \times SU(2)$ gauge theory coupled by
  two bifundamentals is a double cover over the G-curve of 
the $SO(4)$ gauge theory coupled to four hypers.} \label{fig:USOGcurve}
\end{center}
\end{figure}

Let us again interpret the $\IZ_2$ twist-lines as branch-cuts. 
As illustrated in Figure~\ref{fig:USOGcurve}, this time the branched double cover
of the G-curve is a torus with two punctures. The
punctures on the torus project to
two of the four branch-points of the double covering, which means that there must be a
$SU(2)$-flavor symmetry attached to them. This is precisely the 
 the same structure as that of the Gaiotto curve of the $SU(2)\times SU(2)$
gauge theory coupled to two bifundamentals. 

However, note that the $SU(2)\times SU(2)$ G-curve has two complex  
structure parameters corresponding to the two marginal couplings
$q_{U(2),1}$ and $q_{U(2),2}$, whereas the $SO(4)$
G-curve has only a single complex structure parameter corresponding to the
marginal coupling $q_{SO(4)}$ of the $SO(4)$ theory. The
$\IZ_2$-symmetry on the cover curve implies that we should
identify $q_{U(2)} = q_{U(2),1}=q_{U(2),2}$ in order to relate the
$SU(2) \times SU(2)$ Gaiotto curve to the $SO(4)$ Gaiotto curve. 

In equation~(\ref{eqn:SOUVUV}) we found that the relation between the marginal 
couplings of the $SO(4)$ and the $SU(2)\times SU(2)$-theory is given by
the modular lambda mapping  
\begin{align}\label{eqn:UVUVso4u2}
\boxed{q_{SO(4)} = \lambda (2 \tau_{U(2)} )\,.}
\end{align}
The above covering relation between their Gaiotto curve explains the
appearing of this lambda mapping geometrically, as it relates the
complex structure parameter of the torus to the complex structure
parameter of the four-punctured sphere, when we identify  $\tilde{q}_{SO(4)} = q_{SO(4)}$.

Mainly for future convenience, let us make the covering map
explicitly. Take a torus $T^2$ with half-periods 
$\omega_1$ and $\omega_2$ and consider the map $T^2 \to \IC\IP^2$
given by 
\begin{align}
(\wp(z):\wp'(z):1), \notag
\end{align}
where $\wp(z)$ is the Weierstrass $\wp$-function. Since the Weierstrass
$\wp$-function satisfies
\begin{align}
\wp'(z)^2 = 4 \wp(z)^3 - g_2 \wp(z) - g_3, \notag
\end{align}
it defines a branched double cover over the sphere $\IP^1$. The
branch points are determined by the zeroes and poles 
$\{0, \w_1, \w_2, \w_3 = \w_1 + \w_2 \}$
of the derivative $\wp'(z)$ of the Weierstrass $\wp$-function. The
double covering is thus given by the equation
\begin{align}
t^2 = 4 (s-e_1)(s-e_2)(s-e_3) \notag
\end{align}
with $e_i = \wp(\w_i)$. The points $\w_3 = \tau$ and $2 \tau$
on the torus $T^2$ map onto the two branch points $e_3$ and $\infty$
on the sphere $\IP^1$. Let us call $q_{SO(4)}$ the cross-ratio of the
four branch points $0,e_1,e_2,e_3$.  By the shift 
$s \mapsto s+\frac{(1+q_{SO(4)})}{3}$
we simply bring the curve into the form  
\begin{align}
t^2 = 4 s (s-1) (s-q_{SO(4)}). \notag
\end{align}
The mapping between the $SU(2) \times SU(2)$ Gaiotto curve and the
$SO(4)$ Gaiotto curve is thus a combination of the Weierstrass
map $\wp$ and a simple M\"obius transformation. This
indeed determines (\ref{eqn:UVUVso4u2}) as the mapping between the
respective complex structure parameters.

\section{Conformal blocks and $\CW$-algebras}\label{sec:cft}

Compactifying the six-dimensional superconformal $(2,0)$ theory of type
$ADE$ on either a two-dimensional Riemann surface or a four-manifold,
suggests that there should be a correspondence between the following two
systems. The first system is a four-dimensional
superconformal $\CN=2$ gauge theory with $ADE$ gauge group and whose
Gaiotto curve is equal to the Riemann surface. The second system is a 
two-dimensional field theory that lives on the Riemann
surface and should be characterized by an $ADE$ type.

Remember that the tubes of the
Gaiotto curve are associated to the ADE gauge group of the $\CN=2$
gauge 
theory, and the punctures on the G-curve to
matter.  Decomposing the G-curve into pairs of pants 
suggests that the marginal gauge couplings $\tau_{\rm   UV}$ should be
identified with the sewing parameters $q=\exp(2 \pi i \tau_{\rm UV})$
of the curve. The symmetry of the two-dimensional theory
should be related to the ADE gauge group. Furthermore, two-dimensional
operators that are
inserted at punctures of the G-curve should encode the flavor
symmetries of the corresponding matter multiplets.

A particular instance of such a 4d--2d
connection was discovered in \cite{AGT}. It was found that
instanton partition functions  in the $\Omega$-background
$\IR^4_{\e_1,\e_2}$  for linear and cyclic $U(2)$ quiver gauge
theories are closely
related to Virasoro conformal blocks of the pair of pants decomposition
of the corresponding G-curves. In
this so-called AGT correspondence the central 
charge of the Virasoro algebra is determined by the value of the two
deformation parameters $\e_1$ and $\e_2$ as 
\begin{align}
c &= 1 + \frac{6(\e_1 + \e_2)^2}{\e_1 \e_2}.
\end{align}
The conformal weights of the
vertex operators at the
punctures of the G-curve are specified by the masses of the 
hypermultiplets in
the quiver theory, and the conformal weights of the fields
in the internal channels are related in the same way to the Coulomb 
branch parameters.

\begin{figure}[htbp]
\begin{center}
\includegraphics[width=2.0in]{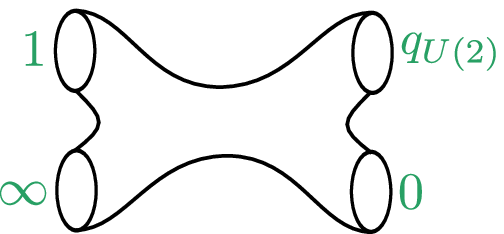}
\caption{The AGT correspondence relates the instanton
  partition function of the $U(2)$ gauge theory coupled to four 
  hypermultiplets to a Virasoro conformal blocks on the four-punctured
  sphere with vertex operator insertions at the four punctures.} \label{fig:U2AGT}
\end{center} 
\end{figure}

More precisely, the instanton partition function can be 
written as the product of the Virasoro conformal block
times a factor that resembles a $U(1)$ partition function.
The interpretation for this is that the single Casimir
of degree 2 of the subgroup $SU(2) \subset U(2)$ corresponds to the 
energy-stress tensor of the CFT, and the overall
$U(1)$ factorizes. This picture was made more explicit in
\cite{Alday:2010vg}. 

For general gauge groups with more Casimirs we expect
the CFT to have a bigger symmetry group. Since the
symmetries of a CFT are captured by its so-called $\CW$
or chiral algebra, we expect that the Nekrasov partition
functions should be identified with $\CW$-blocks
instead of Virasoro blocks. 
For $SU(N)$, this picture was proposed and checked in
\cite{Wyllard:2009hg}. The relation between Hitchin systems and
$\CW$-algebras was also discussed in
\cite{Bonelli:2009zp}. Before continuing let us first introduce $\CW$-algebras 
and some other concepts in CFT.

\subsection{$\CW$-algebras, chiral blocks and twisted representations}
The concept of a conformal block can be generalized for
theories with bigger symmetry groups.
The symmetries of a two-dimensional conformal field theory are given
by its $\CW$ (or chiral) algebra. The algebra $\CW$ always contains the
energy-stress tensor $T$, a distinguished copy
of the Virasoro algebra describing the behavior
under conformal coordinate transformations.  The fields of the
theory decompose into highest weight 
representations of the $\CW$-algebra.
For an introduction to $\CW$ algebras, see \cite{Bouwknegt:1992wg}.

There is no complete classification all $\CW$ algebras, but many
examples are known and have been studied.
One particular family of examples are the the so-called Casimir
algebras, which are based on simply laced Lie algebras. 
Its generators are constructed from the
$\mathfrak{g}$-invariant contractions of the current field $J(z)$ 
of the affine Lie algebra $\mathfrak{g}$. The series of
$\CW_N$-algebras, for instance, is related to the $A_N$ Lie
algebras. In \cite{Wyllard:2009hg} $\CW_N$-blocks have been related to the
instanton partition functions corresponding to $U(N)$ gauge theories. It
is natural to expect that also the other Casimir algebras appear as
dual descriptions of instanton counting.

Since the spectrum of the CFT decomposes into representations
of the $\CW$ algebra, we can use generalized 
Ward identities to relate correlation functions
of ($\CW$-)descendant fields to correlation functions of
($\CW$-)primary fields. In the case of the Virasoro
algebra, we can always reduce them to functions
of primary fields only. For general $\CW$-algebras
this is only possible if one restricts to primary
fields on which the $\CW$-fields 
satisfy additional null relations.

We can make use of this property by computing
chiral blocks.
For a given configuration of a punctured
Riemann surface,
we define the chiral block by
picking a representation $\phi$ for
every tube, inserting the projector on the representation 
$P_{\CH_\phi}$ at that point in the correlator, 
and dividing by the product of all three point
functions of the primary fields. By the above remarks
the result is then independent of the three-point
functions of the theory, \ie it only depends
on the kinematics of the theory.

In the simplest configuration, the sphere with four punctures,
the chiral block is thus given by
\be
\CF =
\frac{\langle V_1(\infty)V_2(1) P_{\CH_\phi} V_3(q)V_4(0)\rangle}
{\langle V_1(\infty)V_2(1)|\phi\rangle\langle\phi|V_3(1)V_4(0)\rangle}\ .
\ee
Note that this definition differs slightly from the usual
definition, as we have not divided out a factor $q^{h_\phi-h_3-h_4}$.
Our definition will be slightly more convenient to work with.
On the gauge theory side it corresponds to the partition function
including the perturbative contribution.
Also, let us take the convention in what follows that whenever we write
a correlator, we assume that it is divided by the appropriate 
primary three point functions.
The projector is usually written as
\be
P_{\CH_\phi}=\sum_{I,J} |\phi_{I}\rangle\langle\phi_{J}|(K^{-1})_{I,J}
\ee
where $I=(i_1,i_2,\ldots)$ denotes the $\CW$ descendants, such that
$\phi_{I}= W_{-i_1}W_{-i_2}\cdots \phi$ is a
$\CW$ descendant. $K$ is the
inner product matrix and the sum runs over all $\CW$ descendants
of $\phi$.

A representation $\phi$ of $\CW$ is called untwisted if
it is local with respect to $\CW$, so that one
can freely move $\CW$-fields around it. The
$\CW$-fields then have integer mode representations
around that representation.

More generally, the $\CW$-fields can pick up phases
when circling around $\phi$, so that the correlation
function has a branch cut extending from $\phi$.
Such $\phi$ are called twisted representations. Because
of the phase $\alpha$ picked up by the $\CW$-fields,
their modes are no longer integer, but given by $r\in \IZ + \alpha$.

A particular case of twisted representations can appear
when the $\CW$ has an outer automorphism such as
a $\IZ_N$-symmetry. Let us say that by circling
around a twisted representation the algebra $\CW$ gets
mapped to an image under $\IZ_N$, such as
\be
W_k \mapsto W_{k+1}\ , \qquad k=1,\ldots N\ .
\ee
By choosing
linear combinations $W^{(k)}$ of the modes $W_k$
that are eigenvectors under the automorphisms,
the $W^{(k)}$ indeed pick up phases
$2\pi i k/N$. For $N=2$, the case that we are interested
in below, the $\CW$-algebra thus decomposes
into generators $W^{+}, W^{-}$ of 
integer and half-integer modes respectively. 

Let us finally note that in the case of Liouville
theory the $\CW$-algebra is simply the 
Virasoro algebra. Example of conformal field theories with bigger
$\CW$-algebras are Toda 
theories.

\subsection{The $SO(4)$ and $Sp(1)$ AGT correspondence}

Remember that the $\CN=2$ geometry is characterized by a ramified
Hitchin system on the Gaiotto curve. 
For conformal $SO(2N)$ and
$Sp(N-1)$ gauge theories the Hitchin system is described in 
terms of the differentials $\phi_{2k}$ (for $k=1,\ldots,N-1$) and
$\tilde\phi_N$ that can be constructed out of the $D_N$-invariants $\Tr
\left(\Phi^{2k} 
\right)$ and $\textrm{Pfaff}(\Phi)$, respectively. 
In the
six-dimensional $(2,0)$ theory these differentials appear as a set of
chiral operators whose conformal weights are equal to the 
exponents of the Lie algebra. When we reduce the six-dimensional
theory over a four-manifold we expect these operators to turn into the
Casimir operators $\CW^{(2k)}$ and $\tilde \CW^{(N)}$ of the
$\CW(D_N)$-algebra.  
The $\mathbb{Z}_2$-automorphism of the $D_N$-algebra translates to an additional $\mathbb{Z}_2$-symmetry on the level of the CFT, and we
thus expect a relation to a twisted 
$\CW(D_N)$-algebra.  In other words,  
 we expect that the Lie algebra underlying the Hitchin
system is precisely reflected in
the Casimir operators of the corresponding $\CW$-algebra on the
Gaiotto curve.  

Let us now put all the pieces together and formulate the
AGT correspondence for $SO(4)$ and $Sp(1)$.
Since the definition of the Gaiotto curve for both theories
  involves the $SO(4)$-invariants $\phi_2$ and $\tilde{\phi}_2$ of
  degree two, we
  expect this CFT to have an underlying  $\CW(D_2)$-algebra.  We will
  denote this algebra by $\CW(2,2)$, as it contains two Casimir
  operators of weight two. In fact those operators correspond
  to two copies $T^A$ and $T^B$ of the Virasoro algebra.

Similar to the correspondence between $U(2)$ instanton partition
function and Virasoro conformal blocks, the new correspondence 
is between
$SO(4)/Sp(1)$ instanton partition functions and twisted
$\CW(2,2)$-algebra blocks. The configuration of the block is given in
the following way: At the full punctures of the $SO(4)/Sp(1)$
G-curve insert untwisted vertex operators, whose weights correspond
to the masses of the $Sp(1)$ fundamental hypers.
At the half punctures, insert twisted vertex operators. 
Whenever a half-puncture
lifts to a regular point on the cover, we should insert the
vacuum of the twist sector $\sigma$, which we will
describe later on. 
For any half-puncture that lifts to a puncture on the
cover we may insert a general twisted field, whose
single weight corresponds to the mass of the $SO(4)$ fundamental
hyper.  

\begin{figure}[htbp]
\begin{center}
\includegraphics[width=3.1in]{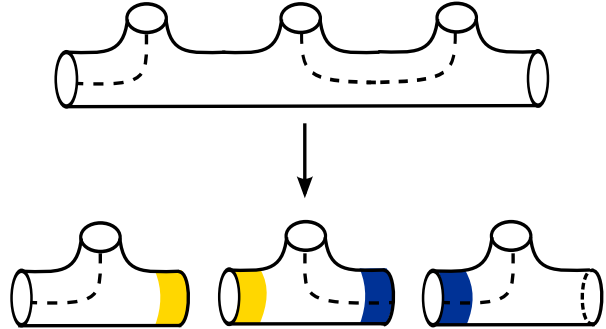}
\caption{Twisted $\CW(2,2)$-algebra blocks can be computed by decomposing
  the Riemann surface into pairs of pants. Internal tubes
  that have a $\IZ_2$ twist line (blue) correspond to an
  $Sp(1)$ gauge group and carry twisted representations
  of $\CW(2,2)$. Internal tubes without a twist line (yellow)
  correspond to $SO(4)$ and carry two copies of the Virasoro algebra.} 
  \label{CFTdecompose}  
\end{center}
\end{figure}

When decomposing the Gaiotto curve into pair of pants, we cut
tubes with or without twist lines (see Figure~\ref{CFTdecompose}). A
tube with a twist line 
corresponds to a $Sp(1)$ gauge group, and the weight of
the twisted primary in the channel 
corresponds to its single Coulomb branch parameter $a$.
A tube without a twist line corresponds
to a $SO(4)$ gauge group, and the two weights
of the untwisted primary in the channel correspond
to the two Coulomb branch parameters $a_1$ and $a_2$.
All of this is summarized in table~\ref{AGTidentification}.
\begin{table}
\begin{equation*}
\renewcommand{\arraystretch}{1.3}
\begin{tabular}{|@{\quad}c@{\quad}|@{\quad}c@{\quad}| }
\hline  {\bf \CN =2 gauge theory} & {\bf CFT}
\\
\hline
\hline  
$SO(4)/Sp(1)$ quiver & $SO(4)/Sp(1)$ Gaiotto curve \\
\hline 
$Sp(1)$ fund. hyper $(\mu_1,\mu_2)$& untwisted $\CW(2,2)$ representation $(h_{\mu_1},h_{\mu_2})$  \\
\hline
$SO(4)$ fund. hyper $\mu$& twisted $\CW(2,2)$ representation $h_{\mu}$ \\
\hline 
$Sp(1)-SO(4)$ bifund. hyper & twist vacuum $\sigma$\\
\hline 
$Sp(1)$ Coulomb par. $a$& weight of twisted int. channel $h_a$\\
\hline
$SO(4)$ Coulomb pars. $(a_1,a_2)$ & weights of untwisted int. channel
$(h_{a_1},h_{a_2})$
\\
\hline
\end{tabular}
\label{closed_params}
\end{equation*}
\caption{The AGT correspondence for $SO(4)/Sp(1)$. \label{AGTidentification}}
\end{table}

The detailed identification of parameters can be found in
the examples we will work out. These examples will show that
$Sp(1)/SO(4)$ instanton partition functions agree with the 
twisted $\CW(2,2)$ up to a spurious factor that is independent of
the Coulomb and mass parameters of the gauge theory.
Note in particular that, unlike in the original $U(2)$ case, no $U(1)$ prefactor
appears, which is exactly what one would expect.

We can also use this correspondence to explain the
relation between $Sp(1)/SO(4)$ and $U(2)$ theories.
More precisely, given an $Sp(1)/SO(4)$ Gaiotto curve, we first
map the corresponding chiral block to its double cover.
This is in fact a well-known method to compute
twisted correlators.
The resulting configuration can then be mapped to
a $U(2)$ configuration by a suitable conformal coordinate transformation.
We will argue below that such a coordinate transformation
only introduces a spurious factor. It thus follows that
the partition function of the $U(2)$ configuration
agrees up to a spurious factor with the partition function
of the $Sp(1)/SO(4)$ configuration once expressed in terms
of the same coupling constants. To put it another way, 
the difference between $U(2)$ and $Sp(1)/SO(4)$ partition
functions is indeed only a reparametrization of the moduli
space caused by choosing a different renormalization scheme.

\subsection{Correlators for the $\CW(2,2)$ algebra and the cover
  trick}
As mentioned above, the algebra $\CW(2,2)$ contains two Casimir
  operators of weight two.
 These operators can be identified with two
  Virasoro tensors $T^A(z)$ and $T^B(z)$. The $\CW$-algebra thus
  decomposes into two copies of the Virasoro algebra. This reflects the
  decomposition of the Lie algebra $\mathfrak{so}(4) \cong
  \mathfrak{su}(2)_A\times \mathfrak{su}(2)_B$.  Geometrically,
the fact that we find
  two copies of the Virasoro algebra follows simply from the
  double covering that relates the $U(2)$ and the $Sp(1)/SO(4)$
  Gaiotto curves.  
A single
  copy of 
  the Virasoro algebra associated to the cover $U(2)$ Gaiotto curve
  descends 
  to two copies of the Virasoro algebra on the base $Sp(1)/SO(4)$
  Gaiotto curve. 

\begin{figure}[h!]
\begin{center}
\includegraphics[width=3.5in]{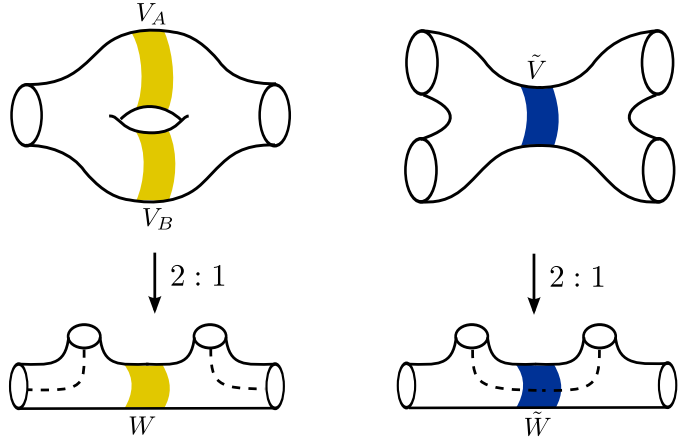}
\caption{On the left (right): Illustration of the branched
    double covering of the $SU(2)$ G-curve over the
  $SO(4)$ G-curve ($Sp(1)$ G-curve). The yellow (blue) tubular
  neighborhoods $W$ and $\tilde{W}$
  on the base  curves are part of internal tubes without (with) a
  $\IZ_2$ twist line. On the 
  cover the yellow (blue) patches illustrate their respective inverse
  images. The $\CW$-algebra modes associated to
  both base tubes lift to a single copy of the
  Virasoro algebra on their inverse images.}  \label{SOSpGcurveCFT} 
\end{center}
\end{figure}

As illustrated on the left in Figure~\ref{SOSpGcurveCFT}, this is in
particular the case for an internal tubular neighborhood of the $SO(4)$
Gaiotto 
curve. The single copy of the Virasoro algebra on its inverse image
descends to two copies of the Virasoro algebra on the tubular neighborhood 
itself. The $SO(4)/Sp(1)$ Gaiotto curves additionally contain
$\mathbb{Z}_2$ twist-lines. When 
crossing such a twist-line the two copies of the Virasoro algebra get
interchanged. 
We thus propose an underlying twisted
  $\CW(2,2)$-algebra. The actual energy-stress
  tensor $T^+(z)=T^A(z)+T^B(z)$ is of course invariant, whereas
  $T^-(z)=T^A(z)-T^B(z)$ picks up a minus sign. 

To compute the corresponding chiral blocks, we decompose the Gaiotto  curves 
into pair of pants and sum over all $\CW$-descendants 
of a given channel. The only difference with Virasoro correlators is that
there are now two types of tubes to cut,
those with $\IZ_2$ twist-lines and those without. This is illustrated in
Figure~\ref{CFTdecompose}.

When cutting open a tube without a $\mathbb{Z}_2$ twist-line, 
the intermediate fields are given by $L^A$ and $L^B$-descendants
of an untwisted representation, 
characterized by the conformal weights $(D_A,D_B)$ under $L^A_0$ and
$L^B_0$. 
We can therefore associate the Hilbert space
\begin{align}
\CH_{SO(4)} = \{ L^{A}_{-m_1} \cdots |\phi_A\rangle  \otimes 
L^{B}_{-n_1}\cdots |\phi_B \rangle\ 
: m_i \in \IN, n_i \in \IN \},
\end{align}
where $|\phi_{A/B} \rangle$ has weight $D_{A/B}$, to a tube without a
$\mathbb{Z}_2$ twist-line. 
On the other hand, if we cut a tube with a $\mathbb{Z}_2$ twist-line,
the intermediate fields are in a twisted representation
of the $\CW$-algebra. It is then most convenient
to describe them in terms of descendants of $L^+$
and $L^-$,
\begin{align}
\CH_{\widetilde{Sp}(1)} = \{ L^{+}_{-m_1}\cdots L^{-}_{-r_1}\cdots |\phi_C \rangle\ 
: m_i \in \IN, r_i \in \frac{1}{2}+\IN \}\ .
\end{align}
Note that since $L^-$ has no zero mode, the representation
$\phi_C$ is characterized by just a single weight.

To actually compute three point functions with twist fields,
we can
use the well-known cover trick, which is nicely explained in \eg 
\cite{Dixon:1986qv,Cardy:2007mb} : We find a function
that maps the punctured Riemann surface with
branch cuts to a cover surface which does not
have any branch cuts. 
Since the theory is conformal, we know
how the correlation functions transform
under this map.
On the cover we can
then evaluate a correlation function with
no twisted fields and no branch cuts in the
usual way. In order for this to work,
we need to find a cover map that
has branch points where the twist fields
are inserted.
The precise map from the base to the cover thus depends
on the positions of the branch cuts. 
On the cover there is then only a single
copy of the Virasoro algebra.

To illustrate all of this, let us take the following
simple model as a
map from the cover to the base:
\be
\tilde z \mapsto z=\tilde z^2\ .
\ee
This particular map has branch cuts at 0 and $\infty$,
and is thus suitable to deal with correlation functions
that have twist fields at those two points.
We can relate the stress-energy tensor on the cover
$T(\tilde z)$ to the two copies on the base in
the following manner.
The stress-energy
tensor on the cover transforms to 
\be
T(\tilde z)= \left(\frac{d\tilde z}{dz}\right)^{-2}\left[T(z)-\frac{c}{12}\{\tilde z;z\}\right]\ ,
\ee
where the Schwarzian derivative given by
\be
\{\tilde z;z\}=\frac{\tilde z'''}{\tilde z}-\frac{3}{2}\left(\frac{\tilde z''}{\tilde z'}\right)^2\ 
\ee
appears because $T$ is not a primary field. Around the branch point 0 on the base
we can then define two $\CW$-fields $T^+$ and $T^-$ by picking out the even and
odd modes of $T$,
\begin{eqnarray}
L^+_n &=& 2\oint \frac{d z}{ z^{n-1}} T( z) =
\frac{1}{2}L_{2n}+\frac{3c}{48}\delta_{n,0} \qquad (\mathrm{for} \, n \in \IZ), \\
L^-_r &=& 2\oint \frac{d z}{ z^{r-1}} T( z) = \frac{1}{2}L_{2r} \qquad (\mathrm{for} \, r \in \frac{1}{2}+\IZ).
\end{eqnarray}
The $L^+$ then form a Virasoro algebra with central charge $2c$, and
$T^-$ is a primary field of weight 2.
As discussed above, the twisted field $\phi$ at the point 0 is a 
twisted representation of 
$T^-$ and $T^+$ which has only one weight, namely
the eigenvalue of $L^+_0$. This means that
on the cover point there sits a field
$\phi$ which is an untwisted representation of the
Virasoro algebra of the corresponding weight, 
and its $L^+$ and $L^-$ descendants are
given by even and odd $L$ descendants.

There is one special twist field $\sigma$ 
which has
the property that its lift to the cover gives the
vacuum. It has the lowest possible conformal weight
for a twist field and serves in some sense as the
vacuum of this particular twist sector.

Around any other puncture on the base that is not a branch point, we simply
obtain two independent copies $L^A$ and $L^B$ of the Virasoro algebra,
coming from the two pre-images of the punctures on the cover.
As long as we stay away from branch points,  
the Virasoro tensor $T(\tilde z)$ on the cover is given by $T^A(\tilde z)$ on
the first and by $T^B(\tilde z)$ on the second sheet of the cover.
Since $T^A$ and $T^B$ commute on the base, 
a field $\phi^{A,B}$ on the base factorizes
into representations of $T^A$ and $T^B$,
$\phi^{A,B}=\phi^A\otimes \phi^B$ with
conformal weights $(D_A,D_B)$ under both
copies.
On the cover this leads to two untwisted fields
$\phi^A$ and $\phi^B$ sitting at the two images
of the cover map, both of which are
again untwisted representations of the Virasoro algebra.

Let us now to turn to some more technical points.
The map from the base to the cover in general introduces
corrections to the three point functions.
In particular since one or more of those fields
are descendants, they will exhibit more complicated
transformation properties than we are used
to from primary fields. Let us therefore briefly discuss
how conformal blocks behave under coordinate transformations.

When dealing with descendants fields, it will be
useful to use the notation $\phi(z)=V(\phi,z)$,
which we shorten to $V_i(z)$ if $\phi_i(z)$ is 
a primary field. 
The transformation of a general descendant field $\phi$ 
under a general coordinate transformation $z\mapsto f(z)$ 
is given by \cite{Gaberdiel:1994fs} 
\be\label{CFTtrafo}
D_f V(\phi,z) D_f^{-1}= V \left( f'(z)^{L_0}\prod_{n=1}^\infty
  e^{T_n(z)L_n} \phi ,f(z) \right) ,
\ee
where the operator $D_f$ is given by
\be
D_f= e^{f(0)L_{-1}} f'(0)^{L_0} \prod_{n=1}^\infty e^{T(0)_n L_n}\ .
\ee
Here we take all products to go from left to right.
The functions $T_n(z)$ are defined recursively. The first two
are given by
\be
T_1(z)=\frac{f''(z)}{2f'(z)}\ , \qquad T_2(z)=\frac{1}{3!}\left(\frac{f'''(z)}{f'(z)}
-\frac{3}{2}\left(\frac{f''(z)}{f'(z)}\right)^2\right)\ .
\ee
First note that if $\phi$ is a primary field, (\ref{CFTtrafo})
reduces to the standard expression $\phi \mapsto (f'(z))^h \phi$.
For general descendants however, the result will be a linear combination of
correlators of lower descendant fields.
Also note that $T_2(z)$ is in fact a multiple of the Schwarzian derivative.
It is actually true that all higher $T_n(z)$ are sums of products
of derivatives of the Schwarzian derivative. Since the Schwarzian
derivative of a M\"obius transformation vanishes, those transformations
lead to much simpler expressions.

In some cases however we can avoid having
to transform descendant fields.
Assume that we want to compute the chiral block of
a configuration for which we know the base to cover
map $f$. When we go to the cover, we can use
the fact that
$D_f$ is a function of the Virasoro modes $L_n$ only.
This means that it does not mix different representations,
so that, more formally, 
\be
D_f^{-1} P_{\CH_\phi} D_f = P_{\CH_\phi}\ .
\ee
From this it follows that conformal block has the same
transformation properties as the underlying correlation function,
as can be seen \eg in the simplest case
\be \label{trafoBlock}
&&\langle V_1(z_1)V_2(z_2) P_{\CH_\phi} V_3(z_3)V_4(z_4)\rangle \\
&&\qquad =\prod_{i=1}^4(f(z_i))^{h_i}\langle V_1(f(z_1))V_2(f(z_2))
P_{\CH_\phi} V_3(f(z_3))V_4(f(z_4))\rangle\ . \notag
\ee
Note that what we have said here is strictly speaking true only
for global coordinate transformations $f$, \ie for
M\"obius transformations
\be
z\mapsto \gamma(z)= \frac{az+b}{cz+d}\ , \qquad a,b,c,d\in \IC\ .
\ee
Other transformations, in particular also cover maps, 
must be treated with more caution,
as they can introduce new singularities. On a technical level
this means that at some points $f$ is no longer locally invertible
and $D_f$ does no longer annihilate the vacuum.

From these remarks it follows that conformal blocks exhibit the same
behavior as correlation functions under coordinate
transformations. This does not
mean, however, that their behavior under channel
crossing is the same. In particular, the full
partition function must be crossing symmetric,
whereas individual conformal blocks will transform
into each others in a very complicated manner. 
More precisely, if we expand the analytic continuation
of the full partition function around 0 or $\infty$,
then the resulting power series has essentially the
same form as the original expansion. This is simply a consequence
of covariance under M\"obius transformations and the
fact that we can change the order of operators in the
correlation function, as they are mutually local.
In contrast, even though the conformal block
still transforms nicely under coordinate changes,
the projector in it is not local, so that we cannot
change the order of the fields at will, which means
that expansions around different points will look
different.

Coming back to the computation of the conformal block, 
if we do not know the full cover map, then we need
to decompose the conformal block into three point
functions with twist fields. We then evaluate these three
point functions by mapping them to their appropriate 
covers. Note that in that case the cover maps
are different for the individual three point functions,
and no longer defined for the entire configuration.
This means that the above arguments no longer apply,
and that we must take into account the transformation
properties of the descendant fields.

Let us make one more remark concerning prefactors in
the AGT correspondence. From (\ref{trafoBlock}) we 
see that any coordinate transformation on the
G-curve leads to a product of prefactors of the form
$(f')^h$. From the way $h$ is related to the 
gauge theory masses, it follows that this factor
does not depend on the Coulomb branch parameters,
and that it only contributes to $\CF_0$ and $\CF_1$.
Nevertheless the structure of the exponent of the $U(1)$ prefactor
found in \cite{AGT} is different, so that it cannot
be transformed away in this way, which is in line
with what was expected on physical grounds.

\subsection{Examples}

We proceed to verify the correspondence in detail in a few examples,
the $Sp(1)$ gauge theory coupled to four hypermultiplets and the
$SO(4)$ gauge theory coupled to two hypermultiplets.  

\subsubsection{$Sp(1)$ versus $U(2)$ correlators}

\begin{figure}[htbp]
\begin{center}
\includegraphics[width=3.5in]{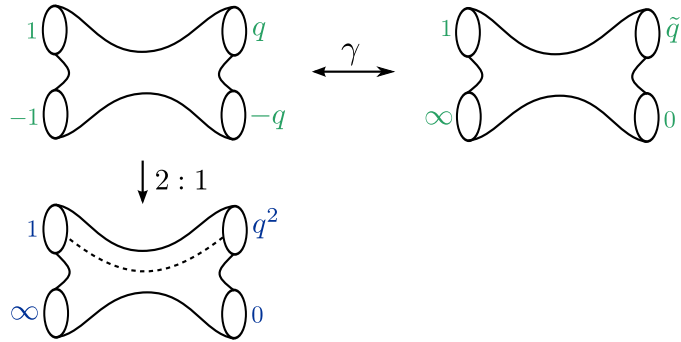}
\caption{On the left, the G-curve of the $Sp(1)$ gauge theory coupled to
4 hypers and its double cover. The M\"obius transformation
$\gamma$ relates the double cover to the $SU(2)$ G-curve. }\label{USpCFT}
\end{center}
\end{figure}

Recall that the G-curve for the $Sp(1)$ gauge theory coupled to
four massive hypers is given by a four-punctured sphere, as
illustrated on the bottom left of Figure~\ref{USpCFT}. The two 
half-punctures at 1 and $\qCFT^2$ are connected by a branch 
cut. As we have found in section~\ref{sec:N=2geometry}, the cross-ratio $\qCFT^2$ of the  four punctures can be expressed in
terms of the $Sp(1)$~instanton coupling $q_{Sp(1)}$ as
\begin{align}\label{eq:Sp1UV}
\qCFT^2 = \left( \frac{q_{Sp(1)}}{4} \right)^2.
\end{align}

The chiral block we need to evaluate is obtained by cutting
the tube with the twist line, so that
\begin{align}\label{Sp14ptBase}
\CF_{Sp(1)}(\qCFT) = \langle V_1^{A,B}(\infty)\sigma(1)\PH
\sigma(\qCFT^2) V_2^{A,B}(0) \rangle  
\end{align}
where the vertex operators $V^{A,B}_{1,2}$ factorize into representations of
$T^A$ and $T^B$ of weight $(h_1,h_2)$ and $(h_3,h_4)$,
\begin{align*}
V_1^{A,B} &= V_1^{A}V_1^{B}\\
V_2^{A,B}  &= V_2^{A}V_2^{B}.
\end{align*}
Here we identified the half-integral mode $L_{-\frac{1}{2}}^-$ of the
twisted $\CW(2,2)$  
algebra with the one-instanton modulus $q_{Sp(1)}$ of the $Sp(1)$ theory.

To evaluate the correlator with twist fields, we want to
go to the double cover.
The base has half-punctures at $1$ and $\qCFT^2$, so that we 
map it to the double cover by 
\be\label{Sp1covermap}
z\mapsto \tilde z = \pm \sqrt{\frac{z-\qCFT^2}{z-1}}\ .
\ee
This maps has indeed branch points at 1 and $\qCFT^2$, 
and it maps the fields at 0 and $\infty$ to $\pm \qCFT$
and $\pm 1$.
The block (\ref{Sp14ptBase}) on the base
thus becomes the block on the cover (up to some constant prefactor)
\be\label{Sp14ptCover}
\CF_{Sp(1)}(\qCFT)=\left(1-\qCFT^2\right)^{\sum_i h_i}\langle V^B_1(-1) V^A_1(1)\PH V^A_2(\qCFT)
V^B_2(\qCFT)\rangle_C\ .
\ee
Note that since we know the full base-cover map, we
were able to make use of (\ref{trafoBlock}) without
worrying about descendant fields.
To evaluate (\ref{Sp14ptCover}), we write it as
a sum over three point functions in the usual
manner. 
Let us therefore define three point coefficients on the cover by
\be 
\Ccover^{h_1,h_2;h_3}_{I_1,I_2,I_3} = \langle V^1_{I_1}(1)
V^2_{I_2}(-1) V^3_{I_3}(0) \rangle \notag
\ee
where $I_i$ gives the Virasoro descendants acting
on $V^i$, which we will usually denote by Young diagrams.
The conformal block can then be evaluated as
\begin{multline}
\CF_{Sp(1)}(\qCFT)
=\left(1-\qCFT^2\right)^{\sum_i h_i}
\sum_{I_a,J_a}\Ccover^{h_1,h_2;h_a}_{\bullet,\bullet,I_a}\Ccover^{h_3,h_4;h_a}_{\bullet,\bullet,J_a} 
(\langle V^{a}_{I_a}| V^{a}_{J_a}\rangle)^{-1} \qCFT^{h_a+|I_a|}\\
= \qCFT^{h_a}\left(1 + \frac{2 (h_1-h_2) (h_3-h_4) }{h_a}\qCFT+  \ldots\right) \ .
\end{multline}
Using the identification of parameters
\begin{align*}
h_{i} &= \frac{ 1 }{\e_1 \e_2} \left( \frac{Q^2}{4} - m_i^2 \right), \\
h_{a} &= \frac{1}{\e_1 \e_2} \left( \frac{Q^2}{4} - a^2 \right), 
\end{align*}
where $Q = \e_1 + \e_2$ and the momenta $m_i$ are related to the mass parameters
$\mu_i$ as 
\begin{align*}
\begin{array}{lll}
m_1 = \frac{\mu_1 + \mu_2}{2} & \quad & m_3 =
\frac{\mu_3 + \mu_4 }{2} \\
m_2 = \frac{\mu_1 - \mu_2 }{2} & \quad & m_4
=\frac{\mu_3 - \mu_4}{2},
\end{array}
\end{align*} 
we find that equation~(\ref{Sp14ptCover}) is indeed equal to the $Sp(1)$ instanton 
partition function up to a spurious factor independent of $a$ and $\mu_i$,\footnote{We checked this result up to order 6 in the instanton parameter.}
\be
Z^{Sp(1)}(\qSp) = \left(1-\left(\frac{\qSp}{4}\right)^2\right)^{-\frac{1}{16} (c+1) }\CF_{Sp(1)}(\qSp) \ .
\ee
Note in particular that this spurious factor is independent of
the masses of the hypermultiplets, in contrast to the spurious
factor in the AGT correspondence for unitary gauge groups. This is 
indeed as expected, as the latter should come from the decoupled
$U(1)$ in the $U(2)$, whereas there is no such $U(1)$ in the
$Sp(1)$ setup. This fact will also be important for
 extending the $Sp(1)$-correspondence to linear $Sp(1)-SO(4)$
quivers.

\subsubsection*{Relation to the $U(2)$ correlator}

We already knew that the full $Sp(1)$ and $U(2)$ Nekrasov partition
function are related by the change of parameters (\ref{eq:Sp1UV}).
To understand this better from the conformal field theory perspective,
let us study the relation 
between the $Sp(1)$ conformal block (\ref{Sp14ptBase}) and the $U(2)$
conformal block. 

We have used the fact that the conformal 
block on the base (\ref{Sp14ptBase})
can be related to the block on the cover (\ref{Sp14ptCover}).
The block on the cover is obviously very closely
related to the original $U(2)$ configuration
depicted on the right of Figure~\ref{USpGcurve}.
We can map one to the other using the
M\"obius transformation $\gamma$ given by
equation~(\ref{eqn:mobiusSp1}). 
This is of course only possible provided that
we make their cross ratios agree by identifying
\begin{align*}
\qU=\frac{\qSp}{\left(1+\frac{\qSp}{4}\right)^2}\ ,
\end{align*}
which is exactly the relation found on the gauge theory side.

From equation~(\ref{trafoBlock}) we also know that this transformation
only introduces 
an overall prefactor which does not depend on the weight
of the intermediate channel, so that it is an $a$-independent
prefactor.
If we were interested in the relation between instanton partition
functions without the perturbative part, 
we would need to divide both (\ref{Sp14ptCover}) and
the $U(2)$ block by $q^{h_a-h_1-h_4}$. The $a$-dependent part of the
ratio of the two is then 
\begin{align*}
\left(\frac{\qU}{\qSp}\right)^{h_a} .
\end{align*}
On the gauge theory side, this factor originates from the difference in the instanton part of $F_0$. By construction,
$F_0^{\rm inst}-\tilde F_0^{\rm inst} = -a^2(\log \qU - \log \qSp)$,
which agrees with the above factor (for $Q=0$).

\subsubsection{$SO(4)$ versus $U(2) \times U(2)$ correlators}\label{ss:SO4block}

\begin{figure}[h!]
\begin{center}
\includegraphics[width=4in]{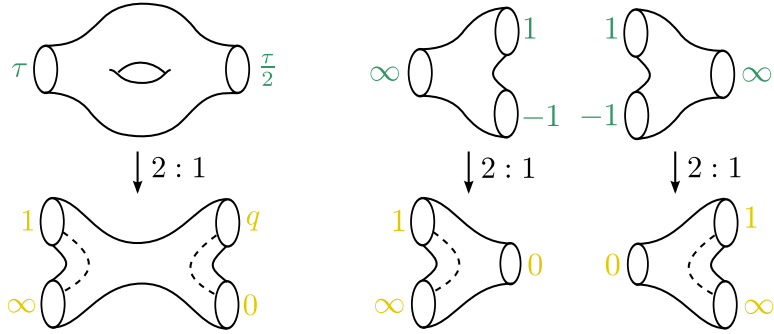}
\caption{The G-curve for the $SO(4)$ coupled to two hypermultiplets and
its double cover. The left picture illustrates the global mapping between the $SO(4)$
G-curve and its double cover, whereas the right picture illustrates the local
mappings that we use to compute the twisted $\CW(2,2)$ conformal
block on the $SO(4)$ G-curve.}\label{USOGcurve} 
\end{center}
\end{figure}

Let us now turn to the $SO(4)$ case. The G-curve for the $SO(4)$
theory with two hypers in the fundamental is given in
the lower left of Figure~\ref{USOGcurve}.
The chiral block we need to evaluate
is
\be\label{SO4CFTblock}
\CF_{SO(4)}(q_{SO(4)}) =
\langle \hat V_1(\infty) \sigma(1)\PH\sigma(\qSO) \hat V_2(0)\rangle\ .
\ee
Note that we identified the integral modes $L^{A/B}_{-1}$ with the
one-instanton parameter $q_{SO(4)}$. 

Similarly to the $Sp(1)$ example
that we discussed previously, there is an elegant way 
of obtaining the chiral block that makes use of the
fact that we know the double cover map of the full 
configuration (\ref{SO4CFTblock}). This double covering was
described in section~\ref{sec:so4geometry}.
In particular it maps the punctures
\begin{align*}
(\infty,1,\qSO,0) \mapsto (1/2,0,\tau/2,(1+\tau)/2)\ .
\end{align*}
The configuration on the cover is a torus
with two punctures at 0 and $\qU$. The conformal
block for this configuration has been computed in \cite{AGT}
and agrees with the $U(2)\times U(2)$ instanton
partition function. Up to spurious prefactors
introduced by the mapping to the cover, (\ref{SO4CFTblock})
is thus given by the conformal block of the
two punctured torus expressed in terms of 
$\qSO$.

However, in more general examples (\ie the ones that we encounter in
section~\ref{sec:linearquiver}) it will be much harder to find the global
mapping between the $SO/Sp$ Gaiotto curve and its double cover. We
thus need to develop a method  that doesn't require this global
information, and computes the twisted $\CW(2,2)$ block from a
simple decomposition of the Gaiotto curve into pairs of pants. Let us
exemplify this for the $SO(4)$ Gaiotto curve.  

Evaluating the $SO(4)$-block (\ref{SO4CFTblock}) is more complicated
than the $Sp(1)$-block we considered previously. Since it 
has two branch cuts, there is no longer a simple square root
map that maps the block~(\ref{SO4CFTblock}) to its double cover,
the torus. What we will do instead is to first decompose
the block into three point functions, and then map those
three point functions individually to their covers,
as depicted on the right side of Figure~\ref{USOGcurve}.

Let us also define twisted three point coefficients on the base as
\be \label{SO4CorrBase}
\Cbase^{h_A,h_B;h_1}_{I_A,I_B,\bullet}:=\langle \hat V_1(\infty) \sigma(1)V^{A,B}_{I_A,I_B}(0)\rangle,
\ee
so that
\begin{align*}
\CF_{SO(4)}(q_{SO(4)}) =\sum_{I_A,J_A,I_B,J_B} \Cbase^{h_A,h_B;h_1}_{I_A,I_B,\bullet}\Cbase^{h_A,h_B;h_2}_{J_A,J_B,\bullet} 
(\langle V^{2,3}_{I_A,I_B}| V^{2,3}_{J_A,J_B}\rangle)^{-1} \qSO^{h_A+|I_A|+h_B+|I_B|}\ .
\end{align*}
Note that we have used that $\sigma(1)$ is a primary field, so that we can
exchange the fields at 0 and $\infty$ at will.
Our task is now to evaluate (\ref{SO4CorrBase}). This we do
by mapping to it the double cover.
Since 1 and $\infty$ are branch points,
we use the map 
\be \label{SO4CoverBaseMap}
z\mapsto \tilde z = \pm(1-z)^{-1/2}\ , 
\ee
which maps (\ref{SO4CorrBase})  
to three point functions on the cover of the form
\be \label{SO4CorrCover}
\langle V^A_{I_A}(1) V^B_{I_B}(-1) \hat V^1(0) \rangle\ .
\ee
To find the precise relation between (\ref{SO4CorrBase}) 
and (\ref{SO4CorrCover}) however
we need to take into account the transformation
properties of all the fields under the map
from the transformation (\ref{SO4CoverBaseMap}).
This is no issue for $\sigma$ and $\hat V^1$, since
those fields are always primary fields,
so that any overall prefactors will always
be cancelled once we divide by the primary
three point function. In what follows, we will
always omit these factors. It is however an issue for $V^A$ and $V^B$,
since those fields are descendants.
Using~(\ref{CFTtrafo})
we can thus express the field on the
base by the field on the cover as
\be\label{SO4coverfields}
&& V^{A,B}(0) = \\
&& =V\left( \left(\frac{1}{2}
  \right)^{L_0}e^{3L_1/4}e^{L_2/16}\cdots\phi^A_{I_A},1\right) 
V\left(\left(-\frac{1}{2}\right)^{L_0}e^{3L_1/4}e^{L_2/16}\cdots
  \phi^B_{I_B},-1 \right) \notag
\ee
where we have only included terms that are relevant up to second level
descendants.

Let us show how to compute the first order term of the chiral block.
To fix the normalization, we use that the primary three point function transforms as
\begin{align*}
\Cbase^{h_A,h_B;h_1}_{\bullet,\bullet,\bullet}=\left(
  \frac{1}{2}\right)^{h_A}\left(-\frac{1}{2}\right)^{h_B} 
\Ccover^{h_A,h_B;h_1}_{\bullet,\bullet,\bullet}\ .
\end{align*}
The normalized coefficients for the first level descendants 
can then be computed to be
\begin{eqnarray*}
\Cbase^{h_A,h_B;h_1}_{\square,\bullet,\bullet}
&=&\frac{1}{2}\Ccover^{h_A,h_B;h_1}_{\square,\bullet,\bullet}
+ \frac{3h_A}{2} \\
\Cbase^{h_A,h_B;h_1}_{\bullet,\square,\bullet}
&=&-\frac{1}{2}\Ccover^{h_A,h_B;h_1}_{\bullet,\square,\bullet}
+ \frac{3h_B}{2} 
\end{eqnarray*}
Using
\begin{align*}
\Ccover^{h_A,h_B;h_1}_{\square,\bullet,\bullet}
=\frac{1}{2} (-h_1-3 h_A+h_B)\ ,\qquad 
\Ccover^{h_A,h_B;h_1}_{\bullet,\square,\bullet}
=\frac{1}{2} (h_1-h_A+3 h_B)\ ,
\end{align*}
we obtain
\begin{multline}
\CF_{SO(4)} = \qSO^{h_A+h_B}\Big(1 + \Big(\frac{(3h_A+h_B-h_1)(3h_A+h_B-h_2)}{2h_A}+\\
\frac{(3h_B+h_A-h_1)(3h_B+h_A-h_2)}{2h_B}\Big)
\frac{\qSO}{16}+\ldots\Big)\ .
\end{multline}

Using the identification of parameters
\begin{align*}
h_{i} &= \frac{1}{\e_1 \e_2} \left( \frac{Q^2}{4} - \mu_i^2 \right), \\
h_{A/B} &= \frac{ 1}{\e_1 \e_2} \left( \frac{Q^2}{4} -\beta_{A/B}^2 \right), 
\end{align*}
where $\beta_{A/B} =\frac{b_1 \pm b_2}{2}$,
we have indeed checked up to order 2
that (\ref{SO4CFTblock}) agrees with the $SO(4)$ partition
function up to a spurious prefactor given by
\be
Z_{\rm sp} =  (1-q)^{\frac{3}{8} Q^2} \,.
\ee

\section{Linear $Sp/SO$ quivers}\label{sec:linearquiver}

In this section we discuss the generalization of the 
correspondence for single $Sp$ and $SO$ gauge groups 
to linear quiver gauge theories involving both $Sp$ and $SO$
gauge groups. This process will involve new elements from the instanton
counting perspective, which we introduce in this section. 

\begin{figure}[h!]
\begin{center}
\includegraphics[width=3in]{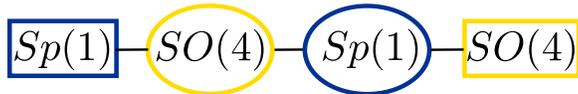} 
\caption{Example of a linear $Sp(1) /SO(4)$
  quiver gauge theory with a single $Sp(1)$ and $SO(4)$ gauge group,
  one $SO(4)$-fundamental hyper, two $Sp(1)$-fundamental hypers and one
  $SO(4) \times Sp(1)$-bifundamental hyper (consisting of eight half-hypermultiplets).} \label{fig:generalSOSpquiver}  
\end{center}
\end{figure}

The $SO/Sp$ correspondence that we studied in the previous sections
can be naturally extended to linear quiver gauge theories with
alternating $Sp$ and $SO$ gauge groups. 
The reason for requiring the  
$SO$ and $Sp$ gauge groups to alternate is that only such gauge
theories can be engineered using an orientifold D4/NS5-brane
set-up.  
These configurations are natural from the gauge theory perspective as
well.  Remember that the flavor symmetry for an
$Sp(N-1)$-fundamental hyper enhances to $SO(2N)$, while the flavor 
symmetry for $SO(N)$-fundamental hyper enhances to $Sp(N-1)$. So, 
for general $N$, only linear quivers with alternating gauge groups
$Sp(N-1)$ and $SO(2N)$ correctly reproduce the flavor symmetry of the
bifundamental fields. 
An example of  a linear $Sp/SO$ quiver is illustrated in
Figure~\ref{fig:generalSOSpquiver}.    
  
Special about linear $Sp/SO$ quivers is that the bifundamental
fields are not full hypermultiplets, but half-hypermultiplets. Let us
discuss this briefly. Usually, a hypermultiplet of representation $R$
of a gauge group $G$ consists of two $\CN=1$ 
chiral multiplets: one chiral multiplet in the representation $R$ and
the other in the complex conjugate representation $\bar{R}$ of 
$G$. When the representation $R$ is pseudoreal, however, a single
chiral superfield already forms an $\CN=2$ hypermultiplet. This is
called a half-hypermultiplet in $R$. 
The half-hypermultiplets must be massless, as it is not possible 
to construct a gauge invariant mass-term
in the Lagrangian for a half-hypermultiplet.  

Even though a half-hypermultiplet is CPT invariant, it is not always
possible to add them to an $\CN=2$ gauge theory due to the Witten
anomaly \cite{Witten:1982fp}. Because an $Sp \times SO$ bifundamental
multiplet contains an even number of half-hypermultiplet
components, we can circumvent the anomaly. Indeed,
the $Sp(N) \times SO(M)$ bifundamental is the tensor product of $2N$
half-hypermultiplets corresponding to the (anti-)fundamental $Sp(N)$
flavor symmetry, and $M$ half-hypermultiplets corresponding to the
fundamental $SO(M)$ flavor symmetry. In total this gives $2NM$
half-hypermultiplets. 

Our goal in this section is to write down Nekrasov contour
integrands for linear $SO/Sp$ quivers and verify the correspondence
with chiral blocks of the $\CW$-algebra. Before getting there, let us
first discuss some of the geometry of linear $SO/Sp$ quivers.

\subsection{G-curves for linear $Sp/SO$ quivers}

\begin{figure}[h!]
\begin{center}
\includegraphics[width=3.5in]{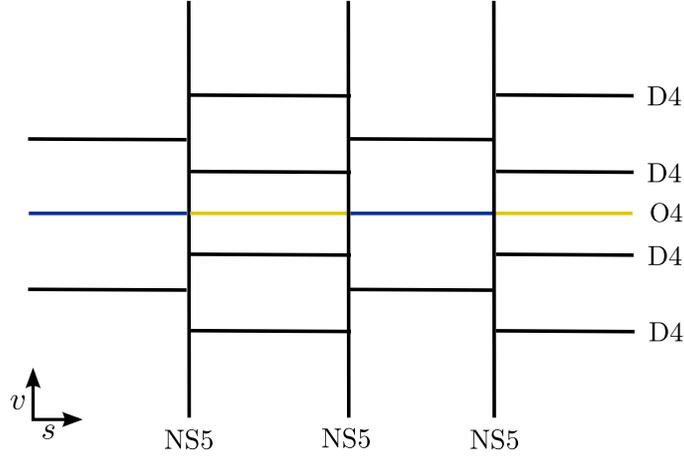} 
\caption{Orientifold D4/NS5-brane embedding of the
  linear $Sp/SO$ quiver theory of Figure~\protect\ref{fig:generalSOSpquiver}.} \label{fig:linearSpSObrane} 
\end{center}
\end{figure}

As illustrated in Figure~\ref{fig:linearSpSObrane}, the orientifold
D4/NS5 brane constructions for $Sp$ and $SO$ gauge 
theories can be naturally extended to any linear quiver theory with alternating
$Sp(N-1)$ and $SO(2N)$ gauge groups by introducing an extra NS5-brane for
every bifundamental field. For this construction to work it is necessary
that the gauge groups alternate as crossing an NS5-brane exchanges
one type of orientifold brane with the other. From this string theory
embedding we can simply read off the Seiberg-Witten curve.

The Seiberg-Witten curve corresponding to a linear quiver
with $Sp(N-1)$ as well as $SO(2N)$ gauge groups
can be written in the Gaiotto-form \cite{Tachikawa:2009rb}   
\begin{align}\label{linearSpSOSWcover}
0= \det(v-\boldsymbol\varphi_{Sp/SO}) = v^{2N} + \varphi_2 v^{2N-2} + \varphi_4
v^{2N-4} + \ldots + \varphi_{2N}.
\end{align}
As before, this equation determines the Seiberg-Witten curve as a
degree $2N$ covering over the Gaiotto curve.  The Hitchin differentials
$\varphi_{2k}$ (for $1 \le k \le N-1$) are of degree $2k$ and encode
the vev's of the Coulomb branch operators $\Tr(\Phi^{2k})$ of the
adjoint scalar $\Phi$ for all 
gauge groups in the linear quiver. On the other hand, the degree $N$ 
differential $\varphi_{\tilde{N}} = \sqrt{\varphi_{2N}}$ encodes the
vev's of the operators $\rm{Pfaff}(\Phi)$ 
for the $SO$ gauge groups in the quiver only. All 
differentials are also functions of the exactly marginal coupling
constants $\tau_{\rm UV}$ and the bare mass parameters, in such a way
that the residue of the matrix-valued differential
$\boldsymbol\varphi_{Sp/SO}$ at each
puncture encodes the flavor symmetry of the corresponding matter
multiplet.   

\begin{figure}[h!]
\begin{center}
\includegraphics[width=2.8in]{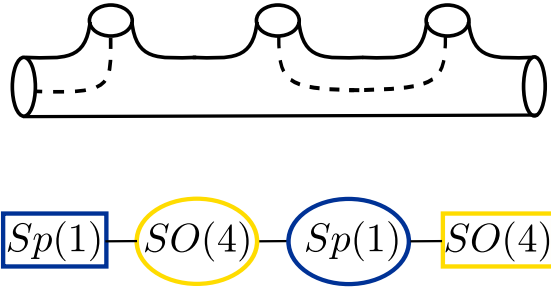} \\
\vspace*{10mm}
\includegraphics[width=2.9in]{generalsospquiver} 
\caption{The Riemann surface on top is the Gaiotto curve corresponding to
  the linear $Sp(1)/SO(4)$ quiver at the bottom.} \label{fig:generalSOSpcurve} 
\end{center}
\end{figure}

It follows from equation~(\ref{linearSpSOSWcover}) that the
Gaiotto curve for a linear $Sp(N-1)/SO(2N)$ quiver theory is a genus zero
Riemann surface with punctures.   
For the $Sp(1)/SO(4)$ theory these punctures can be of two
types. Either the differential $\varphi_{\tilde{2}} =
\sqrt{\varphi_{4}}$ experiences a $\IZ_2$-monodromy when going around
the puncture, or it does not. As before, we call these punctures
half-punctures and full punctures, respectively. The puncture
representing a bifundamental matter field is a half-puncture. 
One way to understand this, is to compare the gauge theory quiver to
the corresponding Gaiotto curve. As is illustrated in
Figure~\ref{fig:generalSOSpcurve}  each $Sp(1)$ gauge group
corresponds to a tube with a $\IZ_2$-twist line on the Gaiotto curve,
whereas each $SO(4)$ gauge group corresponds to a tube without a
$\IZ_2$-twist line. This implies that at each puncture corresponding
to a bifundamental field a $\IZ_2$-twist line has to end.  
Notice that the differential $\boldsymbol\varphi_{Sp/SO}$ should have
a vanishing residue at this half-puncture, since the bifundamental
is forced to have zero mass.
 
\begin{figure}[h!]
\begin{center}
\includegraphics[width=3in]{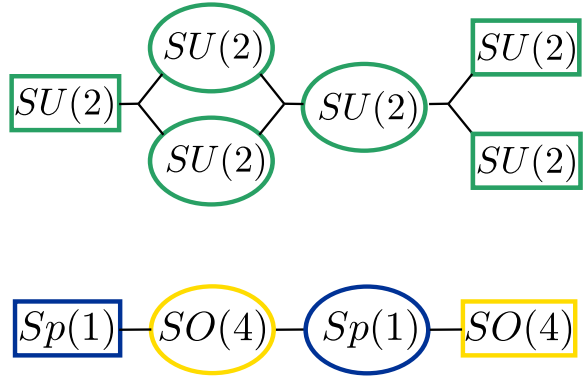} 
\caption{The generalized $SU(2)$
  quiver theory, depicted at the top, has isomorphic gauge and flavor
  symmetries to the linear $SO(4)/Sp(1)$ quiver gauge theory, depicted
  on the bottom. This picture   in particular
  relates  the $SO(4)\times Sp(1)$ bifundamental to the
  $SU(2)^{3}$ trifundamental.} \label{fig:generalUSOSpquiver} 
\end{center}
\end{figure}

A remarkable feature of \emph{linear} $Sp(1)/SO(4)$ quivers is that
they are closely related to \emph{generalized} $SU(2)$ quiver
theories 
\cite{Tachikawa:2009rb}.  We have already seen that an
$Sp(1)$-fundamental can be equivalently represented by an
$SU(2)$-fundamental, and an $SO(4)$-fundamental by an
$SU(2)^{2}$-bifundamental, as their representations are isomorphic.
Even more interestingly, by the same argument an $Sp(1) \times
SO(4)$-bifundamental is closely related to a matter multiplet with flavor
symmetry group $SU(2)^{3}$. The elementary field with this property is
known as the $SU(2)^{3}$-trifundamental, and it consists of eight free
half-hypermultiplets in the fundamental representation of the three
$SU(2)$ gauge groups
\cite{Gaiotto:2009we}.  Since the $Sp(1)\times SO(4)$ bifundamental
contains eight half-hypermultiplets as well, we expect it to be
equivalent to the $SU(2)^{3}$-trifundamental. 
One example of the relation between linear $Sp(1)/SO(4)$ quivers 
and generalized $SU(2)$ quiver theories is illustrated in    
Figure~\ref{fig:generalUSOSpquiver}. 

\begin{figure}[h!]
\begin{center}
\includegraphics[width=2.4in]{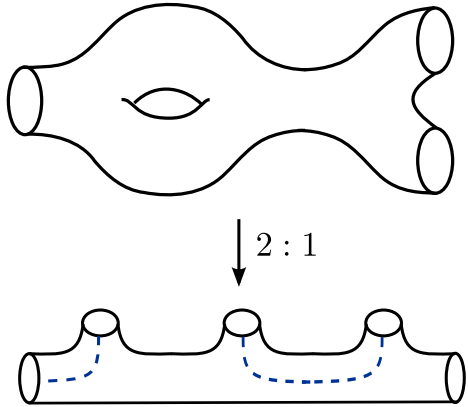}
\caption{The top picture represents the Gaiotto curve of a generalized
  $SU(2)$ quiver
  theory. It is a branched
  double cover over the Gaiotto curve of the linear $SO(4)/Sp(1)$ quiver
  theory illustrated in the bottom.} \label{fig:USOSpcurve}
\end{center}
\end{figure}

Similar to our discussion in
section~\ref{sec:N=2geometry}, we can interpret the $\IZ_2$-twist
lines on the $Sp/SO$ Gaiotto curve as branch cuts. As is illustrated in
Figure~\ref{fig:USOSpcurve}, we find that the Gaiotto
curve for the generalized $A_1$ quiver theory is a double cover of the Gaiotto
curve for the corresponding linear $D_2$ quiver theory. This is
consistent with the fact that the bifundamental
fields cannot carry a 
mass. Indeed, each half-puncture corresponding to a
$Sp(1) \times SO(4)$ bifundamental field lifts to a regular point on the
$SU(2)$ Gaiotto curve. Its flavor symmetry had thus better be trivial.

Let us stress that, according to the arguments of
section~\ref{sec:gauge}, the instanton partition function of a
linear quiver theory that contains the $Sp(1)
\times SO(4)$ bifundamental will be related to the instanton partition
function containing the
$SU(2)^{3}$-trifundamental by a non-trivial mapping of 
marginal gauge couplings. This mapping has a geometric interpretation,
according to section~\ref{sec:N=2geometry}, as it
will relate
the complex moduli of the corresponding Gaiotto curves.   
Studying the instanton partition function of linear  $D_2$ quiver theories thus
sheds light on the understanding of  non-linear $A_1$ 
quiver theories. We leave this for future work~\cite{workinprogress}.

\subsection{Instanton contribution for the $Sp \times SO$
  bifundamental} 

We continue with finding a contour integrand prescription for the
contribution of the $Sp \times SO$ bifundamental matter multiplet.
In general, the instanton partition function for a linear quiver gauge
theory with gauge group $G=G_1 \times \cdots \times G_P$ can be
formulated schematically as 
\be
 Z^{\textrm{inst}}_{k} = \int \prod d\phi_{i} ~
 \mathbf{z}_{\textrm{gauge}}^{k} (\phi_{p,i}, \vec{a}_p ) ~
  \mathbf{z}_{\textrm{bifund}}^k (\phi_{p,i} , \phi_{q,j}, \vec{a}_p, \vec{a}_{q},
  \m)~ \mathbf{z}_{\textrm{fund}}^k (\phi_{p,i}, \vec{a}_{p}), 
\ee
where the three $\mathbf{z}$'s in the integrand refer to the
contribution of gauge multiplets, bifundamental and fundamental matter
fields, respectively. The only missing ingredient needed to compute
the instanton partition 
function for a linear $D_2$-quiver is the contribution of the
$Sp \times SO$ bifundamental half-hypermultiplet. This
contribution has not yet been studied in the literature.  

Remember that a full hypermultiplet consists of two chiral superfields
$Q$ 
and $\tilde{Q}$, where $Q$ and $\tilde{Q}$ are respectively in the
representation 
$R$ and $\bar{R}$ of the gauge group $G$. Since the bifundamental
representation $Sp(1)\times SO(4)$ is pseudo-real, the two chiral
superfields 
in a hypermultiplet actually transform under isomorphic
representations $R = \bar{R}$. 

To find the contour integrand contribution for a half-hypermultiplet,
we start out with 
the same BPS equations as in section~\ref{sec:gauge}. Again we
consider the vector 
bundle of solutions $\CV$ to the Dirac equation over the moduli space
of instantons $\CM_{G,k}$. This time, however, the solutions need to
satisfy an extra reality condition. This implies that the weights of
the equivariant torus on the vector bundle $\CV$ should come in pairs
$\pm w$. We thus want to argue that we can find the contour integrand
for a $Sp(1) \times SO(4)$ bifundamental half-hypermultiplet by taking
the appropriate square-root.

\begin{figure}[htbp]
\begin{center}
\includegraphics[width=3.3in]{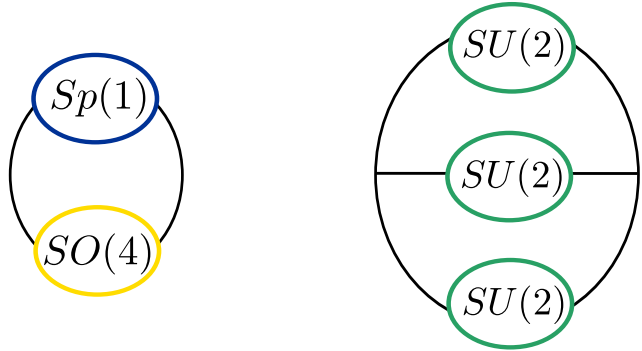}
\caption{On the left: the quiver diagram for a cyclic $Sp(1)/SO(4)$
  quiver gauge theory coupled to two $Sp(1) \times SO(4)$
  bifundamentals.  
  The bifundamentals do not
  have a flavor symmetry group. On
  the 
  right: the corresponding generalized $SU(2)$ quiver including two $SU(2)^3$-trifundamentals.} \label{fig:USOSpquiver}
\end{center}
\end{figure}

\begin{figure}[htbp]
\begin{center}
\includegraphics[width=1.5in]{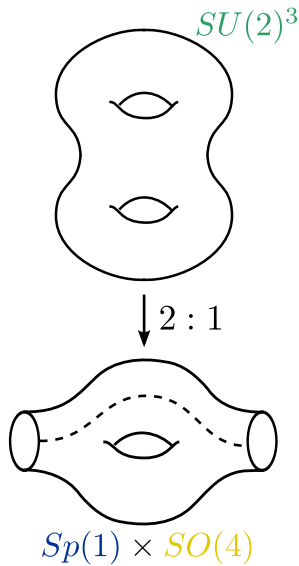}
\caption{The cover and base Gaiotto-curve corresponding to the
  $Sp(1) \times SO(4)$ and $SU(2)$ quiver gauge theories illustrated in
  Figure~\protect\ref{fig:USOSpquiver}. } \label{fig:USOSpGcurve}
\end{center}
\end{figure}

Let us therefore consider the simplest
gauge theory with \emph{two} $Sp(1) 
\times SO(4)$ bifundamental half-hypermultiplets. This theory is
illustrated on the left 
in Figure~\ref{fig:USOSpquiver}. As is illustrated on the bottom of 
 Figure~\ref{fig:USOSpGcurve}, the corresponding Gaiotto curve is a
two-punctured torus with a $\IZ_2$-twist line running between the
punctures. The corresponding $A_1$ theory 
is a generalized quiver theory with genus 2.

Let us define the vector bundles $\CV_{Sp}$ and $\CV_{SO} $ of
solutions to the Dirac equation in the fundamental representation 
of the $Sp$ and $SO$ gauge group, respectively.
Since the double copy of a bifundamental half-hypermultiplet is a
full bifundamental hypermultiplet, its contribution to the instanton
partition function is given by the usual integral
\be
\oint_{\CM_{Sp} \times \CM_{SO}} e(\CV_{Sp} \otimes \CV_{SO} \otimes
\CL \otimes M).
\ee
The integrand is the Euler class of the tensor
product 
$\CV_{Sp} \otimes \CV_{SO}  
\otimes \CL \otimes M$ over the product $\CM_{Sp} \times \CM_{SO}$ of
the instanton moduli spaces.\footnote{Originally, $\CV_{Sp}$ is a 
  vector bundle over the instanton moduli space $\CM_{Sp}$ and
  $\CV_{SO}$ a vector bundle over the instanton moduli space
  $\CM_{SO}$. However, we define both bundles as vector bundles over
  the product $\CM_{Sp} \times \CM_{SO}$, by pulling them back using
  the projection maps $\pi_{Sp/SO}: \CM_{Sp} 
\times \CM_{SO} \to  \CM_{Sp/SO}$.} 
In this expression, $M \cong \IC$ is the flavor
vector space and $\CL$ the half-canonical line bundle over
$\IR^4$. 

Following the same strategy as we lined out in
section~\ref{sec:gauge} for a usual hypermultiplet, we  
obtain a contour integrand 
\be
\mathbf{z}^{Sp, SO}_{k_1, k_2, db}(\phi, \psi, \vec{a}, \vec{b}, m) \notag
\ee
corresponding to the double copy of the bifundamental
half-hypermultiplet. We work this out in
appendix~\ref{app:Instanton}. The parameters $\vec{a}$ and $\vec{b}$
are the 
Coulomb parameters of the $Sp$ and the $SO$ gauge theory,
respectively, whereas $m$ is the mass parameter for the double copy. 
Since the half-hypermultiplets should be massless we substitute
$m=0$. For precisely this value of the mass, the resulting expression
indeed turns out to be a complete square
\be
 \mathbf{z}_{k_1, k_2, \textrm{db}}^{Sp, SO}(\phi, \psi, \vec{a}, \vec{b}, m =0, \e_1, \e_2) = \left( \mathbf{z}_{k_1, k_2, \textrm{hb}}^{Sp, SO} (\phi, \psi, \vec{a}, \vec{b}) \right)^2  . 
\ee
It is thus natural to identify the square-root $\mathbf{z}_\textrm{hb}$
of the double bifundamental with the 
contribution coming from a $Sp \times SO$-bifundamental
half-hypermultiplet. 
We check this prescription with the CFT shortly. 

An interesting observation is that the bifundamental
half-hypermultiplet does not introduce additional poles besides those
coming from the usual gauge
factors $\mathbf{z}_{\rm gauge}$.\footnote{Remember that we did find
  such additional poles for the $Sp(1) \times Sp(1)$ bifundamental
  as well as $Sp(1)$ adjoint hypermultiplet. The existence of
  additional poles 
  might be related to the existence of a string theory embedding.} In analogy to \cite{Alday:2010vg},
we can therefore view the bifundamental half-hypermultiplet as a mapping
\be
 \Phi_{a, b_1, b_2} : \widehat{\CH}_{\widetilde{Sp}(1)}\to
 \widehat{\CH}_{SO(4)} 
\ee
between two vector spaces $\widehat{\CH}_{\widetilde{Sp}(1)}$ and
$\widehat{\CH}_{SO(4)}$, whose bases are parametrized by the poles of
the  
respective gauge multiplet integrands. It is natural to expect that
these vector spaces are related to the $\CW$-representation spaces
$\CH_{\widetilde{Sp}(1)}$ and $\CH_{SO(4)}$ that we encountered 
in the previous section. Note that for $Sp$ and $SO$ gauge groups we 
expect the spaces to be related without any additional $U(1)$
factors. 
For the original $U(2)$ AGT correspondence this
was recently made precise \cite{Alba:2010qc}. The
 structure of the $Sp(1)$ poles is more complicated,
however, and it would be interesting to find the exact mapping
between the two spaces.

\begin{figure}[htbp]
\begin{center}
\hspace*{-3mm}
\includegraphics[width=1.6in]{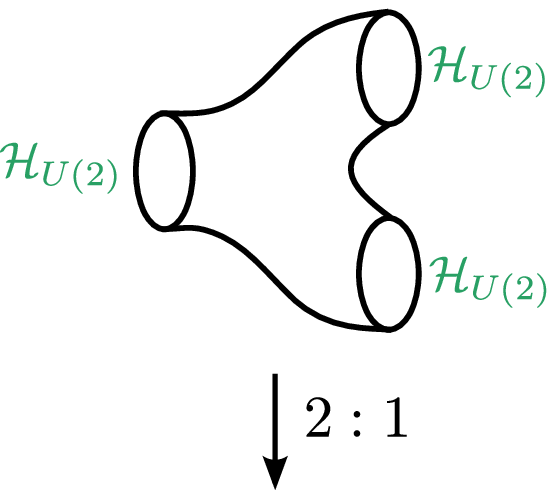}\\
\vspace*{4mm}
\includegraphics[width=2in]{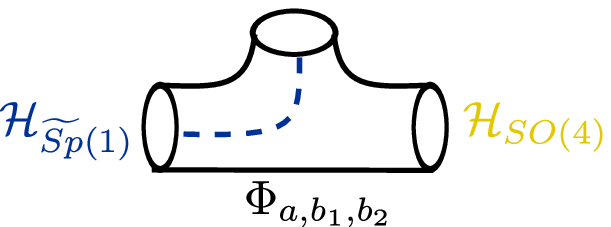}
\caption{The instanton contribution for the $U(2)^3$
  trifundamental can be represented as a linear map $Z^{\rm inst}: \mathcal{H}_{U(2)}
  \to \mathcal{H}_{U(2)} \times \mathcal{H}_{U(2)}$. Similarly, and
  correspondingly, the $Sp(1) \times SO(4)$ 
  bifundamental field defines a linear map $Z^{\rm inst}:
  \widehat{\mathcal{H}}_{\widetilde{Sp}(1)} \to \widehat{\mathcal{H}}_{SO(4)}$.}  
\label{fig:elementarySpSO}
\end{center}
\end{figure}

\subsection{Test of the $Sp(1) \times SO(4)$ AGT correspondence}

Since we know the Gaiotto curve corresponding to alternating
$SO/Sp$ quiver theories, we can extend the correspondence between
$SO/Sp$ gauge groups and $\CW$ blocks. We are now ready to check this 
correspondence.  
At the same time this will also serve as an additional check
that our expression for the half-hypermultiplet is correct.
Consider thus the $SO/Sp$ theory illustrated in 
Figure~\ref{fig:generalSOSpquiver}, with a single 
bifundamental half-hypermultiplet, two fundamental
$Sp(1)$-hypermultiplets and one fundamental
$SO(4)$-hypermultiplet.

\subsubsection{$Sp(1) \times SO(4)$ instantons}

Let us first compute the instanton partition
function of this linear quiver theory. The first
non-trivial term comes from $(k_1, k_2) = (1, 1)$. In the unrefined
case $\e_1 = -\e_2 = \hbar$, it is given by 
\be
Z^{\rm inst}_{1,1}=-\frac{m_1 m_2 b_1 b_2 \left(\left(m_3^2-b_1^2\right)
    \left(-a^2+b_1^2\right) \left(-\hbar ^2+b_1^2\right)+\left( a^2 -
      b_2^2 \right) \left( -m_3^2 + b_2^2 \right) \left(-\hbar^2 +
      b_2^2 \right) \right) }{8 a^2 \hbar^4
  \left(b_1^2-b_2^2\right)^2} ~ \notag
\ee 
where $m_1, m_2$ are the masses of the $Sp(1)$-fundamentals, $m_3$
is the mass of the $SO(4)$-fundamental and $a, b_i$ are the Coulomb
branch parameters of $Sp(1)$ and $SO(4)$ respectively.

\begin{figure}[htbp]
\begin{center}
\includegraphics[width=3in]{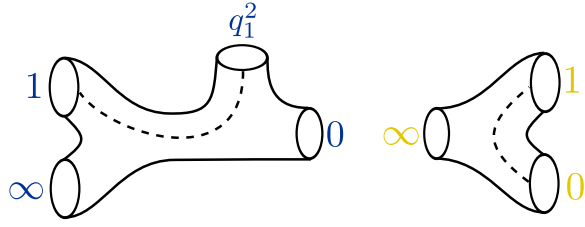}
\caption{Decomposition of the $Sp(1) \times SO(4)$ Gaiotto curve that
  we used for computing the corresponding $\CW(2,2)$-block.}  
\label{fig:SOSpCFT}
\end{center}
\end{figure}

\subsubsection{$Sp(1) \times SO(4)$ correlators}

The correlator that corresponds to the (mirror of the) quiver illustrated in
Figure~\ref{fig:generalSOSpquiver} is the following:
\be
\langle  V_1(\infty)\sigma(1)\sigma(q_1^2)\sigma(q_1^2q_2)\hat V_2(0)\rangle
\ee
For notational simplicity we have introduced the variables
$q_1=\qSp/4$ and $q_2=\qSO$.
A single term in the chiral block is then expanded as (see Figure~\ref{fig:SOSpCFT})
\begin{multline}\label{SpSOexpand}
\langle V_1(\infty)\sigma(1) \sigma(q_1)V^{A,B}(0)\rangle 
\langle V^{A,B}(\infty)\sigma(q_1^2q_2)\hat V_2(0)\rangle \\
=\langle V_1(\infty)\sigma(1) \sigma(q_1^2)V^{A,B}(0)\rangle 
(q_1^2q_2)^{h_A+n_A+h_B+n_B}
\langle V^{A,B}(\infty)\sigma(1)\hat V_2(0)\rangle
\end{multline}
To compute the rightmost correlator,
we can proceed as in subsection~\ref{ss:SO4block}
and map it to a cover correlator of the form
(\ref{SO4CorrCover}). Due to the 
corrections we will get something of the form
\be
\langle V^{A,B}(\infty)\sigma(1)\hat V_2(0)\rangle
= \left(\frac{1}{2}\right)^{n_A+n_B}\Ccover^{h_A,h_B;h_1}_{I_A,I_B,\bullet}
+ \textrm{lower descendant corrections}\ .
\ee
The correlator is thus the same
as the one we computed in the section on $SO(4)$.

The four point correlator on the left we treat
as in the $Sp(1)$ case, \ie we apply the cover map
(\ref{Sp1covermap}). The only difference is then
that $V^{A,B}$ is a descendant field, and thus
like in the $SO(4)$ computation picks up corrections from the map: 
\begin{multline}\label{SO4Sp1coverfields}
V^{A,B}(0)=V\left(\left(\frac{-1+q_1^2}{2q_1}\right)^{L_0}\exp\left[\left(\frac{3}{4}+\frac{1}{4q_1^2}\right)L_1\right]\exp\left[\frac{(-1+q_1^2)^2}{16q_1^4}L_2\right]\cdots\phi^A,q_1 \right)\\ \times
V\left(\left(-\frac{-1+q_1^2}{2q_1}\right)^{L_0}\exp\left[\left(\frac{3}{4}+\frac{1}{4q_1^2}\right)L_1\right]\exp\left[\frac{(-1+q_1^2)^2}{16q_1^4}L_2\right]\cdots\phi^B,-q_1 \right)
\end{multline}
This expression is however different in two ways
from the corresponding $SO(4)$ expression (\ref{SO4coverfields}).
First, the vertex operators are at the positions $\pm q_1$.
We decompose the four punctured correlator on the cover in
usual way, and move them to the standard positions $\pm 1$
using the map $z \mapsto z q_1^{-1}$. This simply leads to an
additional prefactor $q_1^{-L_0}$ in equation~(\ref{SO4Sp1coverfields}).
To pull out the standard prefactor $q_1^{h_{A,B}}$,
it is useful to commute this prefactor all the way to the left, which
we can do by using the identity
\be
x^{L_0}L_n= \frac{L_n}{x^n} x^{L_0}\ .
\ee
This leads to the expression
\begin{align}\label{SO4Sp1coverfields2}
&\left(\frac{-1+q_1^2}{2q_1^2}\right)^{h_A+n_A+h_B+n_B} \times \\
&\langle \phi | V\left(\exp\left[\frac{1+3 q_1^2}{2 (-1+q_1^2)}L_1\right]e^{L_2/4}\cdots\phi^A,1 \right)
V\left(\exp\left[-\frac{1+3 q_1^2}{2
      (-1+q_1^2)}L_1\right]e^{L_2/4}\cdots\phi^B,-1 \right)\rangle\ . \notag
\end{align}
Note that the $q_1^{2(h_A+n_A+h_B+n_B)}$ in the denominator of the prefactor exactly
cancels the corresponding factor of $q_1$ in equation~(\ref{SpSOexpand}).
The numerator of the prefactor on the other hand is the same prefactor that we
already found in the $Sp(1)$ computation.

Let us now actually compute the first few terms of the chiral block. 
The terms where of order zero in either $\qSp$ or $\qSO$ are
are simply the same as in the $SO(4)$ and $Sp(1)$ computation.
We therefore consider the simplest new term, $\qSp \qSO$.
This means in particular that we can neglect all terms of 
order $q_1^2$ in the expression~(\ref{SO4Sp1coverfields2}), so that
the vertex 
operators no longer depend
on $q_1$. Since we would like to rewrite this expression 
in terms of the three point coefficients $\tilde{\mathbf{C}}$
defined above, we need move the field $\phi$ from $\infty$ to 0 by
applying 
the map $z \mapsto z^{-1}$. Note that we pick up some additional
corrections due to the fact that the fields at $\pm 1$ are 
descendant fields. In total (\ref{SO4Sp1coverfields2}) thus becomes
\be
(2^{-n_A-n_B})
\langle V\left(e^{3L_1/2}e^{L_2/4}\cdots\phi^A,1 \right)
V\left(e^{-3L_1/2}e^{L_2/4}\cdots\phi^B,-1 \right)|\phi\rangle\ .
\ee
Not surprisingly, this is the same expression that we had found in 
the $SO(4)$ case.
From this, the term of order $\qSp\qSO$ is
\begin{align}
\frac{\Ccover^{h_1,h_2;h_a}_{\bullet,\bullet,\square}}{2h_a}\frac{1}{4}
\Bigl( &\frac{(\Ccover^{h_A,h_B;h_a}_{\square,\bullet,\square} + 
3h_A\Ccover^{h_A,h_B;h_a}_{\bullet,\bullet,\square} )
(\Ccover^{h_A,h_B;h_3}_{\square,\bullet,\bullet}+3h_A)}{2h_A}\\
&\qquad \qquad+\frac{(\Ccover^{h_A,h_B;h_a}_{\bullet,\square,\square} - 
3h_B\Ccover^{h_A,h_B;h_a}_{\bullet,\bullet,\square} )
(\Ccover^{h_A,h_B;h_3}_{\bullet,\square,\bullet}-3h_B)}{2h_B}
\Bigr). \notag
\end{align}

Similar computations lead to higher order terms.
We have checked that the instanton partition function and the chiral
block agrees up to order $(k_1, k_2) = (1, 2)$ up to a moduli independent
spurious factor, using the same
identifications of the previous examples.


\acknowledgments

It is a pleasure to thank Matthias Gaberdiel, Sergei
Gukov, Daniel Jafferis, Marcos Mari\~{n}o, Yu Nakayama, Hirosi Ooguri,
Vasily Pestun.  We would like to especially thank Fernando
Alday and Yuji Tachikawa for many useful suggestions, helpful
explanations and enlightening discussions, Fernando Alday for
helping us to find some of the exact series in section 2 and Yuji
Tachikawa for carefully reading the manuscript.  

The work of LH is supported by an NWO Rubicon grant and by NSF grant
PHY-0757647. 
The work of CAK is supported by a John A.~McCone Postdoctoral Fellowship.
The work of JS is supported in part by a Samsung Scholarship. 
This work is in addition supported in part by the DOE grant
DE-FG03-92-ER40701. 

LH thanks the Mathematical Institute at the University of Oxford, the University of
Amsterdam, the Aspen Center for Physics and the KITP at
Santa Barbara for kind hospitality during the process of this
project. 
LH and JS as well thank the organizers of the Eighth Simons Workshop
in Mathematics 
and Physics  at the Stony Brook University. 
JS thanks
the hospitality of Korea Institute for Advanced Study and
CAK thanks the Erwin Schr\"odinger Institute
in Vienna for hospitality.

\vskip 1cm

\vskip 1cm
\centerline{\Large\bf Appendix}
\appendix

\section{Instanton counting} \label{app:Instanton}
In this appendix we summarize and extend methods to derive the
instanton counting formulae for general gauge groups and matter
in several representations. In particular, we find the contour
integral prescription for half-bifundamental $Sp-SO$ hypermultiplets.\footnote{ 
Additional explanations about the ADHM moduli space can for instance
be found in \cite{ADHM,CorriganGoddard, BS00}, about instanton
counting in the physics 
literature
\cite{Losev:1997tp,Moore:1998et,Nekrasov:2002qd,Nekrasov:2003rj,Losev:2003py}, 
and in the mathematics literature in 
\cite{NakajimaHilb,Nakajima:2003uh,Nakajima:2003pg,Braverman:2004vv,
  Braverman:2004cr,LicataSavage,Gottsche:2010},  
and about $Sp/SO$ instanton counting in specific in \cite{Nekrasov:2004vw,
  Shadchin:2005mx,Martens,Marino:2004cn,Fucito:2004gi}.}

\subsection{ADHM construction}

Let $E$ be a rank $N$ complex vector bundle on $\IR^4$ with a
connection $A$ and a framing
at infinity. The framing is an isomorphism of the fiber at infinity 
with $\IC^N$. The ADHM construction studies the moduli space $\CM_k$ of
connections $A$ on the bundle $E$ that satisfy the self-dual
instanton equation $F^+(A)=0$, up to gauge transformations that are
trivial at infinity. It turns out
that this moduli space can be realized as a hyperk\"ahler quotient of
linear data. 

\subsubsection*{$U(N)$ gauge group}

\begin{figure}[h!]
\begin{center}
\includegraphics[width=2.3in]{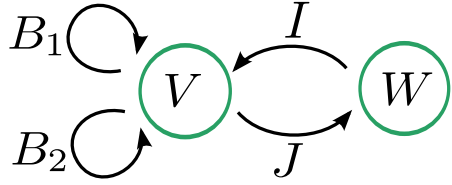}
\caption{Quiver representation of the $U(N)$ ADHM quiver. The vector
  spaces $V$ and $W$
  are $k$ and $N$-dimensional, respectively, with a natural action of
  the dual group $U(k)$ and the 
  framing group $U(N)$. The maps $B_1$, $B_2$, $I$ and $J$ 
are linear. }\label{fig:adhmu}  
\end{center}
\end{figure}

For the gauge group $G=U(N)$ the linear data consists of four linear maps
\be
(B_1, B_2, I,J) \in \mathbf{X} = \mathrm{Hom}(V,V) \oplus \mathrm{Hom}(V,V) \oplus \mathrm{Hom}(W,V) \oplus \mathrm{Hom}(V,W),
\ee
where $V$ and $W$ are two complex vector spaces of dimension $k$ and
$N$, respectively. This linear data is summarized in an ADHM quiver
diagram in Figure~\ref{fig:adhmu}. The vector space $W$ is
isomorphic to the fiber of $E$ (which in our case is of rank $N$). 
It is best thought of as 
the fiber at infinity, as there is a natural action of the
framing group $U(N)$ on it, which physically can be thought
of as the large gauge transformations at infinity.
The tensor product of the vector space $V$ with the half canonical
bundle $K_{\IC^2}^{1/2}$ on $\IC^2 \cong \IR^4$, on the other hand,
can be identified with 
the space of normalizable solutions to the Dirac
equation in the background of the instanton gauge field $A$. 
Since the instanton number $k = 1/8\pi^2 \int F_A \wedge F_A$
is given by the second Chern class of $E$, it follows
from index theorems that this space has dimension $k$,
as we advertised above. In particular it carries in a natural
way the action of the dual group $U(k)$. More algebraically, the
vector space $V$ itself is isomorphic to the cohomology group
$H^1(E)$. 

The framing group $U(N)$ and the dual group $U(k)$ 
thus act naturally on the linear ADHM data. Setting the three real moment maps 
 \be
   \mu_{\IR} &=& [B_1, B_1^\dagger] + [B_2, B_2^\dagger] + I I^\dagger - J^\dagger J \\
   \mu_{\IC} &=& [B_1, B_2] + IJ, 
\ee
to zero gives the so-called \emph{ADHM equations}. The ADHM construction identifies the
instanton moduli space $\CM_k^{U(N)}$ with the hyperk\"ahler quotient
of the solutions $\mathbf{X}$ to those equations by the
dual group,
\be 
\CM^{U(k)}_k = \mathbf{X} // U(k) \equiv \mu^{-1}(0)/U(k). 
\ee

From a physical perspective the ADHM construction can be most
natural understood using D-branes. We can engineer the moduli space of
$k$ instantons in the four-dimensional $U(N)$ theory by putting $k$
D$(p-4)$-branes on top of $N$ D$p$-branes. The D$(p-4)$-branes 
appear as zero-dimensional instantons on the transverse
four-dimensional manifold.  
The maps $(B_1,B_2,I,J)$ can be understood as the zero-modes of
D$(p-4)-$D$(p-4)$), D$p-$D$(p-4)$ and D$(p-4)-$D$p$ open strings,
respectively, and the ADHM equations are the D-term conditions. The
ADHM quotient can thus be identified with the moduli space of the
Higgs branch of the $U(k)$ gauge theory on the D$(p-4)$-branes. 

Since the above quotient is highly singular due to small instantons,
we change it by giving non-zero value to the Fayet-Illiopolous term
$\zeta$. This is equivalent to turning on NS 2-form field on the
D$p$-branes, and the resulting desingularized quotient can be
interpreted as a moduli space of non-commutative
instantons \cite{Nekrasov:1998ss}.   

To perform our computations another, equivalent way of
representing the ADHM construction will be useful. Let us introduce
the spinor bundles $\CS^{\pm}$ of positive and negative chirality on
$\IR^4$, and for brevity denote the half canonical bundle by
$\CL = K_{\IC^2}^{1/2} $.  Consider the sequence
\begin{align}\label{complexU(N)}
 V \otimes L^{-1} \rightarrow^{\hspace*{-3mm}  \sigma}  \hspace{2mm} \begin{array}{c} V \otimes S^- \\ \oplus \\
  W \end{array} \rightarrow^{\hspace*{-3mm}  \tau}
 \hspace{2mm} V \otimes L\, ,
\end{align}
where $S^{-}$ and $L$ are the fibers of the bundles $\CS^{-}$ and
$\CL$, respectively. Although these can all be trivialized, they are non-trivial
equivariantly. We thus need to keep track of them for later. The mappings
$\sigma$ and $\tau$ are defined by   
\be
\sigma = \left( \begin{array}{c} z_1-B_1 \\ z_2-B_2 \\ J \end{array} \right), \qquad 
\tau = \left( \begin{array}{ccc}  -z_2 + B_2, & z_1-B_1, & I \end{array} \right),
\ee
where $(z_1,z_2)$ are coordinates on $\IC^2$. From the ADHM equations
it follows  that $\tau \circ \sigma = 0$,
so that the sequence~(\ref{complexU(N)}) is a chain
complex. Since $\sigma$ is injective and $\tau$
surjective, it is a so-called monad.  

Notice that the
vector space $V \otimes L^{-1}$ at the first position of the
sequence~(\ref{complexU(N)}) fixes the vector spaces at the remainder
of the sequence. The fields $B_1$ and $B_2$ are coordinates on $\IC^2$
and thus map $V \otimes L^{-1} \to V \otimes S^-$. The
fields $I$ and $J$ are the two scalar components of an $\CN=2$
hypermultiplet, that properly speaking transform as sections of the
line bundle~$\CL$. 

To recover the vector bundle $E$, we vary the cohomology
space $(\rm{Ker} \, \tau) / ( \rm{Im}  \, \sigma)$ over $\IC^2$, which
gives indeed a vector bundle whose fiber at infinity is equal to $W$. One can also
show that the curvature of this bundle is self-dual and that it has
instanton number $k$. Even better, every solution of the self-dual
instanton equations can be found in this way. 

We are now ready to construct the main tool in our computation. This is
the \emph{universal bundle} $\CE$ over the instanton moduli space
$\CM^{U(N)}_k \times \IR^4$. The universal bundle is obtained by varying the
ADHM-parameters of the maps in the complex~(\ref{complexU(N)}). It has
the property that 
\be
\CE_{A,z} = E_z,
\ee
\ie its fiber over an element $A \in \CM^{U(N)}_k$ is the total space
of the bundle $E$ with connection~$A$. Remember that the bundle $E$
has fiber $W$ at infinity in $\IR^4$ and that the vector space $V$ of
solutions to the Dirac equations is related to its first cohomology
$H^1(E)$. The vector spaces $V$ and $W$ can be extended to bundles
$\CV$ and $\CW$ over the instanton moduli space. We can then easily
compute the Chern character of the universal bundle $\CE$  
from its defining complex (\ref{complexU(N)}) as
\bea \label{chernuniversal}
\rm{Ch}(\CE) = \rm{Ch}(\CW) + \rm{Ch}(\CV) \left( \rm{Ch} (\CS^-) -
  \rm{Ch}(\CL) - \rm{Ch} (\CL^{-1}) \right).
\ee

\subsubsection*{ $SO/Sp$ gauge groups}
The construction for $SO(N)$ and $Sp(N)$ gauge groups
is very similar. We define $Sp(N)$ to be the
special unitary transformations on $\IC^{2N}$
that preserve its symplectic structure $\Phi_s$,
and $SO(N)$ the special unitary transformations
on $\IC^{N}$ that preserve its real structure $\Phi_r$.

\begin{figure}[h!]
\begin{center}
\includegraphics[width=2.6in]{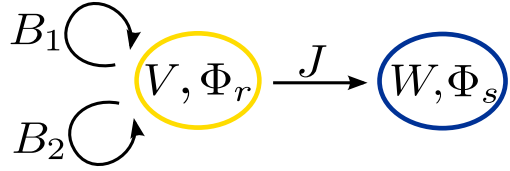}
\caption{Quiver representation of the $Sp(N)$ ADHM quiver. The vector
  spaces $V$ and $W$  are $k$ and $2N$-dimensional, respectively. $V$
  has a real structure $\Phi_r$ and a natural action of
  the dual group $SO(k)$, whereas $W$ has a symplectic structure
  $\Phi_s$ and a natural action of the  framing group $Sp(N)$. The
  maps $B_1$, $B_2$ and $J$ are linear.}\label{fig:adhmsp}  
\end{center}
\end{figure}

For $Sp(N)$ the linear data that is needed to define the ADHM
complex consists of 
\be
(B_1, B_2, J) \in \mathbb{Y} = \mathrm{Hom}(V,V) \oplus
\mathrm{Hom}(V,V) \oplus \mathrm{Hom}(V,W),
\ee
where $V$ and $W$ are a complex $k$ and $2N$-dimensional vector space,
resp., together with a real structure $\Phi_r$ on $V$ and a symplectic
structure $\Phi_s$ on $W$. This is illustrated as a quiver diagram in
Figure~\ref{fig:adhmsp}. The dual group is given by $O(k)$, so that
the moduli space of $Sp(N)$
instantons is given by 
\be
 \CM^{Sp(N)}_k = \{ (B_1, B_2, J) \, | \, \Phi_r B_1,  \Phi_r B_2 \in S^2 V^*,
 \, \Phi_r [B_1, B_2] - J^* \Phi_s J = 0 \} / O(k) . 
\ee 

\begin{figure}[h!]
\begin{center}
\includegraphics[width=2.6in]{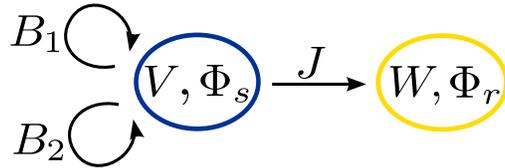}
\caption{Quiver representation of the $SO(N)$ ADHM quiver. The vector
  spaces $V$ and $W$  are $2k$ and $N$-dimensional, respectively. $V$
  has a symplectic structure $\Phi_s$ and a natural action of
  the dual group $Sp(k)$, whereas $W$ has a real structure
  $\Phi_r$ and a natural action of the  framing group $SO(N)$. The
  maps $B_1$, $B_2$ and $J$ are linear.}\label{fig:adhmso}  
\end{center}
\end{figure}

For $SO(N)$ we just need to replace $V$ and $W$ by a complex $2k$ and
$N$-dimensional vector space, resp.,  as well as change the role of
symplectic structure and 
the real structure. This is illustrated as a quiver diagram in
Figure~\ref{fig:adhmso}. The dual group is given by $Sp(k)$, so that 
the moduli space of $SO(N)$
instantons is given by 
\be
 \CM^{SO(N)}_k = \{ (B_1, B_2, J) \, | \, \Phi_s B_1,  \Phi_s B \in \wedge^2 V^*,
 \, \Phi_s [B_1, B_2] - J^* \Phi_r J = 0 \} / Sp(k) . 
\ee 

A subtle issue for the above moduli spaces is that there is no
appropriate Gieseker desingularization which resolves the singularity
due to the zero-sized instantons (as in the case of $U(N)$). One way
to understand this is by considering the string theory embedding. The
above ADHM constructions can be obtained by considering D$p$-D$(p-4)$
system and also adding an $O^{\pm}p$ plane on the top of the D$p$
branes. In the case of $U(N)$, the non-commutativity parameter that we
introduce is coming from the NS 2-form field on the $Dp$-brane. But,
the orientifold makes it impossible to turn on the background NS
2-form field. So we cannot resolve the singularity in the same way. An
alternative way to resolve the singularity was studied by
\cite{Kirwan85}, but a physical understanding of this procedure is
still lacking.  
Nevertheless, we will see that the equivariant volume of the moduli space can be
obtained without explicitly resolving the singularity
\cite{Nekrasov:2004vw}. This formula is verified mathematically using
Kirwan's formula of the equivariant volume of the symplectic quotient~\cite{Martens}. 

We can then again represent
any $Sp(N)$ instanton solution $E$ as the
cohomology bundle of the sequence
\begin{align}\label{complexSp(N)}
 V \otimes L^{-1} \longrightarrow^{\hspace*{-4mm}  \sigma}  \hspace{3mm} \begin{array}{c} V \otimes S^- \\ \oplus \\
  W  \end{array} \longrightarrow^{\hspace*{-6mm}  \sigma^* \beta^*}
 \hspace{4mm} V^* \otimes L,
\end{align}
where the mappings $\sigma$ and $\beta$ are defined by 
\be
\sigma = \left( \begin{array}{c} z_1 - B_1 \\ z_2 -
    B_2 \\ J \end{array} \right), \qquad 
\beta = \left( \begin{array}{ccc} 0 & \Phi_r & 0 \\ - \Phi_r & 0 & 0 \\ 0
    & 0 & \Phi_s \end{array} \right).
\ee
The ADHM equations ensure that $\sigma^* \beta^* \sigma =
0$, and the sequence~(\ref{complexSp(N)}) is another monad.
Analogously to the $U(N)$ example, when varying the ADHM-parameters in 
the complex~(\ref{complexSp(N)}) we find the universal bundle
$\CE_{Sp(N)}$. $V$ is the $k$-dimensional solution space of the $Sp(N)$ 
Dirac operator on $\IC^2$, which carries a real structure, and $W$ is
the fiber of $E$ at infinity, and hence carries a symplectic
structure. 

Similarly, any $SO(N)$ instanton solution $E$ can be represented as the
cohomology bundle of the complex~(\ref{complexSp(N)}) as well, once 
we exchange $\Phi_r$ with $\Phi_s$ in the definition of the map $\beta$. 
The resulting complex is a short exact sequence, since 
according to the ADHM equations $\sigma^* \beta^* \sigma =
0$. By varying the ADHM-parameters we find the $SO(N)$ universal
bundle $\CE_{SO(N)}$. Note that 
$V$ is the $2k$-dimensional solution space of the $SO(N)$  
Dirac operator on $\IC^2$, which carries a symplectic structure, whereas $W$ is
the fiber of $E$ at infinity, and hence carries a real structure.

\subsection{$\Omega$-background and equivariant integration}

Let us start with a supersymmetric $\CN=2$ gauge theory without
matter. This theory can be topologically twisted (using the so-called
Donaldson twist), so that its BPS equation is the instanton equation
$F_A^+=0$.  The instanton partition function is given by 
an integral over the moduli space $\CM^{\rm inst}_k$, 
\be 
Z^{\rm inst} = \sum_k q^k \oint_{\CM^{\rm inst}_k} 1\ ,
\ee
where $\oint 1$ indicates the formal volume of the hyperk\"ahler quotient. 

If we add matter multiplets to the gauge theory, say a single
$\CN=2$ hypermultiplet, the BPS equations turn into the monopole equations
\begin{align}
F_{A,\mu \nu}^+ + \frac{i}{2} \ \overline{q}_{\alpha}
\Gamma^{\hspace{3.5mm}\alpha}_{\mu \nu \hspace{2mm} \beta}
q^{\beta}  &= 0, \\  
\sum_{\mu} \Gamma^{\mu}_{\dot{\alpha} \alpha} D_{A,\mu} q^{\alpha} &=0.  \notag
\end{align}
Here, $\Gamma^{\mu}$ are the Clifford matrices and $\sum_{\mu}
\Gamma^{\mu} D_{A,\mu}$ 
is the Dirac operator in the instanton background for
the connection $A$. Furthermore, $q^{\alpha}$ is the lowest component of 
the twisted hypermultiplet. The representation of the connection $A$
in the connection $D_A$ is determined by the representation of the
gauge group that the hypermultiplet is in.  As is argued in section
3.4 of \cite{Shadchin:2005mx}, it is possible to deform the action in a
Q-exact way such that the first equation gets an extra factor 
\begin{align}
F_{A,\mu \nu}^+ + \frac{i}{2 t} \ \overline{q}_{\alpha}
\Gamma^{\hspace{3.5mm}\alpha}_{\mu \nu \hspace{2mm} \beta}
q^{\beta}  &= 0,, \notag
\end{align}
for an arbitrary value of $t$. Taking the limit $t \to \infty$ reduces
this BPS equation to the selfdual instanton equation. 
Given an instanton solution $A$, the remainder of the action is forced
to localize onto solutions of the Dirac equation in the background
of $A$. The kernel of the Dirac operator thus forms a fiber over 
the instanton moduli space $\CM^{\rm inst}$. 

Let us start out by adding a single hypermultiplet in the
fundamental representation of the gauge group.
Remember that
the corresponding kernel was already encoded in the vector bundle $\CV
\otimes \CL$ 
in the original ADHM construction. 
$N_f$ hypers are simply described by the tensor product of $N_f$
vector bundles $\CV$. Each individual factor 
then carries the usual action of 
the dual group. Moreover, there is now
also a natural action of the flavor symmetry group.
By general arguments (see for
example \cite{Cordes:1994fc}) this partition function then 
localizes to the integral\footnote{Here (and elsewhere in the paper)
  $\CL$ is really just the fiber of the half-canonical bundle $\CL$ at
  the origin of $\IR^4$. More formally, we take  the cup product of
  the bundle $\CV \otimes \CL \otimes M$ over $\CM^{\rm inst}_k \times
  \IR^4$ with the push-forward $i_*$ of the fundamental class $[\CM^{\rm
  inst}_k]$, where $i$ embeds the moduli
space $\CM^{\rm inst}_k$ in the product  $\CM^{\rm inst}_k \times
  \IR^4$ at the origin of $\IR^4$. }
\be \label{instpartfunctionlocappendix}
Z^{\rm inst} = \sum_k q^k \oint_{\CM^{\rm inst}_k} e(\CV \otimes \CL \otimes M)
\ee
of the Euler class of the vector bundle $\CV \otimes
\CL$ of solutions to the Dirac equation over the moduli space
$\CM^{\rm inst}$. The vector space $M = \IC^{N_f}$ encodes the number
of flavors.

Let us now discuss how to compute this partition function.
First,
 however, note that (\ref{instpartfunctionlocappendix}) diverges. We will
 implicitly take care of the UV divergence in the next steps
 by computing a holomorphic character \cite{Nekrasov:2004vw}, but 
 we also need to deal with IR divergences: the
 instanton moduli space has flat directions where the instantons move
 off to infinity. One way to regularize the instanton partition
 function is to introduce the so-called $\Omega$-background, where we
 use equivariant integration with respect to the torus 
\be
\mathbf{T}^2_{\e_1,\e_2} = U(1)_{\e_1} \times  U(1)_{\e_2},    
\ee
that acts on $\IR^4 = \IC \oplus \IC $ by a rotation $(z_1, z_2) \to
(e^{i\e_1} z_1, e^{i \e_2} z_2)$ around the origin. This forces the
instantons to be localized at the origin 
of $\IR^4$. The resulting $\Omega$-background is denoted by
$\IR^4_{\e_1,\e_2}$.  
The partition function in the 
$\Omega$-background is defined by equivariantly integrating with
respect to the $\mathbf{T}^2_{\e_1,\e_2}$--action.

We have already introduced the other components of the 
equivariance group: The torus
$\mathbf{T}^N_a$ of the gauge group $G$ acting on the fiber $W$ with weights $a_l$, 
the torus $\mathbf{T}^k_{\phi_i}$ of the dual group acting 
on $V$ with weights $\phi_i$, and lastly the torus $\mathbf{T}^{N_f}_m$ 
of the flavor symmetry group acting on
the flavor vector space $M$ with weights $ m_j$.  
In total, we perform the equivariant integration with respect to the
torus 
\be
\mathbf{T}= \mathbf{T}^2_{\e_1,\e_2} \times \mathbf{T}^N_a  \times
\mathbf{T}^k_{\phi} \times
\mathbf{T}^{N_f}_m\ .
\ee

 The
instanton partition function in the $\Omega$-background is thus
defined as the equivariant integral  
\be \label{instpartfunctionloc1}
Z(a,m,\e_1,\e_2) = \sum_k q^k \oint_{\CM_k} e_{\mathbf{T}} (\CV
\otimes \CL
\otimes M)\, ,
\ee
where $e_{\mathbf{T}}$ is the equivariant Euler class with respect to the
torus $\mathbf{T}$. 
We will evaluate (\ref{instpartfunctionloc1}) in two steps,
using the fact that $\CM_k$ is given by the solutions $\mu^{-1}(0)$
to the ADHM equations quotiented by the dual group $G^D_k$.
We will thus first perform the equivariant integral
over $\mu^{-1}(0)$, and then take care of the quotient
by integrating out $G^D_k$, which gives
a multiple integral over $\phi_i$. 

To perform the first part of the above integral, we apply the famous
equivariant localization 
theorem, which tells us that the integral only depends
on the fixed points of the equivariant group and its weights
at those points. This then leads to a rational function
in all the weights.

More precisely, suppose that the action of the element $t \in
\mathfrak{t}$ on the integration space $\CM$ (which is represented by a vector
field $V_{t}$) has a discrete number of fixed points $f$. Then
the equivariant localization theorem says
that   
\be \label{equivlocalization}
\int_{\CM} \alpha= \sum_{f} \frac{\i^*\alpha}{\prod_{k} w_k[t] (f)}, 
\ee
where $\i$ embeds the fixed point locus in $\CM$ and
where $w_k[t](f)$ are the weights of the action of the vector field $V_t$
on the tangent space to the fixed point $f \in \CM$.
If we apply the localization theorem to the integral~(\ref{instpartfunctionloc1}), the
denominator of the resulting expression contains a product of weights
of the torus action on the 
tangent bundle to the instanton moduli space. Its numerator is given
by another product of weight of the torus action on the bundle
$\CV \otimes \CL \otimes M$ of Dirac zero modes.  

Let us start with computing the weights in the numerator, and for
convenience restrict the matter content to a single hypermultiplet in
the fundamental representation of the gauge group. Since
the bundle in the numerator is the kernel of the Dirac operator, we can
equally well obtain these weights from the equivariant index
$\textrm{Ind}_{\mathbf{T}} 
= \sum_k n_k e^{i w_k}$ of the Dirac operator. For
the purpose of (\ref{equivlocalization}), the sum over weights
can be translated into a product by the formula  
\be \label{SumToProd}
 \sum_k n_k e^{i w_k} \to \prod_i (w_k)^{n_k}.
\ee
To compute the equivariant index $\textrm{Ind}_{\mathbf{T}}$ of the Dirac
operator coupled to the instanton background, we make
use of the equivariant version of Atiyah-Singer index theorem. It is given by
\be \label{indexfunddirac}
 \textrm{Ind}_{\mathbf{T}} = \int_{\IC^2} \textrm{Ch}_{\mathbf{T}}
 (\CE \otimes \CL) \, \textrm{Td}_{\mathbf{T}} 
 (\IC^2) = \frac{\textrm{Ch}_{\mathbf{T}}(\CE \otimes \CL)
   |_{z_1=z_2=0}}{(e^{i\e_1} - 
   1)(e^{i\e_2} -1)},
\ee
where $\CE$ is the universal bundle over the instanton moduli space
$\CM_k$ that we constructed in the previous section. 
Remember that the fiber of $\CE$ over an element $A$ in the
instanton moduli space is given by the total space of the instanton
bundle $E$ with connection $A$. The second 
equality is obtained by applying the equivariant 
localization theorem and using the equivariant Todd class of $\IC^2$ equals
\be 
\textrm{Td}_{\mathbf{T}} (\IC^2)  = \frac{\e_1 \e_2}{(e^{i\e_1} -
   1)(e^{i\e_2} -1)},
\ee
where the weights of the action of $\mathbf{T}_{\e_1,\e_2}$ on $\IC^2$
are $\e_1$ and $\e_2$. 

The purpose of all of this was to reduce everything to
the equivariant Chern character of the universal bundle $\CE$,
for which have found the simple expression~(\ref{chernuniversal}) 
in terms of the Chern characters of the vector bundles $\CW$, $\CV$, $\CL$
and $\CS$. We can
easily obtain the weights of the torus $\mathbf{T}$ on these bundles,
so that we can compute the contribution of a fundamental
hypermultiplet. We will write down explicit expressions in a
moment, but let us first explain how to obtain the weights for other
representations. 

If we instead wish to extract the weights for an anti-fundamental
hyper we just need replace the equivariant character for universal bundle $\CE$ by
its complex conjugate $\CE^*$. Other representations that are 
tensor products of fundamentals and anti-fundamentals (or symmetric
or antisymmetric combinations thereof) can be obtained
similarly. For instance, the adjoint representation for a classical
gauge group can be expressed as some product of the fundamental and
the anti-fundamental representation. This product is the tensor
product for $U(N)$, the anti-symmetric product for $SO(N)$ and the 
symmetric product for $Sp(N)$. Note that in those cases
we also obtain the representations of the dual groups.
We thus obtain the weights for an
adjoint hypermultiplet by computing the character of the appropriate
product of the universal bundle and its complex conjugate. 
The weights
for the gauge multiplet are the same as for an adjoint hypermultiplet,
but end up in the denominator of 
the contour integral instead of the numerator. This is consistent with
the localization 
formula~(\ref{instpartfunctionloc1}), as the tangent space to the
instanton moduli space can be expressed as the same product of the
universal bundle and its dual.

Once we obtain the index, we can extract the equivariant weights from
it by using the rule \eqref{SumToProd}. 
Finally, we need to integrate out the dual group $G^D_k$.
This leads to a multiple integral over $d\phi_i$ along
the real axis. We will absorb factors appearing from
this integration such as the Vandermonde determinant
of the Haar measure and the volume of the dual group
into the contribution of the gauge multiplet $\mathbf{z}^k_{\rm gauge}$.
The resulting integrand actually has poles on the real
axis, which we cure by giving small imaginary parts
to the equivariance parameters. We will describe this
in more detail once we turn to the actual evaluation
of such integrals.
Since the integrand obtained is a rational function in the 
parameters of $G^D_k$, we can convert
the integral into a contour integral around the poles of the integrand.
In total the equivariant 
integral~(\ref{instpartfunctionloc1}) over the moduli space thus reduces to   
\be \label{finalcontourint}
 Z_k (a, m, \e_1, \e_2 ) =  \oint \prod_i d\phi_i \prod_R \mathbf{z}^k_{R} (\phi_i , a, m, \e_1, \e_2),
\ee
where $\mathbf{z}^k_{R}$ are the integrands that represent the matter
content of the gauge theory and where the $\phi_i$'s parametrize the
dual group. We will discuss which poles (\ref{finalcontourint})
is integrated around shortly.

\subsubsection*{Equivariant index for $U(N)$ theories}

Let us see how this works out explicitly for gauge group $U(N)$. We
first have a look at the weights of the torus
$\mathbf{T}^2_{\e_1,\e_2} $ at the fibers of the half-canonical bundle $\CL$
and the spinor bundles $\CS^{\pm}$ at the origin of $\IR^4$. Remember
that the torus 
$\mathbf{T}^2_{\e_1,\e_2} $ acts on the coordinates $z_1$ and $z_2$
with weights $\e_1$ and $\e_2$ respectively. It thus acts on local sections $s$ of the
half-canonical bundle as
\bea
 & s \in \CL:&  \quad s  \mapsto e^{i \e_+} s \, , \nn 
\ee
with $\e_\pm = \frac{e_1 \pm e_2}{2}$. Local sections of the four-dimensional spinor
bundles $\CS^{\pm}$ can be written in terms of those of the 
two-dimensional spinor bundles on $\IR^2$. Since the weights of
the torus $\mathbf{T}^1_{\e_j}$ on the 
local sections of the two spinor bundles on $\IR^2$ are $ \pm
\frac{\e_j}{2}$, the torus
$\mathbf{T}^2_{\e_1,\e_2} $ acts on
local sections $\psi_{\pm}$ of the four-dimensional spinor bundle as
\bea
 & \psi_{\pm} \in \CS^{\pm}:&  \quad \psi_{\pm}  \mapsto
 \textrm{diag}(e^{i \e_{\pm}}, e^{-i \e_{\pm}}) \, \psi_{\pm}  \, . \nn
\ee

Let us continue with the weights of the equivariant torus $\mathbf{T}^N_a \times
\mathbf{T}^k_{\phi} $. The equivariant torus then acts on
the linear ADHM data as 
\bea
&  v \in V:&  \quad v  \mapsto \textrm{diag}\, (e^{i\phi_1} , \cdots, e^{i
   \phi_k}) \, v \,, \nn \\
 & w \in W: & \quad w  \mapsto \textrm{diag}\, (e^{i a_1}, \cdots, e^{i
   a_N}) \, w \, . \nn
\ee

Combining all weights and using the formula~(\ref{chernuniversal}),
we find that the equivariant Chern character of the universal bundle
$\CE_{U(N)}$ is given by    
\be
 \textrm{Ch}_{\mathbf{T}} (\CE_{U(N)})|_{z_1=z_2=0} 
 &=& \sum_{l=1}^n e^{ia_l} - (e^{i \e_1} -1)(e^{i \e_2} - 1) 
 \sum_{i=1}^k e^{i\phi_i-i\e_+}. 
\ee
Using the index formula~(\ref{indexfunddirac}) we have now computed the
contribution for a fundamental massless hypermultiplet. As we
explained before we can easily generalize this to other 
representations. In particular, we can give
the hypermultiplet a mass by introducing a weight $m$ for the flavor
torus $\mathbf{T}^{1}_m$. This will act on the linear ADHM data as 
\bea
&  v \in V:&  \quad v  \mapsto \textrm{diag}\, (e^{im} , \cdots, e^{im}) \, v \,, \nn \\
 & w \in W: & \quad w  \mapsto \textrm{diag}\, (e^{im}, \cdots, e^{i
   m}) \, w \, . \nn
\ee

For gauge group $U(N)$ the poles of the
resulting contour integral~(\ref{finalcontourint}) can be labeled by a set of colored
Young diagrams $\mathbf{Y} = (Y_1, Y_2, \cdots Y_N)$
\cite{Nekrasov:2002qd, Nakajima:2003pg, NakajimaHilb}. Therefore, the
partition function can be written as   
\be
 Z(a, m, \e_1, \e_2) = \sum_{\mathbf{Y}} q^{|\mathbf{Y}|} \prod_R
 \mathbf{z}_{R,|\mathbf{Y}|}  (\mathbf{Y}; a, m, \e_1, \e_2) \ .  
\ee   
When the gauge group is a product of $M$ factors, the instanton
partition function can be written as
a sum over $M$ colored Young diagrams $\mathbf{Y}$. For
$SO/Sp$ gauge groups, we will see that the structure of the contour
integral is similar. However, the poles are no longer labeled by
a simple set of colored Young diagrams.

\subsubsection*{Equivariant index for $SO/Sp$ gauge theories}

For $Sp(N)$ the weights of the equivariant torus action on the vector spaces
$V$ and $W$ are given by 
\bea
 v \in V: & \quad v &\mapsto \textrm{diag}\, (e^{i\phi_1} , \cdots, e^{i
   \phi_n}, (1)\, , e^{-i\phi_1}, \cdots, e^{-i \phi_n}) \, v \nn \\
 w \in W: & \quad w &\mapsto \textrm{diag}\, (e^{i a_1}, \cdots, e^{i
   a_N}, e^{-i a_1}, \cdots, e^{-i a_N}) \, w \, , \nn
\ee
where $k = 2n + \chi$, with $n = [k/2]$ and $\chi \equiv k
~(\textrm{mod} ~ 2)$. The $(1)$ is inserted when $\chi = 1$ and
omitted when $\chi = 0$. The equivariant character of the universal
bundle is thus given by 
\be
 \Ch_{\mathbf{T}} (\CE_{Sp}) |_{z_1=z_2=0}  &=& \, \sum_{l=1}^N \left( e^{ia_l} + e^{-i a_l} \right)
 \\
&& - \, (e^{i\e_1} - 1)(e^{i\e_2} - 1)  \left( \sum_{i=1}^n \left(
    e^{i\phi_i -i\e_+ } + e^{-i\phi_i -i\e_+} \right) + \chi\,
  e^{-i\e_+}\right). \nn
\ee

For $SO(N)$ the weights are given by 
\bea
 v \in V: & \quad v  &\mapsto \textrm{diag} \, (e^{i\phi_1} , \cdots, e^{i
   \phi_k},  e^{-i\phi_1}, \cdots, e^{-i \phi_k}) \, v \nn \\
 w \in W: & \quad w &\mapsto \textrm{diag}\, (e^{i a_1}, \cdots, e^{i
   a_n}, (1)\, , e^{-i a_1}, \cdots, e^{-i a_n}) \, w \, , \nn
\ee
where $N = 2n + \chi$, such that $n = [N/2]$ and $\chi \equiv N~
(\textrm{mod}~ 2)$. Again, $(1)$ is inserted when $\chi = 1$ and
omitted when $\chi = 0$. The equivariant character of the universal
bundle is therefore equal to
\be
 \Ch_{\mathbf{T}} (\CE_{SO}) |_{z_1=z_2=0}  &=& \, \sum_{l=1}^n \left( e^{ia_l} + e^{-i a_l} \right)
 + \chi \\ 
&& - \,  (e^{i\e_1} - 1)(e^{i\e_2} - 1) \sum_{i=1}^k \left(
   e^{i\phi_i - i \e_+} + e^{-i\phi_i -i \e_+}\right). \nn 
\ee
Building from these expressions we can obtain the instanton partition functions
of quiver gauge theories containing matter fields in various representations.  

\subsection{Contour integrals for $SO/Sp$ matter fields}
Let us collect various contour integrands for $SO/Sp$ instanton
counting, starting with well-known expressions and ending with new
expressions for half-bifundamental hypermultiplets. 

\subsubsection*{Fundamental of $Sp(N)$}
The equivariant index for a fund. $Sp(N)$ hypermultiplet of mass $m$ is given by
\be
\textrm{Ind}_{\mathbf{T}} = &&  \int_{\IC^2} \Ch_{\mathbf{T}}
(\CE_{\rm Sp} \otimes \CL \otimes M) \, \Td_{\mathbf{T}} (\IC^2)\\
= &&   \frac{1}{(e^{i\e_1} - 1)(e^{i \e_2} - 1) } \sum_{l=1}^N \left(
  e^{ia_l + im + i \e_+} + e^{-i a_l+i m + i \e_+} \right)  \notag \\ && 
- \sum_{i=1}^n \left( e^{i\phi_i +im} + e^{-i \phi_i +im} + \chi e^{im} \right), \notag
\ee
where $k = 2n + \chi$ and $\e_+ = \frac{\e_1 + \e_2}{2}$. Here, we
tensored the universal bundle $\CE_{\rm Sp}$ by the vector space
$M\cong \IC$ on whose elements the flavor symmetry $U(1)_m$ acts by
$v \mapsto e^{im} v$.   
Since the first term in the index computes perturbative terms in the free energy, the contour integrand for the instanton contribution to the free energy is 
\be
 \mathbf{z}^N_k = m^\chi  \prod_{i=1}^n (\phi_i +m )(\phi_i -m ). 
\ee 

\subsubsection*{$Sp(N)$ gauge multiplet}

The equivariant index for an $Sp(N)$ gauge multiplet is given by 
\be 
\textrm{Ind}_{\mathbf{T}} = && - \int_{\IC^2} \Ch_{\mathbf{T}} (
\mathrm{Sym}^2 \CE_{Sp(N)}) \, \Td_{\mathbf{T}} (\IC^2)
\ee
The resulting contour integral is
\be
 Z_k &=& \frac{(-1)^n}{2^n n!} \left( \frac{\e}{\e_1 \e_2} \right)^n \left[\frac{-1}{2\e_1 \e_2 P(\e_+)} \right]^\chi  \\
 &{ }& \times \oint \left( \prod_{i=1}^n \frac{d\phi_i}{2\pi i } \right) \frac{\Delta(0) \Delta(\e)}{\Delta(\e_1) \Delta(\e_2)}
  \prod_{i=1}^n \frac{1}{(2\phi_i^2 - \e_1^2) (2 \phi_i^2 - \e_2^2)
    P(\phi_i + \e_+) P(\phi_i - \e_+)} \nn
\ee
with $\e = \e_1+\e_2$ and
\be
 P(x) &=& \prod_{l=1}^N (x^2 - a_l^2) \ , \nn \\
 \Delta(x) &=& \left[ \prod_{i=1}^n (\phi_i^2 - x^2) \right]^\chi
 \prod_{i < j} \left( (\phi_i + \phi_j)^2 - x^2 \right) \left( (\phi_i
   - \phi_j)^2 - x^2 \right) \nn \ . 
\ee
We describe a way to enumerate the poles of the above contour integral
 in appendix \ref{app:instPoles}, using what we call generalized
 Young diagrams.\footnote{The cases where $\e_1 = -\e_2$ were derived and
  discussed in \cite{Marino:2004cn, Fucito:2004gi}. But their
  derivation can not be easily generalized to the general $\e_1,
  \e_2$. } 

\subsubsection*{Bifundamental of $Sp(N_1) \times Sp(N_2)$}
The equivariant index of a bifund. $Sp(N_1) \times Sp(N_2)$ hyper with mass $m$ is given by
\be
 \Ind_{\mathbf{T}} = \int_{\IC^2} \Ch_{\mathbf{T}} (\CE_{Sp}^1 \otimes
 \CE_{Sp}^2 \otimes \CL \otimes M)\, \Td_{\mathbf{T}} (\IC^2).
\ee
where $M \cong \IC$ is acted upon by the flavor symmetry group
$U(1)_m$. Here we extended the universal bundles $\CE_{Sp}^1$ and
$\CE_{Sp}^2$ over the product $\CM_{Sp(N_1), k_1} \times \CM_{Sp(N_2),
  k_2} \times \IR^4$ by pulling them back using the respective
projection maps $\pi_i:  \CM_{Sp(N_1), k_1} \times \CM_{Sp(N_2),
  k_2} \to \CM_{Sp(N_i), k_i} $. Define
\begin{align*}
 P_1(x, a) &= \prod_l^{N_1} (x^2 - a_l^2), \\
 P_2(x, b) &= \prod_m^{N_2} (x^2 - b_m^2), \\
 \Delta(x) &= \prod_{i, j=1}^{n_1, n_2} ((\phi_i + \tilde{\phi}_j)^2 - x^2)((\phi_i - \tilde{\phi}_j)^2 - x^2), \\
 \Delta_1 (x) &= \prod_i (\phi_i^2 - x^2), \\
 \Delta_2 (x) &= \prod_i (\tilde{\phi}_i^2 - x^2),
\end{align*}
where $k_i = 2n_i + \chi_i$. 
Then the instanton contour integrand is given by
\be
 \mathbf{z}^{N_1, N_2}_{k_1, k_2}
 &=& \prod_i^{n_1} P_2 (\phi_i + m) P_2(\phi_i + m +\e ) 
 \prod_j^{n_2} P_1 (\tilde{\phi}_j + m) P_1(\tilde{\phi}_j + m+\e)  \\
 &{}& \times \prod_l^{N_1} (a_l^2 - m^2)^{\chi_2} \prod_k^{N_2} (b_k^2 - m^2)^{\chi_1} \nn \\
 &{}& \times \left(\frac{\Delta( m-\e_-) \Delta(m+\e_- )}{\Delta(m-\e_+ ) \Delta(m+\e_+ )} \right)
 \left(\frac{\Delta( m-\e_-) \Delta(m+\e_- )}{\Delta(m-\e_+ ) \Delta(m+\e_+ ) }\right)^{\chi_2} \nn \\
 &{}& \times \left(\frac{\Delta( m-\e_-) \Delta(m+\e_- )}{\Delta(m-\e_+ ) \Delta(m+\e_+ ) }\right)^{\chi_1} 
 \left(\frac{(m-\e_-) (m+\e_-) }{(m-\e_+)
    (m+\e_+) }\right)^{\chi_1 \chi_2} \nn, 
\ee
where  $\e = \e_1+\e_2$.
Note that there are additional poles that involve the mass parameter $m$. The contour prescription is to assume $\e_3 = -m-\e_+$ and $\e_4 = m-\e_+$ to have a positive imaginary value. This is the same prescription as for the massive adjoint hypermultiplet in the $\CN=2^*$ theory \cite{Bruzzo:2002xf}.

\subsubsection*{Fundamental of $SO(N)$}
The equivariant index of a fund. $SO(N)$ hypermultiplet of mass $m$ is given by
\be
\textrm{Ind}_{\mathbf{T}} = &&  \int_{\IC^2} \Ch_{\mathbf{T}}
(\CE_{\rm SO} \otimes \CL \otimes M)\, \Td_{\mathbf{T}} (\IC^2)\\
= &&  \frac{1}{(e^{i\e_1} - 1)(e^{i \e_2} - 1) } \left( \chi e^{i\e_+}+ \sum_{l=1}^n \left( e^{ia_l+im+i\e_+} + e^{-i a_l+im+i\e_+ }  \right) \right)  \nn
\\ && - \sum_{i=1}^k \left( e^{i\psi_i +im } + e^{-i \psi_i +im} \right), \notag
\ee
where $N = 2n + \chi$. The corresponding instanton integrand is
\be
 \mathbf{z}_k^N = \prod_{i=1}^k (\psi_i +m ) (\psi_i -m  ) . 
\ee

\subsubsection*{$SO(N)$ gauge multiplet}

The equivariant index for an $SO(N)$ gauge multiplet is given by 
\be 
\textrm{Ind}_{\mathbf{T}} = && - \int_{\IC^2} \Ch_{\mathbf{T}} ( \wedge^2
\CE_{SO(N)} ) \, \Td_{\mathbf{T}} (\IC^2)
\ee
The resulting contour integral is
\be
 Z_k = \frac{(-1)^{k(N+1)}}{2^k k!} \left(\frac{\e}{\e_1 \e_2} \right)^k \oint \left( \prod_{i=1}^k \frac{d\psi_i}{2\pi i} \right) \frac{ \Delta(0) \Delta(\e)}{\Delta(\e_1) \Delta(\e_2)} \frac{\psi_i^2(\psi_i^2 - \e_+^2)}{P(\psi_i + \e_+) P(\psi_i - \e_+)}
\ee
where $\e = \e_1 + \e_2$ and 
\be
 P(x) &=& x^\chi \prod_{l=1}^n \left( x^2 - a_l^2 \right) \ ,  \nn \\
 \Delta(x) &=& \prod_{i < j} \left( (\psi_i - \psi_j)^2 - x^2 \right)
 \left( (\psi_i + \psi_j)^2 - x^2 \right) \ . \nn 
\ee
When $\e_1 = - \e_2 $ the pole structure is
 simplified and described by a set of $N$-colored Young
diagrams like for the gauge group $U(N)$.\footnote{The cases where $\e_1 =
  -\e_2$ were derived and discussed in \cite{Marino:2004cn,
    Fucito:2004gi}. But their derivation can not be easily generalized
  to the general $\e_1, \e_2$. }  

\subsubsection*{Double $Sp-SO$ half-bifundamental}
Let us consider the equivariant index
\be \label{indexSpSOdoublebifund}
 \Ind_{\mathbf{T}} = \int_{\IC^2} \Ch_{\mathbf{T}} (\CE_{Sp} \otimes
 \CE_{SO} \otimes \CL \otimes M) \Td_{\mathbf{T}} (\IC^2) = \frac{
   \Ch_{\mathbf{T}} (\CE_{Sp} \otimes \CE_{SO} \otimes \CL \otimes M )}{(e^{i\e_1} - 1)(e^{i\e_2} - 1)}. 
\ee
Suppose that we add this contribution to the instanton partition
function for a quiver with an $Sp(N_1)$ and an $SO(N_2)$ node, which
quiver gauge theory do we describe?   The contour integrand
corresponding to the above equivariant index equals  
\be \label{zSpSOdoublebifund}
 \mathbf{z}_{k_1, k_2}
 &=& \prod_{l=1}^{n_2} \Delta_1 ( m \pm  b_l)  \prod_{k=1}^{N_1} \Delta_2 (  m \pm  a_k) \\
 && \times \left( \frac{\Delta(m - \e_-) \Delta(m + \e_- ) }{\Delta(m
     - \e_+ ) \Delta(m + \e_+)} \right)  \left( \frac{\Delta_2 (m - \e_-)\Delta_2 (m + \e_-)}{\Delta_2 (m - \e_+)\Delta_2 (m+\e_+)} \right)^{\chi_\phi}  \nn \\
&& \times \,
  \Delta_1 ( m)^{\chi_b} P_2 ( m)^{\chi_\phi} (m )^{\chi_b \chi_\phi },  \nn
\ee
where $\pm$ is again an abbreviation for a product over both terms.
Here $k_1 = 2n_1 + \chi_\phi$ and $N_2 = 2 n_2 + \chi_b$, whereas
\begin{align*}
 \Delta_1 (x) &= \prod_{i=1}^{n_1} (\phi_i^2 - x^2) \\
 \Delta_2 (x) &= \prod_{j=1}^{k_2} (\psi_j^2 - x^2) \\
 \Delta (x) &= \prod_{i, j=1}^{n_1, k_2} \left( (\phi_i + \psi_j )^2 - x^2 \right) \left( (\phi_i - \psi_j )^2 - x^2 \right) \\
 P_1 (x, a) &= \prod_{k=1}^{N_1}  (a_k^2 - x^2) \\
 P_2 (x, b) &= \prod_{l=1}^{n_2 } (b_l^2 - x^2). 
\end{align*}
What information can we extract from the terms in
equation~(\ref{zSpSOdoublebifund})? Notice that if we decouple the
$SO(N_2)$ gauge group, the contour integrand~(\ref{zSpSOdoublebifund}) reduces
to that for $2n_2$ 
fundamental $Sp(N_1)$ hypers with masses $m \pm b_l$. If we instead
decouple the $Sp(N_1)$ gauge group, the contour integrand reduces that for
$2N_1$ fundamental $SO(N_2)$ hypers with masses $m \pm a_k$. In other
words, the equivariant index~(\ref{indexSpSOdoublebifund}) contains twice
the degrees of freedom of a half-bifundamental coupling between the
$Sp(N_1)$ and the $SO(N_2)$ gauge group. 

Furthermore,  adding this
contour integral to the contribution for a pure $Sp(N-1)$ and a pure $SO(2N)$
theory, yields a total contour integral with as many terms in the
numerator as in the denominator. The corresponding quiver gauge
theory is therefore conformal. This implies that
equation~(\ref{indexSpSOdoublebifund}) describes two 
copies of the $Sp(N_1)-SO(N_2)$ half-bifundamental.    

If we specify to the $Sp(1)-SO(4)$ interaction we have $\chi_b = 0$, $N_1 = 1$ and $n_2 = 2$. Coupling this to a $Sp(1)$ and $SO(4)$ gauge group gives a quiver with two $Sp(1)-SO(4)$ half-bifundamentals. Explicitly, the instanton integrand is given by 
\be
 \mathbf{z}_{k_1, k_2, \mathrm{db}}^{Sp(1), SO(4)} &=& 
\, \prod_{i=1}^{n_1} \prod_{l=1}^2 \left( \phi_i^2 - ( m + b_l)^2 \right) \left( \phi_i^2 - ( m - b_l)^2 \right)  \\
 &{ }&\times \prod_{j=1}^{k_2} \left( \psi_j^2 - ( m + a)^2 \right) \left( \psi_j^2 - (  m - a)^2 \right) \nn \\
 &{ }&\times \left( \frac{\Delta(m - \e_-) \Delta(m + \e_- )
   }{\Delta(m - \e_+ ) \Delta(m + \e_+)} \right) \nn \\
 &{ }&\times \left( \prod_{l=1}^2 \left(b_l^2 -  m^2 \right)  \frac{\Delta_2 (m - \e_-)\Delta_2 (m + \e_-)}{\Delta_2 (m - \e_+)\Delta_2 (m+\e_+)} \right)^{\chi_\phi}. \nn 
\ee
Notice that there are additional poles that involve mass parameter $m$ just like in the case of the $Sp(N_1) \times Sp(N_2)$ bifundamental. 

\subsection*{$Sp(1)-SO(4)$ half-bifundamental}
The $Sp-SO$ double bifundamental
contribution~(\ref{zSpSOdoublebifund}) turns into a complete square
when we choose the mass to be $m = 0$. We therefore identify the
square-root of this double half-bifundamental contribution for $m = 0$
with the contour integral contribution of the half-bifundamental
hypermultiplet: 
\be
 \mathbf{z}_{k_1, k_2, \textrm{db}}^{Sp(N_1), SO(N_2)}(\phi, \psi, a,
 b, m =0, \e_1, \e_2) = \left( \mathbf{z}_{k_1, k_2,
     \textrm{hb}}^{Sp(N_1), SO(N_2)} (\phi, \psi, a, b) \right)^2  .  
\ee
 For the $Sp(1)-SO(4)$ gauge theory the half-bifundamental contour
 integrand is explicitly given by 
\be
\mathbf{z}_{k_1, k_2, hb}^{Sp(1), SO(4)} &=& \,
 \prod_{i=1}^{n_1} \left( \phi_i^2 - b_1^2 \right)\left( \phi_i^2 - b_2^2 \right) \prod_{j=1}^{k_2} \left( a^2 - \psi_j^2 \right) \frac{\Delta(\e_-)}{\Delta(\e_+ )}  \left( b_1 b_2   \frac{\Delta_2 (\e_-)}{\Delta_2 (\e_+)}\right)^{\chi_\phi},
\ee
where $k_1 = 2n_1 + \chi$. There're many different choices of $\pm$ signs for each of the parenthesis in the expression, but we can fix the signs by studying the decoupling limit of one of the gauge groups and compare them with single gauge group computation.

\section{Evaluating contour integrals} \label{app:instPoles}

In this appendix we explain in more detail
how to evaluate contour integrals for the $Sp(1)$ gauge
group. (Evaluating $SO(4)$ contour integrals works similarly. It is somewhat simpler because there are no fractional instantons.)  
In the case of $U(N)$ gauge groups,
 \cite{Nekrasov:2002qd, Nakajima:2003pg, NakajimaHilb} found closed 
expressions 
for the contribution of $k$ instantons in terms of
sums over Young diagrams.

Unfortunately the pole structure of $Sp(N)$ gauge
groups is much more complicated. In the literature
it has mostly only been evaluated up to three instantons,
which only requires to perform one contour integral
\cite{Shadchin:2005mx,Fucito:2004gi}.\footnote{\cite{Marino:2004cn}
  computed up to four instantons, 
but could only give an ad-hoc prescription
for which poles to include.}
It is possible to describe $Sp(N)$ instantons in terms
of orientifolding the $U(N)$ setup.
In general, there are poles involving Coulomb branch parameters,
as well as poles that just involve the deformation parameters $\e_1,
\e_2$.  
The former \emph{regular} poles are similar to the $U(N)$ instantons,
while the latter \emph{fractional} 
are new. 
In the brane engineering picture, the former can be thought of as an
instanton bound to D4-branes that are separated from the
center at positions $\pm a_n$. The 
latter can be understood as an instanton stuck at the 
orientifold brane at the center. 
When we specialize to the case of $\e_1 + \e_2 = 0$, the analysis of the poles become simpler, 
and it reduces to the ordinary colored Young diagrams plus fractional
instantons  
as in \cite{Marino:2004cn, Fucito:2004gi}. However, this method does not
work for general $\e_1, \e_2$.  

For our analysis we will use the expressions obtained for the
$Sp(1)$ gauge multiplet in \cite{Nekrasov:2004vw,Shadchin:2005mx}.
The 
$k$ instanton contribution is given by the integral over the real
axis of the variables $\phi_i$ of the integrand $z_k$.
Let $n=\lfloor\frac{k}{2}\rfloor$, $\chi = k \mod 2$ such that
$k=2n+\chi$. It is useful to define $\epsilon=\epsilon_1+\epsilon_2$ and
$\epsilon_+=\epsilon/2$.
Define
\bea
\Delta(x) &=& \prod_{i<j\leq n} ((\phi_i+\phi_j)^2-x^2)((\phi_i-\phi_j)^2-x^2)\ ,\\
P(x) &=& x^2-a^2\ .
\eea
Then $z_k$ is given by
\begin{multline}\label{zkgauge}
z_k(a,\phi,\e_1,\e_2)= 
\frac{(-1)^n}{2^{n+\chi}n!}
\frac{\epsilon^n}{\epsilon_1^n\epsilon_2^n}
\left[ \frac{1}{\epsilon_1\epsilon_2(\epsilon_+^2-a^2)}
\prod_{i=1}^n \frac{\phi_i^2(\phi_i^2-\epsilon^2)}{(\phi_i^2-\epsilon_1^2)
(\phi_i^2-\epsilon_2^2)}\right]^\chi\\
\times \frac{\Delta(0)\Delta(\epsilon)}{\Delta(\epsilon_1)\Delta(\epsilon_2)}
\prod_{i=1}^n \frac{1}{P(\phi_i-\epsilon_+)P(\phi_i+\epsilon_+)(4\phi_i^2-\epsilon_1^2)
(4\phi_i^2-\epsilon_2^2)}
\end{multline}
The contribution is obtained by integrating the $\phi_i$ along the
real axis.

Hypermultiplets in the fundamental representation never contribute any
new poles. As pointed out above, hypers in the adjoint representation of 
$Sp(N)$ do introduce new poles. This will not be covered here.

\subsection{The $\epsilon$ prescription}
To render this integral well-defined, we specify that $\epsilon_{1,2} \in \mathbb{R} +i0$.
We can then close the integrals in the upper half plane and simply evaluate all
residues. This prescription can be obtained by \eg
going to the five-dimensional theory, and requiring
that the original integral converge.

At first sight one may be worried that we need additional
information on the imaginary part of the $\epsilon$ if
we want to evaluate the integral. More precisely,
the following situation might arise: Let us take
the residue of $\phi_1$ around the pole $\phi_2 + a$
where $a$ is some linear combination of $\epsilon_{1,2}$.
If the original integrand had a pole $(\phi_3 +\phi_1-b)^{-1}$,
then the resulting expression seems to have
a pole at $\phi_2 = -\phi_3 + b -a$, which would
not longer have clearly defined imaginary part.
To see that this situation never occurs, note that
\begin{align}\label{exint}
\oint_{-\phi_3 + b -a} d\phi_2 &\oint_{\phi_2+a}
d\phi_1 \frac{1}{(\phi_1-\phi_2-a)(\phi_1 +\phi_3
  -b)}F(\phi_1,\phi_2,\phi_3) \\
&= F(-\phi_3+b , -\phi_3 + b-a,\phi_3) \notag \\
&= - \oint_{-\phi_3 + b -a} d\phi_2 \oint_{-\phi_3+b}
d\phi_1 \frac{1}{(\phi_1-\phi_2-a)(\phi_1 +\phi_3
  -b)}F(\phi_1,\phi_2,\phi_3)\ , \notag
\end{align}
\ie when evaluating the poles in both ways (as we must)
the contributions cancel. Note that this argument
is also valid if $\phi_3$ is a constant or zero.
Also note that this does not imply that all residues
vanish. The point is that if either $a$ or $b$ has
negative imaginary part, then by our $\epsilon$ prescription
we only evaluate one residue, which is therefore
not cancelled.
The upshot of this discussion is thus that whenever we
evaluate the residues, we only need to include poles
which have clearly defined positive imaginary part.

The $k$ instanton contribution $Z_k$ is thus a sum over positive poles $(\tilde \phi_i)_{i=1,\ldots,n}$
\be
Z_k = \sum_{(\tilde \phi_i)_{i=1,\ldots,k}}
\oint_{\phi_k=\tilde \phi_n} d\phi_n \ldots \oint_{\phi_1=\tilde \phi_1} d\phi_1\ z_k(a,\phi,\e_1,\e_2)\ ,
\ee
where $\tilde \phi_i$ is a linear combination of $b,\e_1,\e_2$ and possibly $\phi_j, j>i$
with positive imaginary part. Note that different poles can give the same contribution.
In what follows we give an algorithm to obtain those poles and their combinatorial weight.

\subsection{Chains}
If $k$ is even, then $\chi=0$. The possible poles
for $\phi_i$ are
\bea
\phi_i = \pm \epsilon_1/2\ , && \phi_i = \pm \epsilon_2/2 \label{root1}\\
\phi_i =  a \pm \epsilon_+ && \phi_i = -a \pm \epsilon_+ \label{root2} \\
\phi_i = \phi_j \pm \epsilon_{1,2} && \phi_i = -\phi_j \pm \epsilon_{1,2} \label{chainpole}
\eea
We will call poles as in (\ref{root1}) and (\ref{root2}) `roots'. Due to poles
of the form (\ref{chainpole}), the $\phi_i$ will take
values in chains, just as in the $U(N)$ case.
If a chain contains a root, we will call it an anchored chain.
Note that unlike the $U(N)$ case there can also
be chains that have no roots.
If $k$ is odd, then there are the additional poles
\bea
\phi_i = \pm \epsilon_1\ , && \phi_i = \pm \epsilon_2
\eea
Note that the numerator has a double zero for $\phi_i = \phi_j$
and $\phi_i= -\phi_j$.

For a more uniform treatment in the spirit of the
$U(2)$ analysis, we define the set of roots $b_l$
\be\label{broots}
b_l \in \{ a+\epsilon_+, -a+\epsilon_+, \epsilon_1/2, \epsilon_2/2 \}\
\ee
for $n$ even, and similarly for $n$ odd.
A general pole can consist of multiple chains that are independent 
of each other. It is thus possible
to describe our algorithm by using the following
toy model which only contains one root,
\be \label{zToy}
z_k = \frac{\Delta(0)\Delta(\epsilon)}{\Delta(\eo)\Delta(\et)}
\prod_{i=1}^n \frac{1}{\phi_i^2-b^2} \ .
\ee
When going back to the full $Sp(1)$ integrand, one sums
over all decompositions of $k$ into chains with different
roots.
Note that not all roots can appear as anchors for a given pole.
In particular, due to the numerator in (\ref{zkgauge})
there cannot be two chains with roots $a+\epsilon_+$
and $-a+\epsilon_+$ in the same pole.
Also note that one has to be somewhat careful when to exactly specialize
the values $b_l$. To get the correct result,
one has to take the residue of the expression with general $b_l$,
and only afterward specialize to (\ref{broots}).

\subsubsection*{Anchored poles}
Anchored poles can be described in the following way:
First, pick a `generalized Young diagram' with
$n$ boxes. A generalized Young diagram is a set of connected
boxes, one of them marked by $\times$, which we take
to be the origin.
The upper and lower edge of such a diagram
must be monotonically decreasing, \ie must slope from
the upper left to the lower right. To illustrate this,
here are some examples:
\begin{align*}
\textrm{allowed}:&& \young(\hfil,\times\hfil) && \young(\hfil\times\hfil,::\hfil) &&
\young(\hfil\hfil,:\times\hfil)\\
\textrm{not allowed}: && \young(:\times,\hfil\hfil) && \young(\hfil\hfil\times,\hfil)
&& \young(:\hfil,\hfil\times\hfil,:\hfil)
\end{align*}
This is due to the zeros of $\Delta(0)$ and $\Delta(\e)$ in the numerators,
which can only be cancelled by double poles.
To obtain the actual pole corresponding to a diagram,
we consider signed diagrams, \ie diagrams where
each box comes with a sign. Let us denotes
the signed diagrams corresponding to $Y$ by $\tilde Y$,
and let $\tilde Y_0$ be the diagram with all plus signs. The value of
a variable $\phi_i$ for a box with sign $\pm$ is
then
\be
\phi_i = \pm(b+m\eo + n\et)\ ,
\ee
where $m$ and $n$ are the horizontal and vertical positions of
the box, respectively.
The advantage of this description is that the even though 
a given signed diagram can arise from many different poles,
the numerical values of the $\phi_i$ are determined by it.
Moreover, because the integrand is invariant under 
$\phi_i \leftrightarrow -\phi_i$, the contribution of a signed
diagram is the same of the unsigned diagram, up to an overall sign.

More precisely,
each signed diagram contribues with a certain combinatorial
weight, given by the (signed) number of the poles that 
contribute, 
so that the total contribution of
$Y$ is
\be
I_Y = I_{\tilde Y_0}n_Y= I_{\tilde Y_0}\sum_{\tilde Y} n_{\tilde Y}\ .
\ee
It remains to compute the $n_{\tilde Y}$. To do this, write
down all $n!$ numbered diagrams corresponding to $\tilde Y$,
and check which ones give a contribution, \ie are
obtained during the evaluation of the contour integrals. The number $i$
in each box indicates which $\phi_i$ takes this value.
We then perform the contour integral consecutively, 
starting from $\phi_1$.
For a given $\phi_i$, three things can happen: If it
is at the origin, then it can take the value of the root $b$.
This simply means that we the pole comes from the 
factor$\phi_i-b$.
If it is not at the origin, we can connect it 
to one of its neighbors. If this neighbor has not
been evaluated yet, then the pole comes from the
factor $(\phi_i\pm \phi_j-\e_{1,2})$, $i<j$. If
$j<i$, then the pole comes from the same factor,
but we have already plugged in the value $\tilde \phi_j$
for $\phi_j$. 
Finally, if it has already been connected to
other boxes, we can also connect it to neighbors
of those boxes. 

All this is obviously
subject to the constraint that the relative
signs of the two boxes are correct, and that
the imaginary part of the pole be positive.
We can thus deduce some rules on
evaluating numbered diagrams.
In the following, an arrow over the boxes
shows which way we can connect them.
\begin{itemize}
\item The highest number $n$ must always appear in a box of positive sign in the upper right quadrant.
\item We can only connect boxes in the following way: $\overrightarrow{\young(--)}\ , \ \overleftarrow{\young(++)}\ ,
\ \overleftrightarrow{\young(-+)}$.
\item We can never connect the boxes $\young(+-)$.
\item If there is a single rightmost box, its sign must be positive.
\item If there is a single leftmost box in the negative quadrant, its sign must
be negative.
\end{itemize}
The last four rules were stated for horizontally connected boxes. 
Of course equivalent rules also hold for vertically connected ones.

\subsubsection*{Cycles}
A cycle is a chain that contains no root. Let us concentrate
for the moment on its `circular part' of length $n$. We start
by integrating out $\phi_1$, $\phi_2$, and so on, and for
$\phi_i$ we pick the pole
\be\label{eqCycles}
\phi_{i} = \sigma_{i} \phi_{i+1} + \delta_{i}\ , \qquad i=1,\ldots,n-1\ ,
\ee
where periodicity $\phi_{n+1} =\phi_1$ is implied,
and $\sigma_i = \pm 1$, $\delta_i = \epsilon_{1,2}$.
For $\phi_n$ we then pick the pole whose numerical value
is determined in such a way that $\phi_n =\sigma_n\phi_1+\delta_n$.
This value can be determined by noting that the
variable $\phi_l$ then takes the value
\be
\phi_{l} = \left(\prod_{i=1}^{l-1} \sigma_{i}\right) \phi_1 -
\sum_{j=1}^{l-1} \left(\prod_{i=j}^{l-1} \sigma_{i}\right) \delta_{j}
\ee
so that the cycle only gives a contribution if $\prod_{i=1}^n \sigma_i = -1$,
as otherwise there is either no solution, or there is a double pole
which gives no contribution.
The total pole is thus given by
\be
(\tilde \phi)= (\sigma_1\phi_2+\delta_1,\cdots,\sigma_{n-1}\phi_n+\delta_{n-1},
-\frac{1}{2} \sum_{j=1}^{n-1} \left(\prod_{i=j}^{n-1} \sigma_{i}\right) \delta_{j}
+\frac{1}{2}\delta_n )\ .
\ee
Again, this only contributes if the value of the last entry has
a well-defined positive imaginary part.

\subsection{Some examples}
Let us now explain this more explicitly for the first
low lying terms. For $k=0$ and $k=1$ there are no
integrals. For $k=2$ and $k=3$ there is just one
integral, so that one can simply sum over all
poles. This has been treated in
\cite{Shadchin:2005mx,Marino:2004cn,Fucito:2004gi}.

\subsubsection*{Four and five instantons: $n=2$}
Let us consider the case $n=2$. There are four unsigned diagrams,
\begin{align*}
&& (\tilde \phi_1,\tilde\phi_2) && n_Y\\
\young(\hfil,\times) && (b,b+\et) && 3\\
\young(\times\hfil) && (b,b+\eo) && 3\\
\young(\times,\hfil) && (b,b-\et) && -1\\
\young(\hfil\times) && (b,b-\eo) && -1
\end{align*}
To arrive at the combinatorial weights, we first write down
all signed versions of \eg the first diagram:
\be
\young(+,\bp) \qquad \young(+,\bm) \qquad \young(-,\bp) \qquad \young(-,\bm)
\ee
For each signed diagram we then write down
all possible numbered diagrams and see if they are allowed.
From the rules given above it is straightforward to see that only
\begin{align}
\young(+,\bp) && \young(1,\two)\ , \ \young(2,\one) && (\phi_2+\et,b), (b,b+\et) \\
\young(+,\bm) && \young(2,\one) && (-\phi_2+\et,b+\et)
\end{align}
are allowed. It is clear that the first diagram in the 
first line gives the same contribution as $(b,b+\et)$. The diagram can be
reduced to $(-b,b+\et)$ by the same procedure as in (\ref{exint}). Since
we pick up a minus sign in this process, the total 
combinatorial weight is $n_Y=2+1=3$.
Similarly, for the third diagram we obtain
\begin{align*}
\young(\bp,-) && \young(\two,1) && (-\phi_2+\et,b)
\end{align*}
This time we do not pick up a sign, so that $n_Y=-1$.
Let us turn to the cyclic chains. If we choose $\sigma_1 = 1$, then
$\phi_2=\frac{1}{2}(\delta_1-\delta_2)$,
which we know does not contribute. The contributions thus come from
$\sigma_1=-1$, $\sigma_2=1$ and are given by
\begin{align*}
(-\phi_2+\eo,\eo) && (-\phi_2+\et,\et) && (-\phi_2+\eo,\frac{1}{2}(\eo+\et))
&& (-\phi_2+\et,\frac{1}{2}(\eo+\et))\\
\young(-+) && \young(+,-) && \young(-+) && \young(+,-)
\end{align*}
where we have represented the first two cycles by signed
diagrams with root $0$, and the second two
by diagrams with root $\pm\frac{1}{2}(\eo-\et)$.

\subsubsection*{Six instantons: $n=3$}
Let us turn to $n=3$ now. For completeness, we have listed all
generalized Young diagrams,
their values of the $\phi$ and the combinatorial weights
in table~\ref{n3}.
\begin{table}
\begin{align*}
{\rm Diagram}&& (\phi_1,\phi_2,\phi_3) && n_Y\\
\young(\hfil,\hfil,\times) && (b,b+\et,b+2\et) && 15\\
\young(\hfil,\times\hfil) && (b,b+\et,b+\eo) && 10\\
\young(\times\hfil\hfil) && (b,b+\eo+2\eo) && 15\\
\young(\hfil,\times,\hfil) && (b,b+\et,b-\et) &&-6\\
\young(\hfil\times\hfil) && (b,b+\eo,b-\eo) && -6\\
\young(\times,\hfil,\hfil) && (b,b-\et,b-2\et) && 3\\
\young(\hfil\hfil\times) && (b,b-\eo,b-2\eo) && 3\\
\young(\times,\hfil\hfil) && (b,b-\et,b+\eo-\et) && -1\\
\young(\hfil,\hfil\times) && (b,b-\eo,b-\eo+\et) &&-1\\
\young(\hfil\times,:\hfil) && (b,b-\eo,b-\et) && 2\\
\young(\times\hfil,:\hfil) && (b,b+\eo,b+\eo-\et) && -5\\
\young(\hfil\hfil,:\times) && (b,b+\et,b+\et-\eo) && -5
\end{align*}
\caption{Generalized Young diagrams, values of $\phi$, and 
combinatorial weights\label{n3}}
\end{table}

As an example, let us explain how to obtain the combinatorial weight
for some of those cases.
Take for instance the diagram $\young(\times\hfil\hfil)$ and
write down all signed diagrams.
By the rules given above can immediately exclude all diagrams that have
a minus sign in the rightmost box.
The diagram $\tilde Y_0$ gives the same contribution as in the
$U(N)$ case and has weight 6.
The remaining three diagrams are
\begin{align*}
\young(-++) && 4 && \young(123) && (-\phi_2+\eo,b+\eo,b+2\eo) \\
&& && \young(213) && (-\phi_2+\eo,-\phi_3+2\eo,b+2\eo) \\
&& && \young(132)  && (-\phi_3+\eo,\phi_3+\eo,b+\eo) \\
&& &&\young(231) && (\phi_3+\eo,-\phi_3+\eo,b+\eo) \\
\young(--+) && 2 && \young(123) && (\phi_2+\eo,-\phi_3+\eo,b+2\eo) \\
&& && \young(213) && (-\phi_3+\eo,-\phi_3+2\eo,b+2\eo) \\
\young(+-+) && 3 && \young(123) && (b,-\phi_3+\eo,b+2\eo)\\
&& && \young(213) && (-\phi_3+\eo,b,b+2\eo)\\
&& && \young(312) && (-\phi_2+\eo,\phi_3+2\eo,b)
\end{align*}
The top diagram is exactly as in the $U(N)$ case,
so its combinatorial weight is 6. For the other diagrams,
we have listed all numbered diagrams that contribute
together with the precise pole they correspond to.
Note that when converting the poles to the form
of the table, it turns out that
minus signs appear in such
a fashion that all diagrams give positive contribution.
The total combinatorial weight of
$\young(\times\hfil\hfil)$ is thus 15.

Another example is $\young(\hfil\hfil\times)$. The rightmost box
must have a positive sign, and the leftmost box
is in a negative quadrant and must therefore have
a negative sign. This leaves just two possibilities,
\begin{align*}
\young(-++)&& 1&& \young(213) && (-\phi+\eo,-\phi_3+2\eo,b)\\
\young(--+) && 2&& \young(123) && (\phi_2+\eo,-\phi_3+\eo,b)\\
&& && \young(213) && (-\phi_3+\eo,-\phi_3+2\eo,b)
\end{align*}
which give combinatorial weight 3.

Let us briefly describe the cycles now. For 3-cycles we get
\be
(\tilde\phi_1,\tilde\phi_2,\tilde\phi_3)=(\sigma_1\phi_2+\delta_1,\sigma_2\phi_3+\delta_2,
-\frac{1}{2}(\sigma_1\sigma_2\delta_1 +\sigma_2\delta_2)+\frac{1}{2}\delta_3)
\ee
A priori, the allowed solutions are
\begin{align*}
\sigma_1=-1,\sigma_2=1,\sigma_3=1&,\delta_2=\delta_3\\
\sigma_1=1,\sigma_2=-1,\sigma_3=1&,\delta_i=\e_{1,2}\\
\sigma_1=-1,\sigma_2=-1,\sigma_3=-1&,\delta_1=\delta_3
\end{align*}
A closer analysis reveals however that all such solutions
lead to zeros in the numerator due to the factor
$\Delta(0)\Delta(\e)$.

The only contribution thus comes from
poles corresponding to
2-cycles with one attached arm.
This means take the diagrams of the
2-cycles with root $0$ and $\pm\frac{1}{2}(\eo-\et)$
and attach one box to it.
Again, we want to compute the combinatorial
weight of these configurations. Note however that
in this case $\tilde \phi$ cannot be reduced to 
its numerical values.

For the extra root 0,
note that the diagram $\young(\hfil\times\hfil)$ gives
a vanishing contribution.
We thus consider only $\young(\times\hfil\hfil)$. We have
\begin{align*}
\young(-++) && 4 && \young(123) && (-\phi_2+\eo,\eo,2\eo)\\
&& && \young(213) && (-\phi_2+\eo,-\phi_3+2\eo,2\eo)\\
&& && \young(231) && (\phi_3+\eo,-\phi_3+\eo,\eo)\\
&& && \young(132) && (-\phi_3+\eo,\phi_3+\eo,\eo)\\
\young(--+) && 2 && \young(123) && (\phi_2+\eo,-\phi_3+\eo,2\eo)\\
&& && \young(213) && (-\phi_3+\eo,-\phi_3+2\eo,2\eo)
\end{align*}
For $\young(\hfil,\times\hfil)$, the only signed diagram
is $\young(+,-+)$, which has weight 4,
\begin{align*}
\young(3,12) && \young(3,21) && \young(2,13) && \young(1,23) &\\
(-\phi_2+\eo,&\eo,\et) & (-\phi_2+\eo,-&\phi_3+\et,\et) 
& (-\phi_2+\et&,\et,\eo) & (-\phi_2+\et,-\phi_3&+\eo,\eo)
\end{align*}
and for $\young(\times\hfil,:\hfil)$ the only signed diagram $\young(-+,:-)$
has weight 2:
\begin{align*}
\young(13,:2) &&\young(23,:1) &\\
(-\phi_3+\eo,-&\phi_3+\et,\eo) & (-\phi_3+\et,-&\phi_3+\eo,\eo)
\end{align*}
All other configurations can be obtained by exchanging $\eo \leftrightarrow \et$.

For the root $\pm\frac{1}{2}(\eo-\et)$, note that no box can be
attached to the box $\frac{\eo+\et}{2}$ because of the zeros in the numerator. This leaves
just three configurations
\begin{align*}
\young(--+) && 2&& \young(123) && (\phi_2+\eo,-\phi_3+\eo, \frac{\eo+\et}{2})\\
&& && \young(213) && (-\phi_3+\eo,-\phi_3+2\eo,\frac{\eo+\et}{2}) \\
\young(-++) && 1 && \young(213) && (-\phi_2+\eo,-\phi_3+2\eo,\frac{\eo+\et}{2})\\
\young(+,-+) && 1 && \young(1,23) && (-\phi_2+\et,-\phi_3+\eo,\frac{\eo+\et}{2})
\end{align*}
and similarly for their mirror images under $\eo \leftrightarrow \et$.

To obtain the full contribution for the toy model (\ref{zToy}) we also need to include
all poles that consist of a 2-cycle and the root $b$. In total
there are thus 112 poles.

\section{$SU(2)$ Seiberg-Witten curves}\label{app:SW}

Since the discovery of Seiberg-Witten theory a few different (yet
physically equivalent) parametrizations for the $SU(2)=Sp(1)$ Seiberg-Witten
curve have appeared in the literature. Let us summarize these
different approaches here.  

First of all, the Seiberg-Witten curve for the $SU(2)$ theory coupled
to four hypermultiplets can be witten in the hyperelliptic form
\cite{Argyres:1995wt} 
\begin{align}\label{U2SWcurve}
y^2 = P_{U(2)}(w)^2 - f Q,
\end{align}
where
\begin{align*}
P_{U(2)}(w) = w^2 - \tilde{u}, \quad f= \frac{4 q_{U(2)}}{(1+q_{U(2)})^2}, \quad q_{U(2)} =
\frac{\theta_2^4(\tau_{\rm IR})}{\theta_3^4(\tau_{\rm IR})}, \quad Q = \prod_{j=1}^4 (w-\tilde{m}_j).
\end{align*}
We should be careful that the mass parameters $\tilde{m}_j$ are \emph{not}
exactly the hypermultiplet masses $\tilde{\mu}_j$. Instead,
they are 
related to the hypermultiplet masses $\tilde{\mu}_j$ 
as 
\begin{align}\label{eqn:massredef}
\tilde{m}_j = - \tilde{\mu}_j + \frac{q_{U(2)}}{2(1+q_{U(2)})} \sum_k \tilde{\mu}_k.
\end{align}
Indeed, the meromorphic Seiberg-Witten differential
\begin{align}\label{eqn:hypSWform}
\lambda = \frac{-w + \frac{q_{U(2)}}{2(1+q_{U(2)})} \sum_k
  \tilde{\mu}_k}{2 \pi i} d \log \left( \frac{P_{U(2)}(w) -
    y}{P_{U(2)}(w) + y} \right)
\end{align}
has residues $\pm \tilde{\mu_j}$ at the position $w= \tilde{\mu_j}$, so
that the parameters $\tilde{\mu_j}$ are the hypermultiplet 
masses. These are also the parameters that appear in the Nekrasov
formalism. 

Another parametrization is found by D4/NS5 brane engineering in type
IIA. The Seiberg-Witten curve (\ref{U2SWcurve}) can be rewritten in
the MQCD form \cite{Witten:1997sc}   
\begin{align}\label{eqn:SW-Witten}
(w-\tilde{m}_1)(w-\tilde{m}_2)t^2 - (1+q_{U(2)})(w^2-u)t +
q_{U(2)} (w-\tilde{m}_3)(w-\tilde{m}_4)=0
\end{align}
by the coordinate transformation $$t=-
\frac{(1+q_{U(2)})(y-P(w))}{2 (w-\tilde{m}_1)(w-\tilde{m}_2)}.$$ 
In this parametrization the meromorphic Seiberg-Witten
1-form can simply be taken to be
\begin{align}\label{eqn:wittenSWform}
\lambda = w \frac{dt}{t}. 
\end{align}
This differential differs from the one in
equation~(\ref{eqn:hypSWform}) by an exact 1-form. It has first order poles at the
positions $t \in \{0,q_{U(2)},1,\infty\}$. At $t=\infty$ and $t=0$ the
residues are  
given by the mass parameters $\{\tilde{m}_1,\tilde{m}_2\}$ and $\{\tilde{m}_3,\tilde{m}_4\}$
respectively, whereas at $t=1$ and $t=q_{U(2)}$  there is only a single nonzero residue. The mass-parameters at
$t=0,\infty$ parametrize the Cartan of the flavor symmetry group
$SU(2)$, whereas the single residue at the other two punctures is an
artifact of the chosen parametrization (that only sees a $U(1) \subset
SU(2)$). 

To restore the $SU(2)$ flavor symmetry at each of the four punctures,
Gaiotto introduced the parametrization \cite{Gaiotto:2009we} 
\begin{align}\label{eqn:SW-Gaiotto}
\tilde{w}^2 =  \left( \frac{(\tilde{m}_1+\tilde{m}_2)t^2+q_{U(2)}(\tilde{m}_3+\tilde{m}_4) }{4
  t(t-1)(t-q_{U(2)})} \right)^2 + \frac{\tilde{m}_1 \tilde{m}_2 t^2 + (1+q_{U(2)}) \tilde{u} t + q_{U(2)}
\tilde{m}_3 \tilde{m}_4}{t^2(t-1)(t-q_{U(2)})}. 
\end{align}
This is found from equation~(\ref{eqn:SW-Witten}) by eliminating the
linear term in $w$ and mapping $w\mapsto \tilde{w} = t w$. By writing
equation~(\ref{eqn:SW-Gaiotto}) in the form  
\begin{align}\label{U2SWcurvegaiottoform}
\tilde{w}^2 = \varphi_2(t),
\end{align}
it is clear that the Seiberg-Witten curve is a branched double cover
over a two-sphere $\IP^1$ with punctures at $t=0,q_{U(2)},1,\infty$. The
coefficients of $\varphi_2$ at the punctures are given by
\begin{eqnarray}\label{eqn:resphi2}
 t=0: && \varphi_2 \sim \frac{(\tilde{\mu}_3 - \tilde{\mu}_4)^2}{4} \frac{dt^2}{t^2} \\
 t=q_{U(2)}:& &  \varphi_2 \sim \frac{(\tilde{\mu}_3 +   \tilde{\mu}_4)^2}{4} \frac{dt^2}{(t-q_{U(2)})^2} \notag \\ 
 t=1: && \varphi_2 \sim \frac{(\tilde{\mu}_1 + \tilde{\mu}_2)^2}{4} \frac{dt^2}{(t-1)^2} \notag \\
 t=\infty: && \varphi_2 \sim \frac{(\tilde{\mu}_1 - \tilde{\mu}_2)^2}{4}
 \frac{dt^2}{t^2}\ .\notag 
\end{eqnarray}
So if we keep 
\begin{align}
\lambda = \tilde{w} \frac{dt}{t}
\end{align}
as the Seiberg-Witten differential (which is allowed since it only
differs from (\ref{eqn:wittenSWform}) by a shift of the flavor current
by a multiple of the gauge current), we find that its residues are
given by the square-roots of the coefficients of $\varphi_2$ in
equation~(\ref{eqn:resphi2}). 
 
The Seiberg-Witten curve in the Gaiotto parametrization
(\ref{U2SWcurvegaiottoform}) is invariant under M\"obius
transformations, and therefore completely symmetric in all four
punctures. 
This follows automatically as $\lambda$ and $\varphi_2$ are 
respectively a 1-form and a 2-form on the two-sphere~$\IP^1$.

Furthermore, an $Sp(1)$ parametrization of the Seiberg-Witten curve is
given by \cite{Argyres:1995fw}
\begin{align}\label{Sp1N=4curve}
x y^2 = P_{Sp(1)}(x)^2 - g^2 \prod (x-\mu_j^2),
\end{align}
where
\begin{align*}
 P_{Sp(1)}(x)= x(x-u)+g \prod \mu_j, \quad g^2 =
 \frac{4 \tilde{q}_{Sp(1)}}{\left(1+\tilde{q}_{Sp(1)} \right)^2},
\end{align*}
Since $Sp(1) = SU(2)$, the parametrization~(\ref{Sp1N=4curve})
should be equivalent to the curve defined
by~(\ref{U2SWcurve}). Indeed, comparing the 
discriminants 
of these two curves yields non-trivial relations between the Coulomb parameters
and  the masses, that
become trivial in the weak-coupling limit \cite{Argyres:1995fw}. 

In fact, there is a simple relation between the
$Sp(1)$ parametrization~(\ref{Sp1N=4curve}) and the $SU(2)$ curve that
Seiberg and Witten originally proposed \cite{Seiberg:1994aj}. By
expanding equation~(\ref{Sp1N=4curve}) and dividing out the constant
term in $x$, we find an equation of the form $y^2 = x^3 +
\ldots$. After some redefinitions this gives the original $SU(2)$
parametrization \cite{Argyres:1995fw}. 

Let us finally mention that by the coordinate transformation $x=v^2$,
 $y=\tilde{y}/v$ and $$ \frac{2s}{\left(1+ \tilde{q}_{Sp(1)}\right)}  =
-\frac{\tilde{y}-P_{Sp(1)}(v^2)}{(v^2-\mu_1^2)(v^2-\mu_2^2)}, $$ the
$Sp(1)$ Seiberg-Witten curve~(\ref{Sp1N=4curve}) can be
written in the Witten-form   
\begin{align}\label{Sp1N=4curve-Witten}
(v^2-\mu_1^2)(v^2-\mu_2^2) s^2 - (1+\tilde{q}_{Sp(1)})  P_{Sp(1)}(v^2) s +
\tilde{q}_{Sp(1)} (v^2-\mu_3^2)(v^2-\mu_4^2)=0. 
\end{align}
This representation of the $Sp(1)$ SW curve is a double cover over
the original $Sp(1)$ SW curve (\ref{Sp1N=4curve}), because of the
coordinate transformation $x=v^2$. 

The above curve describes the embedding of the $Sp(1)$
gauge theory in string theory using a D4/NS5 brane construction
including orientifold branes \cite{Landsteiner:1997vd}. It should be
viewed as being embedded in the covering space of the orientifold. For
each D4-brane at position $v=v^*$ there is a mirror brane at position
$v=-v^*$. The extra factor $v^2$ in the polynomial $P$ can be identified
with two extra D4-branes that are forced to sit at the orientifold at $v=0$.


\end{document}